\documentclass[12pt]{sourcebook}
\usepackage{graphicx}
\usepackage{epsfig}
\usepackage{fancyhdr}
\usepackage{multind}

\newcommand{\beq}{\begin{equation}}
\newcommand{\eeq}[1]{\label{#1}\end{equation}}
\newcommand{\eeqn}{\end{equation}}
\newenvironment{Eqnarray}{\arraycolsep 0.14em\begin{eqnarray}}{\end{eqnarray}}
\newcommand{\beqa}{\begin{Eqnarray}}
\newcommand{\eeqa}[1]{\label{#1}\end{Eqnarray}}
\newcommand{\eeqan}{\end{Eqnarray}}

\newcommand{\leqn}[1]{(\ref{#1})}
\renewcommand{\bar}[1]{\overline{#1}}
\newcommand{\etal}{{\it et al.}}

\newcommand{\eg}{{\it e.g.}}

\newcommand{\VEV}[1]{\left\langle{ #1} \right\rangle}
\newcommand{\lsim}{\mathrel{\raise.3ex\hbox{$<$\kern-.75em\lower1ex\hbox{$\sim$}}}}
\newcommand{\gsim}{\mathrel{\raise.3ex\hbox{$>$\kern-.75em\lower1ex\hbox{$\sim$}}}}

\renewcommand{\L}{{\cal L}}
\newcommand{\half}{\frac{1}{2}}

\newcommand{\BR}{\mbox{\rm BR}}
\newcommand{\ee}{e^+e^-}
\newcommand{\sstw}{\sin^2\theta_w}

\newcommand{\mz}{m_Z}

\newcommand{\mw}{m_W}
\newcommand{\mt}{m_t}
\newcommand{\mh}{m_h}

\newcommand{\msb}{{\bar{\scriptscriptstyle M \kern -1pt S}}}
\newcommand{\eps}{\epsilon}
\newcommand{\ch}[1]{\widetilde\chi^+_{#1}}
\newcommand{\chm}[1]{\widetilde\chi^-_{#1}}
\newcommand{\neu}[1]{\widetilde\chi^0_{#1}}
\newcommand{\s}[1]{\widetilde{#1}}

\makeatletter
\def\section{\@startsection{section}{0}{\z@}{5.5ex plus .5ex minus
 1.5ex}{2.3ex plus .2ex}{\large\bf}}
\def\subsection{\@startsection{subsection}{1}{\z@}{3.5ex plus .5ex minus
 1.5ex}{1.3ex plus .2ex}{\normalsize\bf}}
\def\subsubsection{\@startsection{subsubsection}{2}{\z@}{-3.5ex plus
-1ex minus  -.2ex}{2.3ex plus .2ex}{\normalsize\sl}}

\renewcommand{\@makecaption}[2]{%
   \vskip 10pt
   \setbox\@tempboxa\hbox{\small #1: #2}
   \ifdim \wd\@tempboxa >\hsize     
       \small #1: #2\par          
     \else                        
       \hbox to\hsize{\hfil\box\@tempboxa\hfil}
   \fi}

 \def\citenum#1{{\def\@cite##1##2{##1}\cite{#1}}}
\def\citea#1{\@cite{#1}{}}
 
\newcount\@tempcntc
\def\@citex[#1]#2{\if@filesw\immediate\write\@auxout{\string\citation{#2}}\fi
  \@tempcnta\z@\@tempcntb\m@ne\def\@citea{}\@cite{\@for\@citeb:=#2\do
    {\@ifundefined
       {b@\@citeb}{\@citeo\@tempcntb\m@ne\@citea\def\@citea{,}{\bf ?}\@warning
       {Citation `\@citeb' on page \thepage \space undefined}}%
    {\setbox\z@\hbox{\global\@tempcntc0\csname b@\@citeb\endcsname\relax}%
     \ifnum\@tempcntc=\z@ \@citeo\@tempcntb\m@ne
       \@citea\def\@citea{,}\hbox{\csname b@\@citeb\endcsname}%
     \else
      \advance\@tempcntb\@ne
      \ifnum\@tempcntb=\@tempcntc
      \else\advance\@tempcntb\m@ne\@citeo
      \@tempcnta\@tempcntc\@tempcntb\@tempcntc\fi\fi}}\@citeo}{#1}}
\def\@citeo{\ifnum\@tempcnta>\@tempcntb\else\@citea\def\@citea{,}%
  \ifnum\@tempcnta=\@tempcntb\the\@tempcnta\else
  {\advance\@tempcnta\@ne\ifnum\@tempcnta=\@tempcntb \else\def\@citea{--}\fi
    \advance\@tempcnta\m@ne\the\@tempcnta\@citea\the\@tempcntb}\fi\fi}
\makeatother

\newcommand{\TeV}{{\rm\,TeV}}
\newcommand{\tev}{{\rm\,TeV}}

\newcommand{\GeV}{{\rm\,GeV}}
\newcommand{\gev}{{\rm\,GeV}}

\newcommand{\sweff}{\sstw^{\mathrm{eff}}}

\newcommand{\mgl}{m_{\tilde{g}}}

\newcommand{\BC}{\begin{center}}
\newcommand{\EC}{\end{center}}

\newcommand{\slashchar}[1]{\setbox0=\hbox{$#1$}           
  \dimen0=\wd0                 
 \setbox1=\hbox{/} \dimen1=\wd1               
 \ifdim\dimen0>\dimen1                        
 \rlap{\hbox to \dimen0{\hfil/\hfil}}      
 #1                                        
 \else                                        
 \rlap{\hbox to \dimen1{\hfil$#1$\hfil}}   
 /                                         
 \fi}

\newcommand{\bit}{\begin{itemize}}         
\newcommand{\eit}{\end{itemize}}

\newcommand{\ifb}{{\rm fb}^{-1}}
\newcommand{\xfb}{\, {\rm fb}}

\newcommand{\what}{\widehat}
\newcommand{\anti}{\overline}
\newcommand{\gam}{\gamma}
\newcommand{\mev}{~\mbox{MeV}}
\newcommand{\fbi}{~\mbox{fb$^{-1}$}}

\newcommand{\rts}{\sqrt s}
\newcommand{\mpl}{M_{\rm PL}}
\newcommand{\tanb}{\tan\beta}
\newcommand{\sinbma}{\sin(\beta-\alpha)}
\newcommand{\cosbma}{\cos(\beta-\alpha)}
\newcommand{\hl}{h^0}
\newcommand{\ha}{A^0}
\newcommand{\hh}{H^0}
\newcommand{\hpm}{H^\pm}
\newcommand{\hp}{H^+}
\newcommand{\hm}{H^-}
\newcommand{\mha}{m_{\ha}}
\newcommand{\mhl}{m_{\hl}}
\newcommand{\mhh}{m_{\hh}}
\newcommand{\mhpm}{m_{\hpm}}
\newcommand{\mhmax}{m_h^{\rm max}}

\newcommand{\mstopa}{M_{\widetilde t_1}}
\newcommand{\mstopb}{M_{\widetilde t_2}}
\newcommand{\SM}{Standard Model}
\newcommand{\phm}{\phantom{-}}

\def\D0{D\O\ \hskip-0.5mm}

\newcommand{\erlpm}{\ensuremath{\widetilde{e}_{R,L}^\pm}}
\newcommand{\chionepm}{\ensuremath{\widetilde{\chi}_1^\pm}}
\newcommand{\asneul}{\ensuremath{\overline{\widetilde{\nu}}_\ell}}
\newcommand{\chionez}{\ensuremath{\widetilde{\chi}_1^0}}
\newcommand{\chitwopm}{\ensuremath{\widetilde{\chi}_2^\pm}}
\newcommand{\elpm}{\ensuremath{\widetilde{e}_L^\pm}}
\newcommand{\erpm}{\ensuremath{\widetilde{e}_R^\pm}}
\newcommand{\murpm}{\ensuremath{\widetilde{\mu}_R^\pm}}
\newcommand{\sneue}{\ensuremath{\widetilde{\nu}_e}}
\newcommand{\murlpm}{\ensuremath{\widetilde{\mu}_{R,L}^\pm}}
\newcommand{\mur}{\ensuremath{\widetilde{\mu}_R}}
\newcommand{\mul}{\ensuremath{\widetilde{\mu}_L}}
\newcommand{\chitwoz}{\ensuremath{\widetilde{\chi}_2^0}}
\newcommand{\chionep}{\ensuremath{\widetilde{\chi}_1^+}}
\newcommand{\chionem}{\ensuremath{\widetilde{\chi}_1^-}}
\newcommand{\topone}{\ensuremath{\widetilde{t}_1}}
\newcommand{\toptwo}{\ensuremath{\widetilde{t}_2}}
\newcommand{\tauonepm}{\ensuremath{\widetilde{\tau}_1^\pm}}
\newcommand{\tautwopm}{\ensuremath{\widetilde{\tau}_2^\pm}}
\newcommand{\etmiss}{\slashchar{E}_T}
\def\dofigs#1#2#3{\centerline{\epsfxsize=#1\epsfbox{#2}%
   \hfil\epsfxsize=#1\epsfbox{#3}}}

\newcommand{\mhsm}{m_{h_{\rm SM}}}

\begin{document}

\pagestyle{fancy}
\thispagestyle{empty}

\setcounter{footnote}{0}
\renewcommand{\thefootnote}{\fnsymbol{footnote}}

\begin{flushright}
{\small
  BNL--52627, 
           CLNS 01/1729, 
           FERMILAB--Pub--01/058-E,  \\
           LBNL--47813,  
           SLAC--R--570,  
           UCRL--ID--143810--DR\\   
                 LC--REV--2001--074--US \\
              hep-ex/0106056\\
             June 2001}  
\end{flushright}

\bigskip
\begin{center}
{\bf\LARGE
 Linear Collider Physics Resource Book\\[1ex] for Snowmass 2001\\[4ex]
Part 2: Higgs and Supersymmetry Studies}
\\[6ex]
{\it American Linear Collider Working Group}
\footnote{Work supported in part by the US Department of Energy under
contracts DE--AC02--76CH03000,
DE--AC02--98CH10886, DE--AC03--76SF00098, DE--AC03--76SF00515, and
W--7405--ENG--048, and by the National Science Foundation under
contract PHY-9809799.}
\medskip
\end{center}

\vfill

\vfill

\begin{center}
{\bf\large
Abstract }
\end{center}

This Resource Book reviews the physics opportunities of a next-generation
$e^+e^-$ linear collider and discusses options for the experimental program.
Part 2 reviews the possible experiments on Higgs bosons and supersymmetric
particles that can be done at a linear collider.

\vfill
\vfill

\newpage
\emptyheads
\blankpage \thispagestyle{empty}
\fancyheads
\emptyheads

 \frontmatter\setcounter{page}{1}

\hbox to\hsize{\null}
\thispagestyle{empty}
\vfill
\begin{figure}[hp]
\begin{center}
\epsfig{file=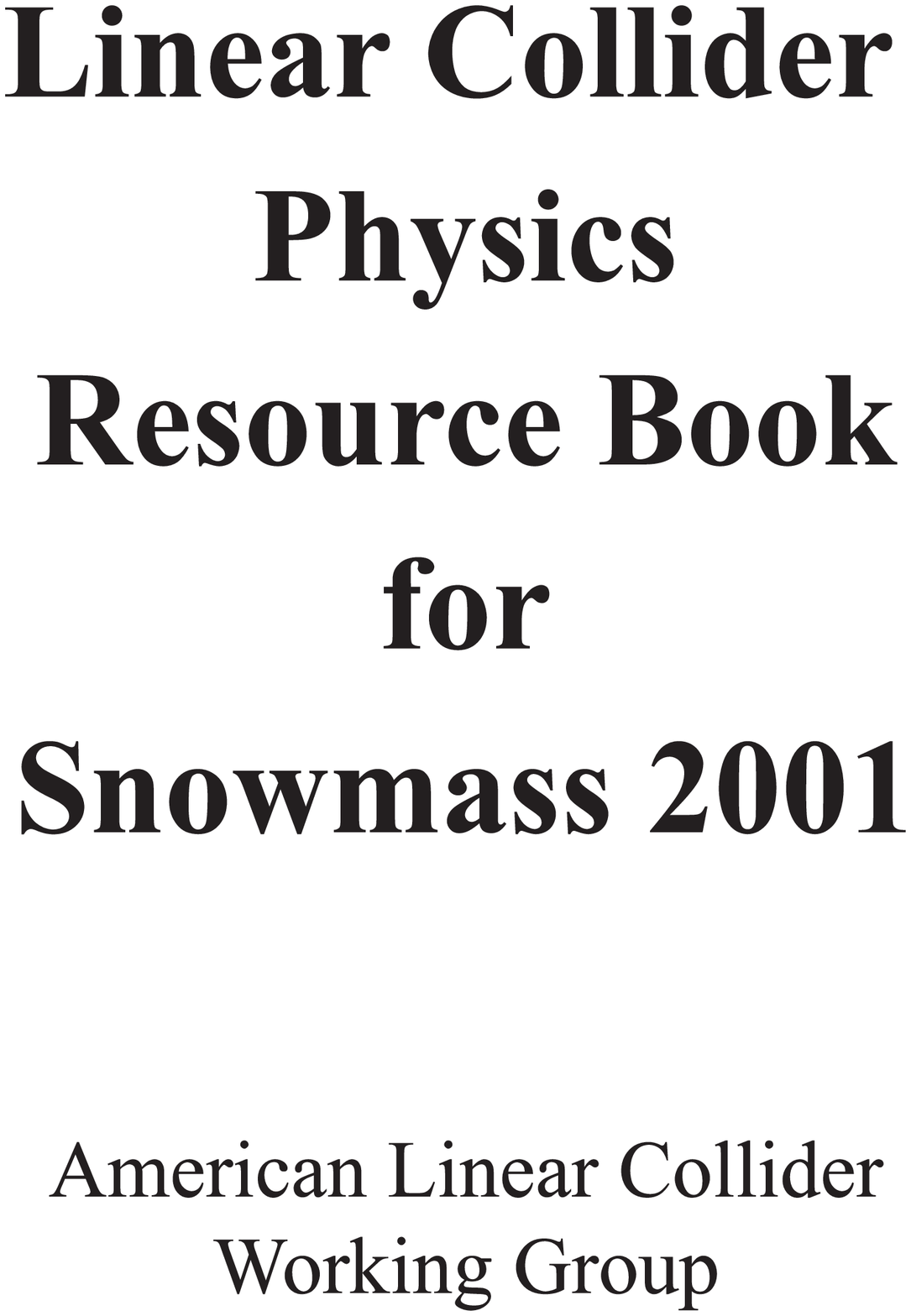,height=6.5in}
\end{center}
\end{figure}
\vfill
\newpage

\begin{center} 
           BNL--52627, 
           CLNS 01/1729, 
           FERMILAB--Pub--01/058-E,  \\
           LBNL--47813,  
           SLAC--R--570,  
           UCRL--ID--143810--DR\\[2ex]   
            LC--REV--2001--074--US \\[2ex]
            June 2001  
\end{center}

\vfill

\begin{center}
\fbox{\parbox{5in}
{This document, and the material and data contained therein, was developed
under sponsorship of the United States Government.  Neither the United
States nor the Department of Energy, nor the Leland Stanford Junior University, 
nor their employees, nor their respective contractors, subcontractors, or
their employees, makes any warranty, express or implied, or assumes any
liability of responsibility for accuracy, completeness or usefulness of any
information, apparatus, product or process disclosed, or represents that its
use will not infringe privately owned rights.  Mention of any product, its
manufacturer, or suppliers shall not, nor is intended to imply approval,
disapproval, or fitness for any particular use.  A royalty-free,
nonexclusive right to use and disseminate same for any purpose
whatsoever, is expressly reserved to the United States and the
University.
}}\end{center}

\vfill

\noindent 
Cover:  Events of $\ee \to Z^0 h^0$, simulated with the Large linear collider
detector \hfill \break described in Chapter 15.
Front cover:  $h^0 \to \tau^+\tau^-$, $Z^0 \to b\bar b$.
Back cover:  $h^0 \to b\bar b$, $Z^0 \to \mu^+\mu^-$.

\vfill

\noindent
Typset in \LaTeX\ by S. Jensen.

\vfill
\noindent
Prepared for the Department of Energy under contract number DE--AC03--76SF00515
by Stanford Linear Accelerator Center, Stanford University, Stanford, California.
Printed in the United State of America.  Available from National Technical
Information Services, US Department of Commerce, 5285 Port Royal Road,
Springfield, Virginia 22161.

  \thispagestyle{empty}

  \begin{center}
{\Large American Linear Collider Working Group}
\end{center}
 
\bigskip\bigskip
\begin{center}

T. Abe$^{52}$,
N.~Arkani-Hamed$^{29}$,
D.~Asner$^{30}$,
H.~Baer$^{22}$,
J.~Bagger$^{26}$,    
C.~Balazs$^{23}$,
C.~Baltay$^{59}$,           
T.~Barker$^{16}$,
T.~Barklow$^{52}$,   
J.~Barron$^{16}$
% ...

\end{center}

\newpage

\centerline{$^{1}$ Argonne National Laboratory, Argonne, IL 60439}
\centerline{$^{2}$ Universitat Autonoma de Barcelona, E-08193 Bellaterra,Spain}
\centerline{$^{3}$ University of British Columbia, Vancouver, BC V6T 1Z1, Canada}
\centerline{$^{4}$ Brookhaven National Laboratory, Upton, NY 11973}
\centerline{$^{5}$ Budker INP, RU-630090 Novosibirsk, Russia}
\centerline{$^{6}$ University of California, Davis, CA 95616}
\centerline{$^{7}$ University of California, Riverside, CA 92521}
\centerline{$^{8}$ University of California at San Diego, La Jolla, CA  92093}
\centerline{$^{9}$ University of California, Santa Cruz, CA 95064}
\centerline{$^{10}$ California Institute of Technology, Pasadena, CA 91125}
\centerline{$^{11}$ Carleton University, Ottawa, ON K1S 5B6, Canada}
\centerline{$^{12}$ Universidad Catolica de Chile, Chile}
\centerline{$^{13}$ CERN, CH-1211 Geneva 23, Switzerland}
\centerline{$^{14}$ University of Chicago, Chicago, IL 60637}
\centerline{$^{15}$ Chonbuk National University, Chonju 561-756, Korea}
\centerline{$^{16}$ University of Colorado, Boulder, CO 80309}
\centerline{$^{17}$ Cornell University, Ithaca, NY  14853}
\centerline{$^{18}$ DESY, D-22063 Hamburg, Germany}
\centerline{$^{19}$ DESY, D-15738 Zeuthen, Germany}
\centerline{$^{20}$ Fairfield University, Fairfield, CT 06430}
\centerline{$^{21}$ Fermi National Accelerator Laboratory, Batavia, IL 60510}
\centerline{$^{22}$ Florida State University, Tallahassee, FL 32306}
\centerline{$^{23}$ University of Hawaii, Honolulu, HI 96822}
\centerline{$^{24}$ Indian Institute of Science, Bangalore, 560 012, India}
\centerline{$^{25}$ Indiana University, Bloomington, IN 47405}
\centerline{$^{26}$ Johns Hopkins University, Baltimore, MD 21218}
\centerline{$^{27}$ Kansas State University, Manhattan, KS 66506}
\centerline{$^{28}$ Kyoto University,  Kyoto 606, Japan}
\centerline{$^{29}$ Lawrence Berkeley National Laboratory, Berkeley, CA 94720}
\centerline{$^{30}$ Lawrence Livermore National Laboratory, Livermore, CA 94551}
\centerline{$^{31}$ Universit\"at Leipzig, D-04109 Leipzig, Germany}
\centerline{$^{32}$ Ludwigs-Maximilians-Universit\"at, M\"unchen, Germany}
\centerline{$^{32a}$ Manchester University, Manchester M13~9PL, UK}
\centerline{$^{33}$ Centre de Physique Theorique, CNRS, F-13288 Marseille, France}
\centerline{$^{34}$ University of Massachusetts, Amherst, MA 01003}
\centerline{$^{35}$ Massachussetts Institute of Technology, Cambridge, MA 02139}
\centerline{$^{36}$ Max-Planck-Institut f\"ur Physik, M\"unchen, Germany}
\centerline{$^{37}$ University of Michigan, Ann Arbor MI 48109}
\centerline{$^{38}$ State University of New York, Buffalo, NY 14260}
\centerline{$^{39}$ State University of New York, Stony Brook, NY 11794}
\centerline{$^{40}$ Northern Illinois University, DeKalb, IL 60115}
\centerline{$^{41}$ University of Notre Dame, Notre Dame, IN 46556}
\centerline{$^{42}$ University of Oregon, Eugene, OR 97403}
\centerline{$^{43}$ Oxford University, Oxford OX1 3RH, UK}
\centerline{$^{44}$ Paul Scherrer Institut, CH-5232 Villigen PSI, Switzerland}
\centerline{$^{45}$ University of Pennsylvania, Philadelphia, PA 19104}
\centerline{$^{46}$ Pennsylvania State University, Mont Alto, PA 17237}
\centerline{$^{47}$ Perkins-Elmer Bioscience, Foster City, CA 94404}
\centerline{$^{48}$ Purdue University, West Lafayette, IN 47907}
\centerline{$^{49}$ RWTH Aachen, D-52056 Aachen, Germany}
\centerline{$^{50}$ University of Rochester, Rochester, NY 14627}
\centerline{$^{51}$ Southern Methodist University, Dallas, TX 75275}
\centerline{$^{52}$ Stanford Linear Accelerator Center, Stanford, CA 94309}
\centerline{$^{53}$ Texas A\&M University, College Station, TX 77843}
\centerline{$^{54}$ Tokoku University, Sendai 980, Japan}
\centerline{$^{55}$ University of Wisconsin, Madison, WI  53706}
\centerline{$^{57}$ Uniwersytet Warszawski, 00681 Warsaw, Poland}
\centerline{$^{57}$ Wayne State University, Detroit, MI 48202}
\centerline{$^{58}$ Universit\"at W\"urzburg, W\"urzburg 97074, Germany}
\centerline{$^{59}$ Yale University, New Haven, CT 06520}

\vfill

\noindent
Work supported in part by the US Department of Energy under
contracts DE--AC02--76CH03000,
DE--AC02--98CH10886, DE--AC03--76SF00098, DE--AC03--76SF00515, and
W--7405--ENG--048, and by the National Science Foundation under
contract PHY-9809799.

  \blankpage  \thispagestyle{empty}

\setcounter{chapter}{2}

\setcounter{page}{71} \thispagestyle{empty}
\renewcommand{\thepage}{\arabic{page}}
\begin{center}\begin{large}{\Huge \sffamily Sourcebook for Linear Collider Physics}
\end{large}\end{center}

\emptyheads
\blankpage \thispagestyle{empty}
\fancyheads

\chapter{Higgs Bosons at the Linear Collider}
\fancyhead[RO]{Higgs Bosons at the Linear Collider}

\section{Introduction}
\label{seca}

This chapter shows how a linear collider (LC) can contribute to our
understanding of the Higgs sector through detailed studies
of the physical Higgs boson state(s).
Although this subject has been reviewed several times in the
past~\cite{hhg,Gunion:1996cn,snow96,Murayama:1996ec,Accomando:1998wt},
there are at least two reasons to revisit the subject.
First, the completion of the LEP2 Higgs search, together with earlier
precise measurements from SLC, LEP, and the Tevatron, gives us a clearer
idea of what to expect.
The simplest explanations of these results point to a light Higgs boson
with (nearly) standard couplings to $W$ and~$Z$.
The key properties of such a particle can be investigated
with a  500~GeV LC.
Second, the luminosity expected from the LC is now higher:
200--300~fb$^{-1}$yr$^{-1}$ at $\sqrt{s}=500$~GeV, and
300--500~fb$^{-1}$yr$^{-1}$ at $\sqrt{s}=800$~GeV.
Consequently, several tens of thousands of Higgs bosons should be
produced in each year of operation.
With such samples, several measurements become more feasible, and the
precision of the whole body of expected results becomes such as to lend insight
not only into the nature of the Higgs boson(s), but also into the dynamics
of higher scales.

There is an enormous literature on the Higgs boson and, more generally,
on possible mechanisms of electroweak symmetry breaking.
It is impossible to discuss all of it here.
To provide a manageable, but nevertheless illustrative, survey of
LC capabilities, we focus mostly on the Higgs boson of the Standard
Model~(SM), and on the Higgs bosons of the minimal supersymmetric
extension of the~SM (MSSM).
Although this choice is partly motivated by simplicity, a stronger
impetus comes from the precision data collected over the past few years,
and some other related considerations.

The SM, which adds to the observed particles a single complex doublet of scalar
fields, is economical.
It provides an impressive fit to the precision data.
Many extended models of electroweak symmetry breaking possess a limit,
called the decoupling limit, that is experimentally almost
indistinguishable from the SM.
These models agree with the data equally well, and even away from the
decoupling limit they usually predict a weakly coupled Higgs boson
whose mass is at most several hundred~GeV.
Thus, the SM serves as a basis for discussing the Higgs phenomenology of
a wide range of models, all of which are compatible with experimental
constraints.

The SM suffers from several theoretical problems, which are either
absent or less severe with weak-scale supersymmetry.
The Higgs sector of the MSSM is a constrained two Higgs doublet model,
consisting of two CP-even Higgs bosons, $\hl$ and $\hh$, a CP-odd
Higgs boson, $\ha$, and a charged Higgs pair, $\hpm$.
The MSSM is especially attractive because the superpartners modify
the running of the strong, weak, and electromagnetic gauge couplings in
just the right way as to yield unification  at about $10^{16}$~GeV
\cite{susyguts}.
For this reason, the MSSM is arguably the most compelling extension of
the SM.
This is directly relevant to Higgs phenomenology, because in the MSSM
a theoretical bound requires that the lightest CP-even Higgs
boson $\hl$ has a mass less than 135~GeV.
(In non-minimal supersymmetric models, the bound can be relaxed to
around 200~GeV.)
Furthermore, the MSSM offers, in some regions of parameter space,
very non-standard Higgs phenomenology, so the full range of 
possibilities in the MSSM
can be used to indicate how well the LC performs in non-standard
scenarios.
Thus, we use the SM to show how the LC fares when there is only one
observable Higgs boson, and the MSSM to illustrate how extra
fields can complicate the phenomenology.
We also use various other models to illustrate important exceptions
to conclusions that would be drawn from these two models alone.

The rest of this chapter is organized as follows.
Section~\ref{secb} gives, in some detail, the argument that one should
expect a weakly coupled Higgs boson with a mass that is probably
below about 200~GeV.
In Section~\ref{secc}, we summarize the theory of the Standard Model Higgs
boson.
In Section~\ref{secd}, we review the expectations for Higgs discovery and the
determination of Higgs boson properties at the Tevatron and LHC.
In Section~\ref{sece}, we introduce the Higgs sector of the minimal
supersymmetric extension of the Standard Model (MSSM) and discuss its
theoretical properties.
The present direct search limits are reviewed, and expectations for
discovery at the Tevatron and LHC are described in Section~\ref{secf}.
In Section~\ref{secg}, we treat the theory of the non-minimal Higgs sector
more generally.
In particular, we focus on the decoupling limit, in which the properties
of the lightest Higgs scalar are nearly identical to those of the
Standard Model Higgs boson, and discuss how to distinguish the two.
We also discuss some non-decoupling exceptions to the usual decoupling
scenario.

Finally, we turn to the program of Higgs measurements that can be
carried out at the LC, focusing on $e^+e^-$ collisions at
higher energy, but also including
material on the impact of Giga-Z operation and $\gamma\gamma$ collisions.
The measurement of Higgs boson properties in $e^+e^-$
collisions is outlined in Section~\ref{sech}.
This includes a survey of the measurements that can be made for a
SM-like Higgs boson for all masses up to 500~GeV.
We also discuss measurements of the extra Higgs bosons that appear
in the MSSM.
Because the phenomenology of decoupling limit mimics, by definition,
the SM Higgs boson, we emphasize how the precision that stems from high
luminosity helps to diagnose the underlying dynamics.
In Section~\ref{seci}, we outline the impact of Giga-Z operation
on constraining and exploring various scenarios.
In Section~\ref{secj}, the most important gains from $\gam\gam$
collisions are reviewed.
Finally, in Section~\ref{seck}, we briefly discuss the 
case of a Higgs sector containing triplet Higgs representations
and also consider the Higgs-like particles that can arise if the
underlying assumption of a weakly coupled elementary Higgs sector is
{\it not} realized in Nature.

\section{Expectations for electroweak symmetry breaking}
\label{secb}

With the recent completion of experimentation at the LEP collider,
the Standard Model of particle physics appears close to final
experimental verification.  After more than ten
years of precision measurements of
electroweak observables at LEP, SLC and the Tevatron, no 
definitive departures from Standard Model predictions have been found
\cite{precision}.  In some cases, theoretical predictions have been
checked with an accuracy of one part in a thousand or better.
However, the dynamics responsible for electroweak symmetry breaking 
has not yet been directly identified.  Nevertheless, this dynamics  
affects predictions for currently observed electroweak processes at
the one-loop quantum level.  Consequently, the analysis of precision
electroweak data can already provide some useful constraints on the nature of
electroweak symmetry breaking dynamics.

In the minimal Standard Model, electroweak symmetry
breaking dynamics arises via a
self-interacting complex doublet of scalar fields, which consists of four
real degrees of freedom.  Renormalizable interactions are
arranged in such a way that the neutral component of the scalar doublet
acquires a vacuum expectation value, $v=246$~GeV, which sets the scale
of electroweak symmetry breaking.
Hence, three massless Goldstone bosons 
are generated that are absorbed by the $W^\pm$ and $Z$, 
thereby providing the resulting massive gauge
bosons with longitudinal components.  
The fourth scalar degree of freedom that remains in the physical spectrum
is the CP-even neutral Higgs boson of the Standard Model.
It is further assumed in the Standard Model that the scalar doublet also 
couples to fermions through
Yukawa interactions.  After electroweak symmetry breaking, these interactions
are responsible for the generation of quark and charged lepton masses.

The global analysis of electroweak
observables provides a superb fit to the Standard Model predictions.
Such analyses take the Higgs mass as a free parameter.  The 
electroweak observables depend logarithmically
on the Higgs mass through its one-loop effects.
The accuracy of the current data (and the reliability of the
corresponding theoretical computations) already provides a significant
constraint on the value of the Higgs mass. 
In \cite{erler,degrassi}, the non-observation of the Higgs boson is
combined with the constraints of the global precision electroweak
analysis to yield $m_{h_{\rm SM}}\lsim 205$--230~GeV at 95\% CL (the quoted
range reflects various theoretical choices in the
analysis).  Meanwhile, direct searches
for the Higgs mass at LEP achieved a 95\% CL limit of
$m_{h_{\rm SM}}>113.5$~GeV.\footnote{The LEP experiments presented evidence for
a Higgs mass signal at a mass of $m_{h_{\rm SM}}=115.0^{+1.3}_{-0.9}$~GeV, with
an assigned significance of $2.9\sigma$ \cite{igo}.   Although
suggestive, the data are not significant enough to warrant a claim of a
Higgs discovery.}

One can question the significance of these results.  After all,
the self-interacting scalar field is only one  model of
electroweak symmetry breaking; other approaches,
based on very different dynamics, are also possible.
For example, one can introduce new fermions and new forces, in
which the Goldstone bosons are a consequence of the strong
binding of the new fermion fields \cite{techni}.  Present
experimental data are not sufficient to identify with certainty the
nature of the dynamics responsible for electroweak symmetry breaking.
Nevertheless, one can attempt to classify alternative scenarios and
study the constraints of the global precision electroweak fits and the
implications for phenomenology at future colliders.  Since electroweak
symmetry dynamics must affect the one-loop corrections to
electroweak observables, the constraints on alternative approaches can
be obtained by generalizing the global precision electroweak fits to
allow for new contributions at one-loop.  These enter primarily
through corrections to the self-energies of the gauge bosons (the
so-called ``oblique'' corrections).  Under the assumption that any new
physics is characterized by a new mass scale $M\gg m_Z$, one can
parameterize the leading oblique corrections by three constants, 
$S$, $T$, and $U$, first introduced by Peskin and 
Takeuchi~\cite{takeuchi}.   In
almost all theories of electroweak symmetry breaking dynamics, $U\ll
S$, $T$, so it is sufficient to consider a global electroweak fit in
which $m_{h_{\rm SM}}$, $S$ and $T$ are free parameters.  (The zero of the
$S$--$T$ plane must be defined relative to some fixed value of the
Higgs mass, usually taken to be 100~GeV.)   New electroweak symmetry
breaking dynamics could generate non-zero values of $S$ and $T$, while
allowing for a much heavier Higgs mass (or equivalent).  Various
possibilities have been recently classified by 
Peskin and Wells \cite{peskinwells}, who
argue that any dynamics that results in a significantly heavier Higgs
boson should also generate new experimental signatures at the TeV scale
that can be studied at the LC, either directly by producing new
particles or indirectly by improving precision measurements of electroweak
observables.

In this chapter, we mainly consider the simplest possible
interpretation of the precision electroweak data, namely, that  there 
exists a light weakly coupled Higgs boson.  Nevertheless, this still
does not fix the theory of electroweak symmetry breaking.  It is easy
to construct extensions of the scalar boson dynamics and generate
non-minimal Higgs sectors.  Such theories can contain charged Higgs
bosons and neutral Higgs bosons of opposite (or indefinite) CP-quantum
numbers.  Although some theoretical constraints exist,
there is still considerable freedom in constructing models which
satisfy all known experimental constraints.  Moreover, in most
extensions of the Standard Model, there exists a large range of
parameter space in which the properties of the lightest Higgs scalar 
are virtually indistinguishable from those of the Standard Model Higgs
boson.  One of the challenges of experiments at future colliders, once
the Higgs boson is discovered, is to see whether there
are any deviations from the 
properties expected for the Standard Model Higgs boson.

Although the Standard Model provides a remarkably successful description of the
properties of the quarks, leptons and spin-1 gauge bosons at energy scales
of ${\cal O}(100)$~GeV and below, the Standard Model is not the
ultimate theory of the fundamental particles and their interactions.
At an energy scale above the Planck scale, $\mpl\simeq 10^{19}$~GeV, quantum
gravitational effects become significant
and the Standard Model must
be replaced by a more fundamental theory that incorporates
gravity.
It is also possible that the Standard Model breaks down at some 
energy scale, $\Lambda$, below the Planck scale.
In this case, the Standard Model degrees of freedom are no longer
adequate for describing the physics above $\Lambda$ and new physics
must enter. 
Thus, the Standard Model is not a {\it fundamental} theory;
at best, it is an {\it effective field theory}~\cite{EFT}.
At an energy scale below $\Lambda$, the Standard Model (with
higher-dimension operators to parameterize the new physics at the scale
$\Lambda$) provides an extremely good description of all observable
phenomena.

\begin{figure}[hbt]
  \centering
\psfig{file=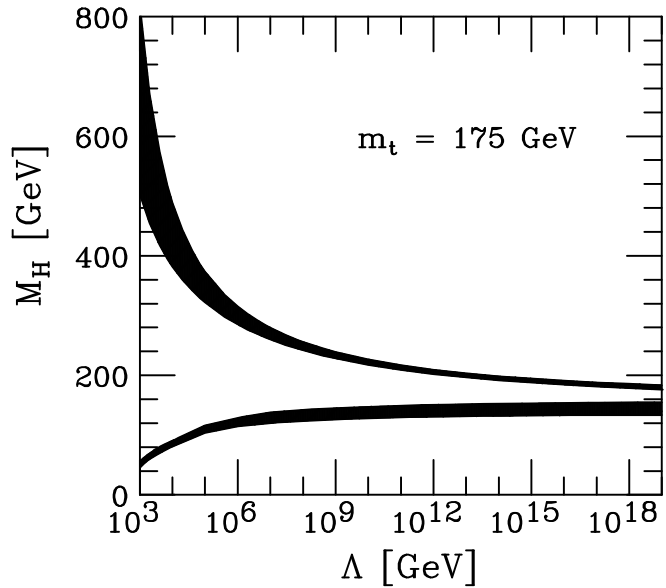,width=9cm}
  \caption[0]{\label{trivial} The upper \protect\cite{hambye} and the
 lower \protect\cite{quiros} Higgs mass bounds as a function of the
energy scale $\Lambda$ at which the Standard Model breaks down,
assuming $m_t=175$~GeV and $\alpha_s(m_Z)=0.118$.  The shaded
areas above reflect the theoretical uncertainties in the
calculations of the Higgs mass bounds.  This figure is taken from
\protect\cite{Riesselmann}.}
\end{figure}

An essential question that future experiments
must address is: what is the minimum scale $\Lambda$ at which new
physics beyond the Standard Model must enter?  The answer to this
question depends on the value of
the Higgs mass, $m_{h_{\rm SM}}$.  If $m_{h_{\rm SM}}$ is too large, then the
Higgs self-coupling blows up at some scale $\Lambda$ below the
Planck scale \cite{hambye}.  If $m_{h_{\rm SM}}$ is too small, then the Higgs
potential develops
a second (global) minimum at a large value of the scalar field of order
$\Lambda$ \cite{quiros}.  Thus, new physics must enter at a scale
$\Lambda$
or below in order that the true minimum of the theory correspond to the
observed SU(2)$\times$U(1) broken vacuum with $v=246$~GeV for scales above
$\Lambda$. Thus, given a value of
$\Lambda$, one can compute the minimum and maximum Higgs mass allowed.
Although the arguments just given are based on perturbation
theory, it is possible to repeat the analysis of the Higgs-Yukawa
sector non-perturbatively~\cite{h:nonpert}.
These results are in agreement  with the perturbative estimates.
The results of this analysis (with shaded bands indicating the
theoretical uncertainty of the result) are illustrated in Fig.~\ref{trivial}.

Although the Higgs mass range 130~GeV~$\lsim m_{h_{\rm SM}}\lsim 180$~GeV appears
to permit an effective Standard Model
that survives all the way to the Planck scale, most theorists consider
such a possibility unlikely.  This conclusion is based on
the ``naturalness'' \cite{naturally} argument as follows.  In
an effective field theory, all parameters of the low-energy theory
({\it i.e.},  masses and couplings) are calculable in terms of parameters
of a more fundamental theory that
describes physics at the energy scale $\Lambda$.  All
low-energy couplings and fermion masses are logarithmically sensitive to
$\Lambda$.  In contrast, scalar squared-masses are {\it quadratically}
sensitive to $\Lambda$.  The Higgs
mass (at one-loop) has the following heuristic form:
\begin{equation} \label{natural}
m_h^2= (m_h^2)_0+{cg^2\over 16\pi^2}\Lambda^2\,,
\end{equation}
where $(m_h^2)_0$ is a parameter of the fundamental theory and $c$ is a
constant, presumably of ${\cal O}(1)$, that depends on the physics of the
low-energy effective theory.  The ``natural'' value
for the scalar squared-mass is $g^2\Lambda^2/16\pi^2$.  Thus, the expectation
for $\Lambda$ is
\begin{equation} \label{tevscale}
\Lambda\simeq {4\pi m_h\over g}\sim {\cal O}(1~{\rm TeV})\,.
\end{equation}
If $\Lambda$ is significantly larger than 1~TeV 
then the only way
for the Higgs mass to be of order the scale of electroweak symmetry
breaking is to have an ``unnatural'' cancellation between the two terms
of Eq.~(\ref{natural}).
This seems highly unlikely given that the two terms of
Eq.~(\ref{natural}) have completely different origins.  

An attractive theoretical framework that incorporates weakly coupled Higgs
bosons and satisfies the constraint of Eq.~(\ref{tevscale}) is that of
``low-energy'' or ``weak-scale'' 
supersymmetry \cite{Nilles84,Haber85}.
In this framework, supersymmetry is
used to relate fermion and boson masses and interaction strengths.  
Since fermion masses are only
logarithmically sensitive to $\Lambda$, boson masses will exhibit the
same logarithmic sensitivity if supersymmetry is exact.  Since no
supersymmetric partners of Standard Model particles have yet been
found, supersymmetry cannot be an exact symmetry of nature.
Thus, $\Lambda$ should be identified 
with the supersymmetry breaking scale.  The
naturalness constraint of Eq.~(\ref{tevscale}) is still relevant.   It
implies that the scale of supersymmetry
breaking should not be much larger than 1~TeV, to preserve the
naturalness of scalar masses.  The supersymmetric extension
of the Standard Model would then replace the Standard Model as the
effective field theory of the TeV scale.  
One advantage of the supersymmetric
approach is that the effective low-energy supersymmetric theory {\it
can} be valid all the way up to the Planck scale, while still being
natural!  The unification of the three gauge couplings at an energy
scale close to the Planck scale, which does not occur in the Standard
Model, is seen to occur in the minimal supersymmetric extension of the
Standard Model, and provides an additional motivation for seriously 
considering the low-energy supersymmetric framework~\cite{susyguts}.
However, the fundamental origin of supersymmetry breaking
is not known at present.  Without a fundamental theory of
supersymmetry breaking, one ends up with an effective low-energy
theory characterized by over 100 unknown parameters that in principle
would have to be measured by experiment.
This remains one of the main stumbling
blocks for creating a truly predictive model of fundamental particles
and their interactions.  Nevertheless, the Higgs sectors of the
simplest supersymmetric models are quite strongly constrained, and
exhibit very specific phenomenological profiles.

\section{The Standard Model Higgs boson---theory}
\label{secc}

In the Standard Model, the Higgs mass is given by
$m_{h_{\rm SM}}^2=\lambda v^2$, where $\lambda$ is the Higgs self-coupling.
Since $\lambda$ is unknown at present, the value of
the Standard Model Higgs mass is not predicted (although other
theoretical considerations, discussed in Section~\ref{secb}, place
constraints on the Higgs mass, as exhibited in Fig.~\ref{trivial}). The
Higgs couplings to fermions and gauge bosons are proportional to
the corresponding particle masses.  As a result, Higgs
phenomenology is governed primarily by the couplings of the Higgs
boson to the $W^\pm$ and  $Z$ and the third generation quarks and
leptons. It should be noted that a $h_{\rm SM} gg$  coupling,
where $g$ is the gluon, 
is induced by the one-loop graph in which the Higgs boson
couples to a virtual $t\bar t$ pair. Likewise, a
$h_{\rm SM}\gamma\gamma$ coupling is generated, although in this case
the one-loop graph in which the Higgs boson couples to a virtual
$W^+W^-$ pair is the dominant contribution. Further details of
Standard Higgs boson properties are given in \cite{hhg}.

\subsection{Standard Model Higgs boson decay modes}
\label{secca}

\begin{figure}[htb]
\centering
  \epsfxsize=0.7\textwidth
  \epsffile{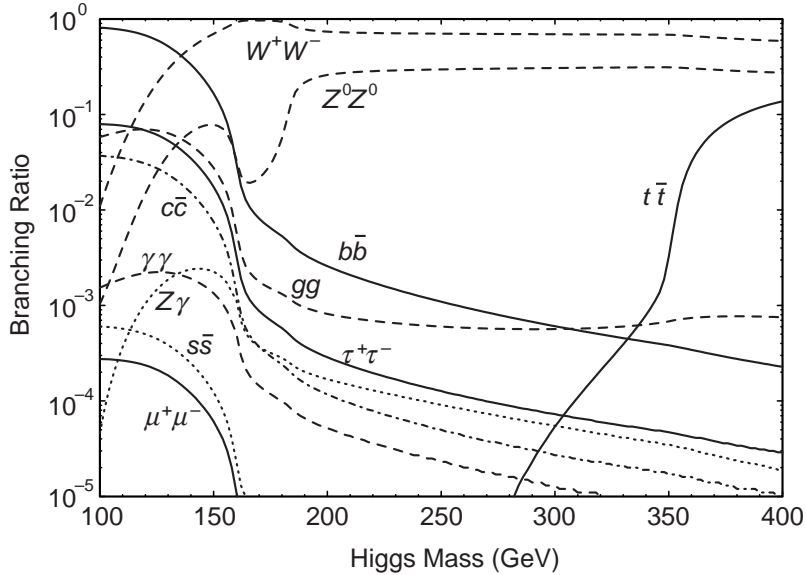}
\caption[0]{\label{fg:1}  Branching ratios of the dominant decay
modes of the \SM\ Higgs boson.  These results
have been obtained with the program HDECAY \cite{hdecay}, and include
QCD corrections beyond the leading order. 
}
\end{figure}

The Higgs boson mass is the only unknown parameter in the Standard
Model. Thus, one can compute Higgs boson branching ratios and
production cross sections as a function of $m_{h_{\rm SM}}$. The branching
ratios for the dominant decay modes of a Standard Model Higgs
boson are shown as a function of Higgs boson mass in Fig.~\ref{fg:1}.
Note that subdominant channels are important to establish a complete
phenomenological profile of the Higgs boson, and to check consistency
(or look for departures from) Standard Model predictions.
For $115~{\rm GeV}\sim m_{h_{\rm SM}}\lsim 2\mw$ many decays modes are large
enough to measure, as discussed in Section~\ref{sech}.

For $m_{h_{\rm SM}}\lsim 135$~GeV, the main Higgs decay mode is
$h_{\rm SM}\to b\bar b$, while the decays $h_{\rm SM}\to \tau^+\tau^-$ and
$c\bar c$ can also be phenomenologically relevant. In addition,
although one--loop suppressed, the decay $h_{\rm SM}\to gg$ is
competitive with other decays for $m_{h_{\rm SM}}\lsim 2\mw$ because of the
large top Yukawa coupling and the color factor.
As the Higgs mass increases above 135~GeV, the branching ratio to
vector boson pairs becomes dominant. In particular, the main Higgs
decay mode is $h_{\rm SM}\to WW^{(*)}$, where one of the $W$'s must be
off-shell (indicated by the star superscript) if $m_{h_{\rm SM}}<2\mw$. For
Higgs bosons with $m_{h_{\rm SM}}\gsim 2m_t$, the decay $h_{\rm SM}\to t\bar t$
begins to increase until it reaches its maximal value of about 20\%.

Rare Higgs decay modes can also play an important role. The
one-loop decay $h_{\rm SM}\to\gamma\gamma$ is a  suppressed
mode.   For
$\mw\lsim m_{h_{\rm SM}}\lsim 2\mw$, ${\rm BR}(h_{\rm SM}\to\gamma\gamma)$ is
above $10^{-3}$.  This decay channel provides an important Higgs
discovery mode at the LHC for $100~{\rm GeV}\lsim m_{h_{\rm SM}}\lsim
150$~GeV.  At the LC, the direct observation of
$h_{\rm SM}\to\gamma\gamma$ is difficult because of its suppressed branching
ratio.  Perhaps more relevant is the partial width
$\Gamma(\hl\to\gamma\gamma)$, which controls the Higgs production
rate at a $\gamma\gamma$ collider.

\subsection{Standard Model Higgs boson production at the LC}
\label{seccb}

\begin{figure}[tbh]
\centering
  \epsfxsize=0.8\textwidth
  \epsffile{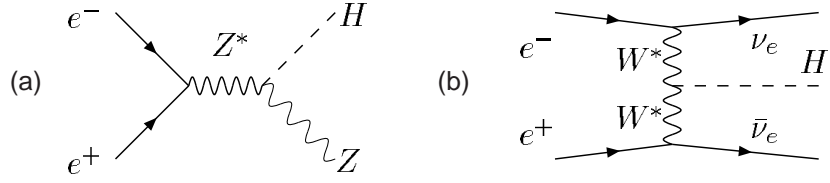}
\caption[higgs:fig:HZdiagram]{Main production processes for Higgs
production in $e^+e^-$ annihilation.
(a) Higgsstrahlung.
(b) $WW$ fusion.}
\label{higgs:fig:HZdiagram}
\end{figure}

In the Standard Model there are two main processes to produce
the Higgs boson in $e^+e^-$ annihilation.
These processes are also relevant in many extensions of the Standard
Model, particularly in nearly-decoupled extensions, in which the
lightest CP-even Higgs boson possesses properties nearly identical to
those of the SM Higgs boson.  
In the  ``Higgsstrahlung'' process, a virtual $Z$~boson
decays to an on-shell~$Z$ and the $h_{\rm SM}$,
depicted in Fig.~\ref{higgs:fig:HZdiagram}(a).
\begin{figure}[thb]
\centering
  \psfig{file=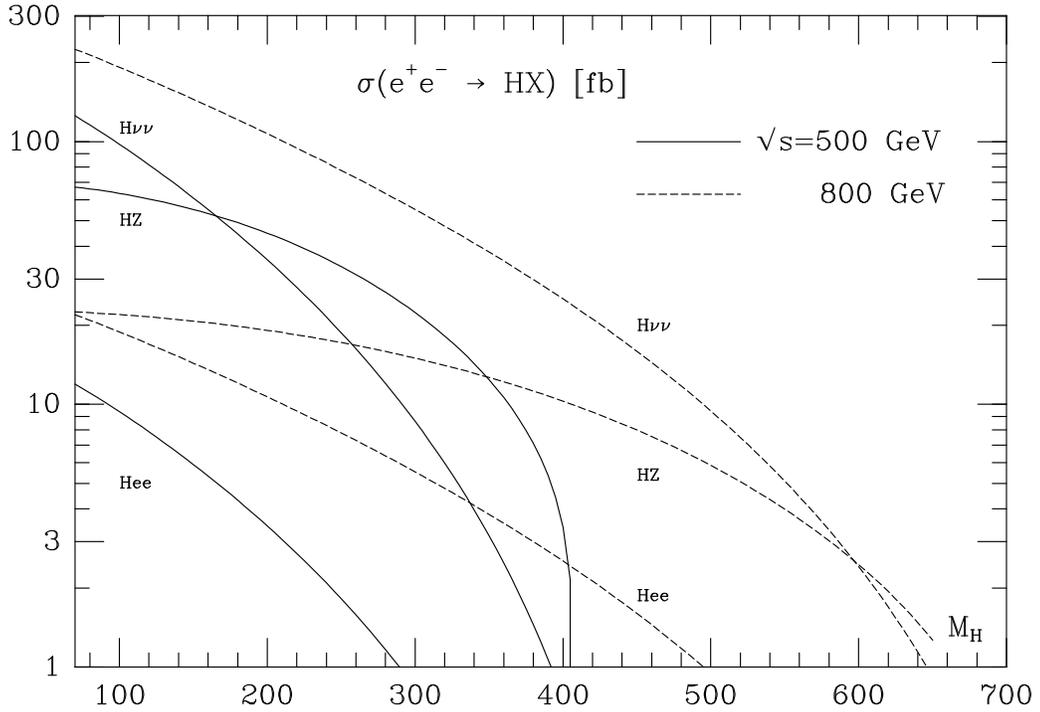,width=0.65\textwidth,angle=-90}
\caption[0]{Cross sections for Higgsstrahlung ($e^+e^-\to Zh_{\rm SM}$)
and Higgs production via $W^+W^-$ fusion ($e^+e^-\to \nu\bar\nu h_{\rm SM}$)
and $ZZ$ fusion ($e^+e^-\to e^+e^-h_{\rm SM}$) as a function of $m_{h_{\rm SM}}$ for 
two center-of-mass energies, $\sqrt{s}=500$ and 800~GeV \cite{Accomando:1998wt}.}  
\label{smhiggsxsec}
\end{figure}
The cross section for Higgsstrahlung rises sharply at threshold to a
maximum a few tens of GeV above $m_h+m_Z$, and then falls off as $s^{-1}$,
as shown in Fig.~\ref{smhiggsxsec}.
The associated production of the $Z$ provides an important trigger 
for Higgsstrahlung events.  In particular, in some theories beyond the
Standard Model, in which the Higgs boson decays into invisible modes,
the Higgs boson mass peak can be reconstructed in 
the spectrum of the missing mass recoiling against the~$Z$.
The other production process is called ``vector boson fusion'', where
the incoming $e^+$ and $e^-$ each emit a virtual vector boson,
followed by vector boson fusion to the $h_{\rm SM}$. 
Figure~\ref{higgs:fig:HZdiagram}(b) depicts the $W^+W^-$ fusion process.
Similarly, the $ZZ$ fusion process corresponds to $e^+e^-\to e^+e^-h_{\rm SM}$.
In contrast to Higgsstrahlung, the vector boson fusion cross section
grows as $\ln s$, and thus
is the dominant Higgs production mechanism for 
$\sqrt{s}\gg m_{h_{\rm SM}}$.
The cross section for $WW$ fusion is about ten times larger than that
for $ZZ$ fusion.  Nevertheless, the latter provides complementary
information on the $ZZh_{\rm SM}$ vertex.   Note that at an $e^-e^-$
collider, the Higgsstrahlung and $W^+W^-$ fusion processes are absent,
so that $ZZ$ fusion is the dominant Higgs production process.

\begin{figure}[thb]
\centering
  \psfig{file=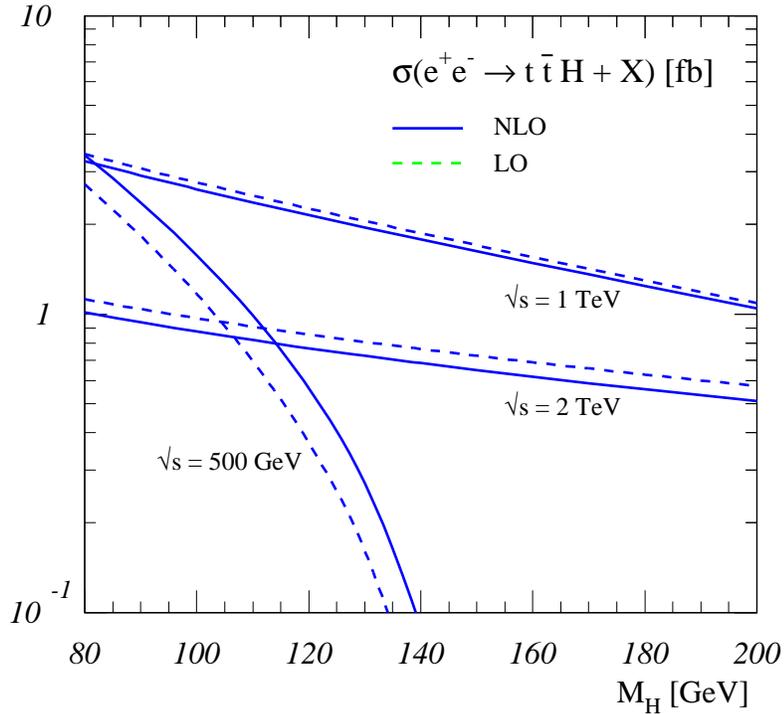,width=0.7\textwidth}
\caption[0]{Cross-sections for $e^+e^-\to t\bar th_{\rm SM}$ in fb for
three choices of center-of-mass energy.  The dashed lines correspond
to the tree-level result \cite{Djouadi:1992tk}, and the solid lines include the
next-to-leading order QCD corrections \cite{Dittmaier:1998dz}.}
\label{ttbarhiggs}
\end{figure}
Other relevant processes for producing Higgs bosons are associated
production with a fermion-antifermion pair,
and multi-Higgs production.
For the former class, only $e^+e^-\to t\bar{t}h_{\rm SM}$ has a
significant cross section, around the femtobarn level in the Standard
Model, as depicted in Fig.~\ref{ttbarhiggs}.
As a result, if $m_{h_{\rm SM}}$ is small enough (or $\sqrt{s}$ is large
enough), this process can be used for determining
the Higgs--top quark Yukawa coupling.  
The cross section for double Higgs production 
($e^+e^-\to Zh_{\rm SM}h_{\rm SM}$) are even smaller, of
order 0.1~fb for $100~{\rm GeV}\lsim m_{h_{\rm SM}}\lsim 150$~GeV and 
$\sqrt{s}$ ranging between 500~GeV and 1~TeV.
With sufficient luminosity, the latter can be used
for extracting the triple Higgs self-coupling.

At the $\gamma\gamma$ collider, a Higgs boson is produced 
as an $s$-channel resonance via the one-loop 
triangle diagram.
Every charged particle whose mass is generated  by
the Higgs boson contributes to this process.
In the Standard Model, the main contributors are the $W^\pm$ 
and the $t$-quark loops.  See Section~\ref{secj} for further discussion.

\section{SM Higgs searches before the linear collider}
\label{secd}

\subsection{Direct search limits from LEP}
\label{secda}

The LEP collider completed its final run in 2000, and presented
tantalizing hints for the possible observation of the Higgs 
boson.  
Combining data from all four LEP collaborations~\cite{igo}, 
one could interpret
their observations as corresponding to the production of a Higgs boson
with a mass of $\mhl=115.0^{+1.3}_{-0.9}$~GeV with a significance of
$2.9\sigma$.  This is clearly not sufficient to announce a discovery
or even an ``observation''.  A more conservative interpretation of the
data then places a 95\%~CL lower limit of $m_{h_{\rm SM}}>113.5$~GeV.

\subsection{Implications of precision electroweak measurements}
\label{decdb}

Indirect constraints on the Higgs boson mass within the SM can be
obtained from confronting the SM predictions with results of electroweak
precision measurements. In the case of the top quark mass,
the indirect determination turned out to be in remarkable agreement
with the actual experimental value. In comparison,  to obtain
constraints on $m_{h_{\rm SM}}$ of similar precision,
much higher accuracy is required for both the 
experimental results and the theory predictions.
 This is due to the fact that
the leading dependence of the precision observables on $m_{h_{\rm SM}}$ is only
logarithmic, while the dominant effects of the top-quark mass enter
quadratically. 

\begin{figure}[b!]
\centerline{
\psfig{figure=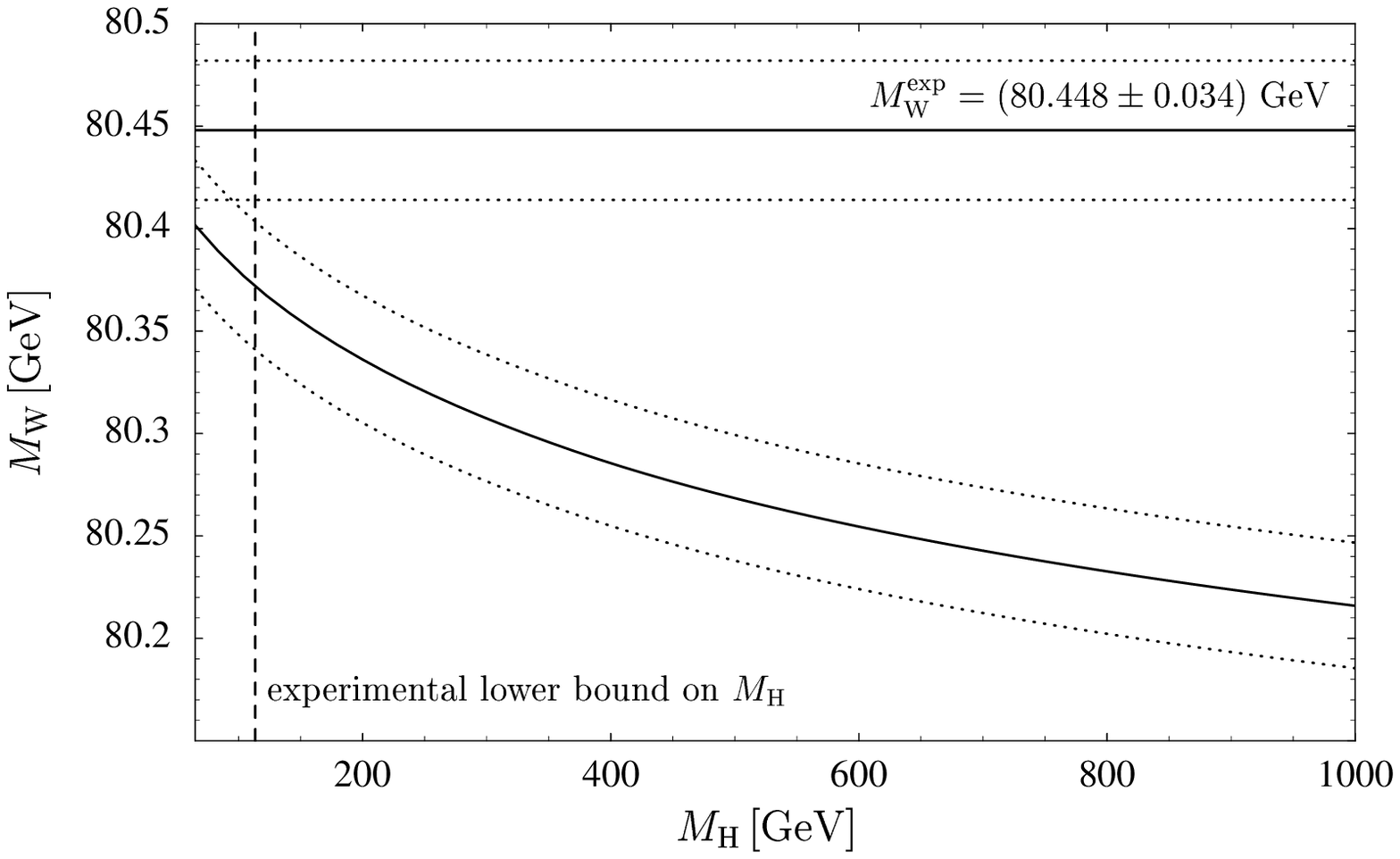,width=8cm,height=6.3cm}
\psfig{figure=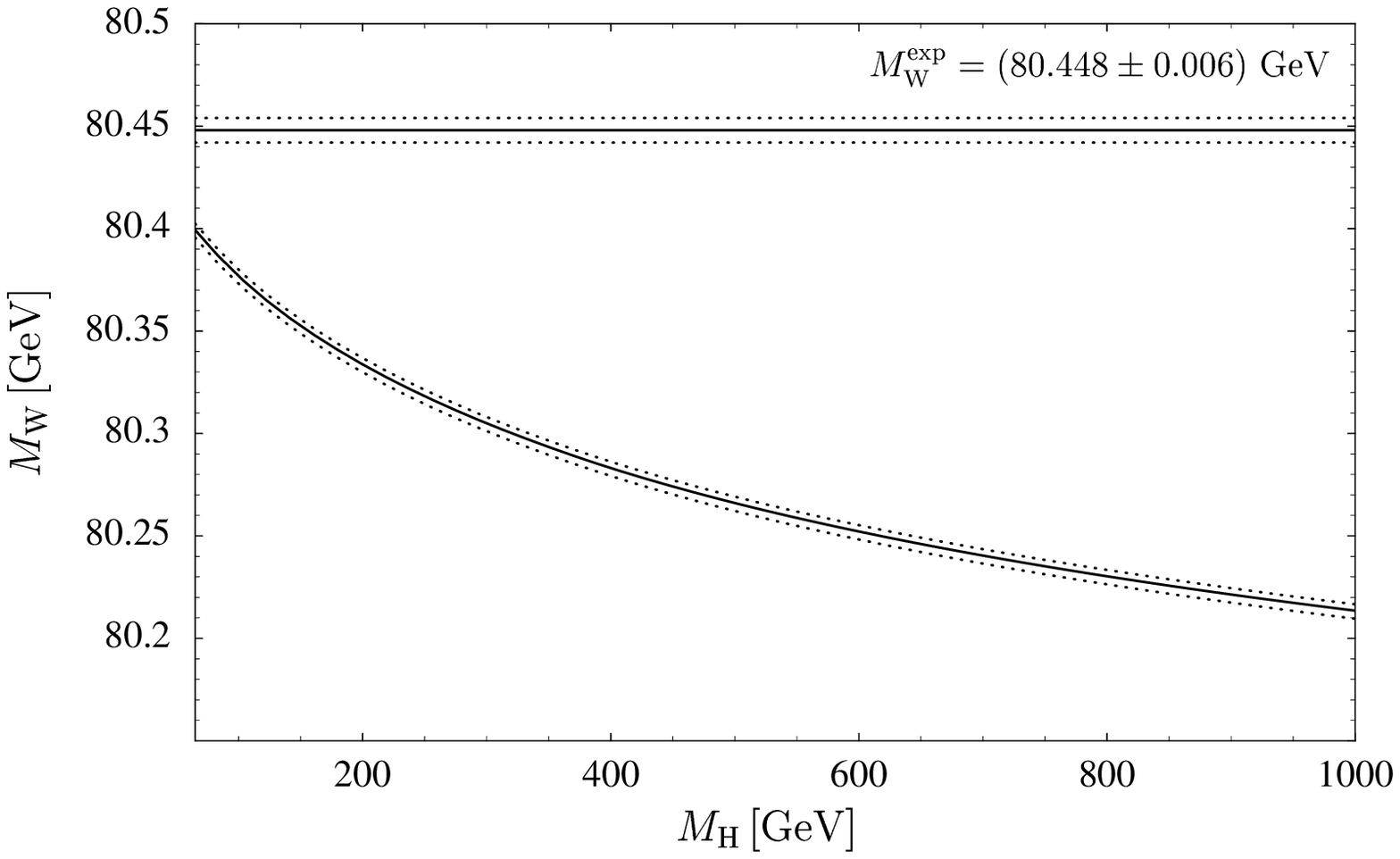,width=8cm,height=6.3cm}
}
\caption[]{
The prediction for $m_W$ as a function of $m_{h_{\rm SM}}$ is compared with the
experimental value of $m_W$ for the current experimental accuracies of 
$m_W$ and $\mt$ (left plot) and for the prospective future accuracies at
a LC with Giga-Z option (right plot, the present experimental central
values are assumed)~\cite{deltaral2}. In the left plot also the present 
experimental 95\% CL\ lower bound on the Higgs-boson mass,
$m_{h_{\rm SM}} = 113.5$~GeV, is indicated.
\label{fig:mw2l}
}
\end{figure}

The left plot of Fig.~\ref{fig:mw2l} shows the currently most precise
result for $m_W$ as function of $m_{h_{\rm SM}}$ in the SM, 
and compares it with the present experimental value of $m_W$.  The
calculation incorporates 
the complete electroweak fermion-loop contributions at 
${\cal O}(\alpha^2)$~\cite{deltaral2}.  Based on this result, the remaining theoretical
uncertainty from unknown higher-order corrections has been estimated to
be about $6 \mev$~\cite{deltaral2}. It is about a factor five smaller than
the uncertainty induced by the current experimental error on the
top-quark mass, $\Delta\mt^{\rm exp} = \pm 5.1 \gev$, 
which presently dominates 
the theoretical uncertainty. The right plot of Fig.~\ref{fig:mw2l} shows the
prospective situation at a future $e^+e^-$ linear 
collider after Giga-Z operation and a threshold measurement
of the $W$ mass (keeping the
present experimental central values for simplicity), which are expected to
reduce the experimental errors to $\Delta m_W^{\rm exp} = 6 \mev$
and $\Delta\mt^{\rm exp} = 200 \mev$. 
This program is described in Chapter 8.
The plot clearly shows the considerable 
improvement in the sensitivity to $m_{h_{\rm SM}}$ achievable at the LC via very precise
measurements of $m_W$ and $\mt$. Since furthermore the experimental
error of $\sweff$ is expected to be reduced by almost a factor of 20 at
Giga-Z, the accuracy in the indirect determination of the Higgs-boson
mass from all data will improve by about a factor of 10 compared to the
present situation~\cite{gigaz}.

\subsection{Expectations for Tevatron searches}
\label{secdc}

The upgraded Tevatron began taking data in the spring of 2001.  This
is the only collider at which the Higgs boson can be produced for the
next five years, until the LHC begins operation in 2006.
The Tevatron Higgs working group presented a detailed analysis of the
Higgs discovery reach at the upgraded Tevatron \cite{tevreport}.
Here, we summarize the main results.  Two Higgs mass ranges were
considered separately: (i) 100~GeV$\lsim m_{h_{\rm SM}}\lsim 135$~GeV and (ii)
135~GeV$\lsim m_{h_{\rm SM}}\lsim 190$~GeV, corresponding to the two different
dominant Higgs decay modes: $h_{\rm SM}\to b\bar b$ for the lighter mass
range and $h_{\rm SM}\to WW^{(*)}$ for the heavier mass range.

In mass range (i), the relevant production mechanisms are $q_i\bar
q_j\to Vh_{\rm SM}$, where $V=W$ or $Z$. In all cases, the dominant
$h_{\rm SM}\to b\bar b$ decay was employed.  The most relevant final-state signatures
correspond to events in which the vector boson decays leptonically
($W\to\ell\nu$, $Z\to\ell^+\ell^-$ and $Z\to\nu\bar\nu$, where
$\ell=e$ or $\mu$), resulting in $\ell\nu b\bar b$, $\nu\bar\nu b\bar
b$ and $\ell^+\ell^- b\bar b$ final states.  In mass range (ii), the
relevant production mechanisms include $gg\to h_{\rm SM}$, $V^* V^*\to h_{\rm SM}$ 
and $q_i\bar q_j\to Vh_{\rm SM}$, with decays $h_{\rm SM}\to WW^{(*)}$,
$ZZ^{(*)}$. 
The most relevant phenomenological signals are those in which two of
the final-state vector bosons decay leptonically, resulting in
$\ell^+\ell^-\nu\bar\nu$ or $\ell^\pm\ell^\pm jjX$, where $j$ is a
hadronic jet and $X$ consists of two additional leptons (either
charged or neutral).  For example, the latter can arise from $Wh_{\rm SM}$
production followed by $h_{\rm SM}\to WW^{(*)}$, where the two like-sign $W$
bosons decay leptonically, and the third $W$ decays into hadronic
jets.  In this case $X$ is a pair of neutrinos.

\begin{figure}[thb]
  \begin{center}
    \parbox{6.0in}{\epsfxsize=\hsize\epsffile[92 341 515 620]
           {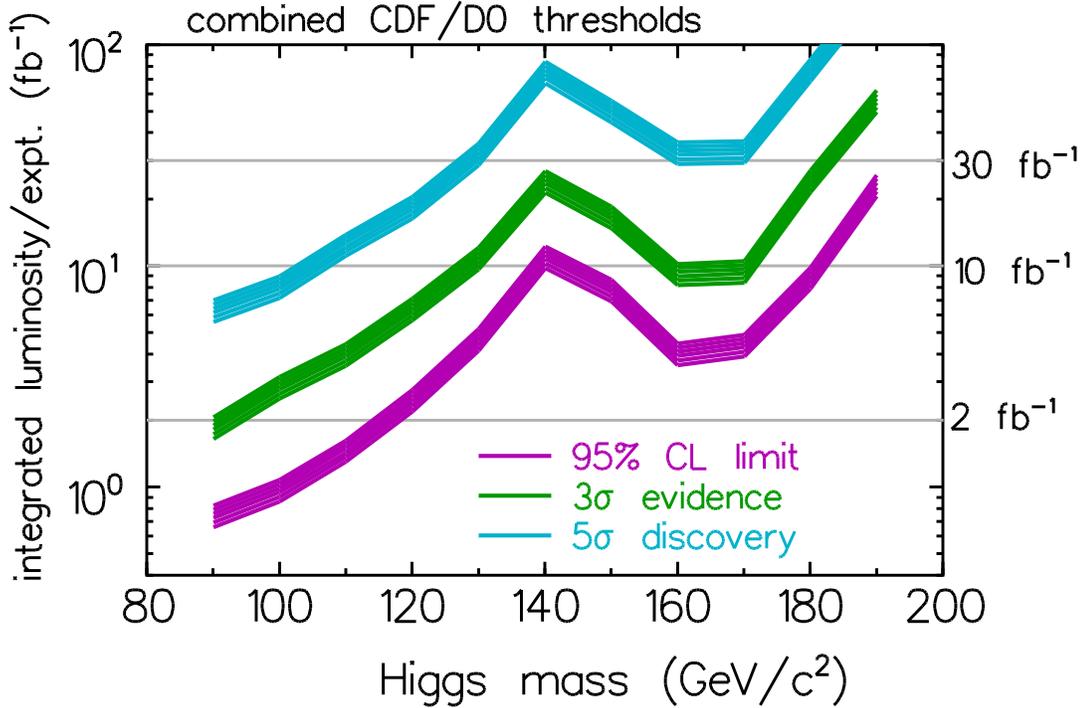}}
  \end{center}
\caption[0]{The integrated luminosity required per experiment, to
            either exclude a SM Higgs boson at 95\% CL or discover it at the
            $3\sigma$ or $5\sigma$ level, as a function of the Higgs mass.  
            These results are based on the combined
            statistical power of both experiments.  The curves shown  
            are obtained by combining the $\ell\nu b\bar b$,
            $\nu\bar\nu b\bar b$ and $\ell^+\ell^-b\bar b$ channels
            using the neural network selection in the low-mass Higgs region
            ($90~{\rm GeV}~\lsim m_{h_{\rm SM}}\lsim 130$~GeV), and the
            $\ell^\pm\ell^\pm jjX$ and $\ell^+\ell^-\nu\bar\nu$ 
            channels in the high-mass Higgs region
            ($130~{\rm GeV}~\lsim m_{h_{\rm SM}}\lsim 190$~GeV).  The lower edge of 
            the bands is the calculated threshold; the bands extend upward 
            from these nominal thresholds by 30\% as an indication of the
            uncertainties in $b$-tagging efficiency, background rate,
            mass resolution, and other effects.}
  \label{f:combined-final}
\end{figure}

Figure~\ref{f:combined-final} summarizes 
the Higgs discovery reach versus the total integrated luminosity
delivered to the Tevatron (and by assumption, delivered to each
detector).
As the plot shows, the required integrated luminosity increases
rapidly with Higgs mass to 140 GeV, beyond which the high-mass
channels play the dominant role.  With
2~fb$^{-1}$ per detector (which is expected after one year of running  
at design luminosity),
the 95\% CL limits will barely extend the
expected LEP2 limits, but with 10 fb$^{-1}$, the SM Higgs boson can be
excluded up to 180 GeV if the Higgs boson
does not exist in that mass range.

Current projections envision that the Tevatron, with further machine
improvements, will
provide an integrated luminosity of 15~fb$^{-1}$ after six years of
running.  If $m_{h_{\rm SM}}\simeq 115$~GeV, as suggested by 
LEP data, then the Tevatron experiments will be able to achieve a $5\sigma$
discovery of the Higgs boson.  If no Higgs events are detected, the LEP
limits will be significantly extended, with a 95\%~CL exclusion
possible up to about $m_{h_{\rm SM}}\simeq 185$~GeV.  Moreover, evidence for a
Higgs boson at the $3\sigma$ level could be achieved up to about
$m_{h_{\rm SM}}\simeq 175$~GeV.  (The Higgs mass region around
140 GeV might require more luminosity, depending on the
magnitude of systematic errors due to uncertainties in $b$-tagging
efficiency, background rate, the $b\bar b$ mass resolution, {\it
etc.})
Evidence for or discovery of a Higgs boson at
the Tevatron would be a landmark in high energy physics.
However, even if a Higgs boson is seen, the Tevatron 
data would only provide a very rough phenomenological profile.
In contrast, the LC, and to a lesser extent, the LHC could measure
enough of its properties with sufficient precision to 
verify that the observed Higgs is truly SM-like. The LHC is also
certain to yield $>5\sigma$ discovery of a SM Higgs boson
over the full range of possible masses, up to 1~TeV.

\subsection{Expectations for LHC searches}
\label{secdd}

\begin{figure}[t!]
\centerline{\psfig{file=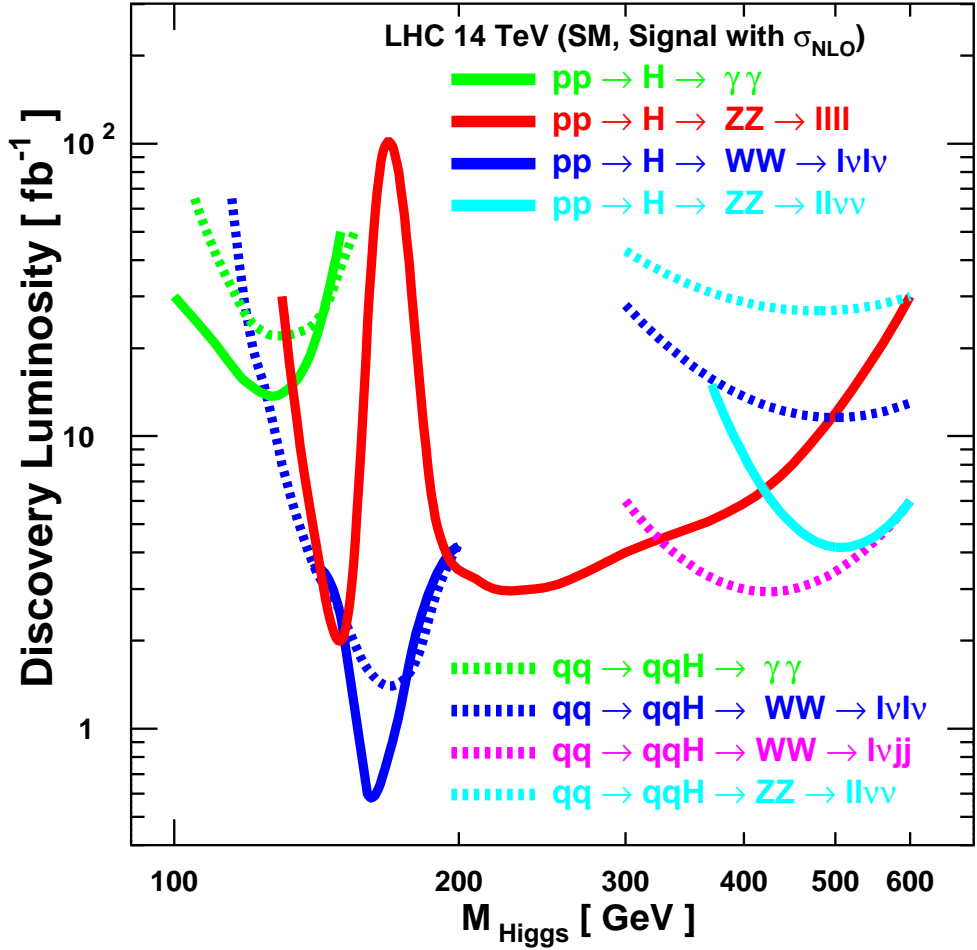,width=14cm}}
\caption[0]{Expected $5\sigma$ SM Higgs discovery luminosity requirements at the LHC,
for one experiment, statistical errors only~\cite{higgs-CMS_Higgs_lumi}.
The study was performed with CMS fast detector simulation.}
\label{f:atlashsm}
\end{figure}

At the LHC, the ATLAS and CMS detectors have been specifically designed
so as to guarantee discovery of a SM Higgs boson, regardless of mass.
The most important production processes for the $h_{\rm SM}$
are the gluon fusion process, $gg\to h_{\rm SM}$, and the vector boson
fusion process, $WW\to h_{\rm SM}$.   In particular,
for $m_{h_{\rm SM}}\lsim 130\gev$ the important discovery modes are 
$gg,WW\to h_{\rm SM}\to\gam\gam$,  $\tau^+\tau^-$.
At high luminosity, $q_i\anti q_j\to W^\pm h_{\rm SM}$
and $gg\to t\anti th_{\rm SM}$ with $h_{\rm SM}\to\gam\gam$ and $h_{\rm SM}\to b\anti b$
should also be visible.
Once $m_{h_{\rm SM}}>130\gev$, $gg\to h_{\rm SM} \to ZZ^{(*)}\to 4\ell$ is extremely
robust except for the small mass region with $m_{h_{\rm SM}}$ just above $2\mw$
in which  $h_{\rm SM}\to WW$ is allowed and $B(h_{\rm SM}\to ZZ^*)$ drops
sharply.  In this region, $gg,WW\to h_{\rm SM}\to WW\to \ell\nu\ell\nu$ provides
a strong Higgs signal. Once $m_{h_{\rm SM}}>300\gev$ ($400\gev$), the final states
$h_{\rm SM}\to WW\to \ell\nu jj$ and $h_{\rm SM}\to ZZ\to \ell\ell\nu\nu$,
where the $h_{\rm SM}$ is produced by a combination of $gg$ and $WW$ fusion, 
provide excellent discovery channels. These latter allow
discovery even for $m_{h_{\rm SM}}\gsim 1\tev$, {\it i.e.},
well beyond the $m_{h_{\rm SM}}\sim 800\gev$
limit of viability for the $h_{\rm SM}\to 4\ell$ mode.  These results are summarized
in Fig.~\ref{f:atlashsm}, from which we observe that the net statistical
significance for the $h_{\rm SM}$, after combining channels, exceeds $10\sigma$
for all $m_{h_{\rm SM}}>80\gev$, assuming accumulated luminosity of
$L=100\fbi$ at the ATLAS detector \cite{atlashsmref}.  Similar results
are obtained by the CMS group \cite{cmshsmref}, the $\gam\gam$ mode being even
stronger in the lower mass region.

\begin{figure}[htb]
\centerline{\psfig{file=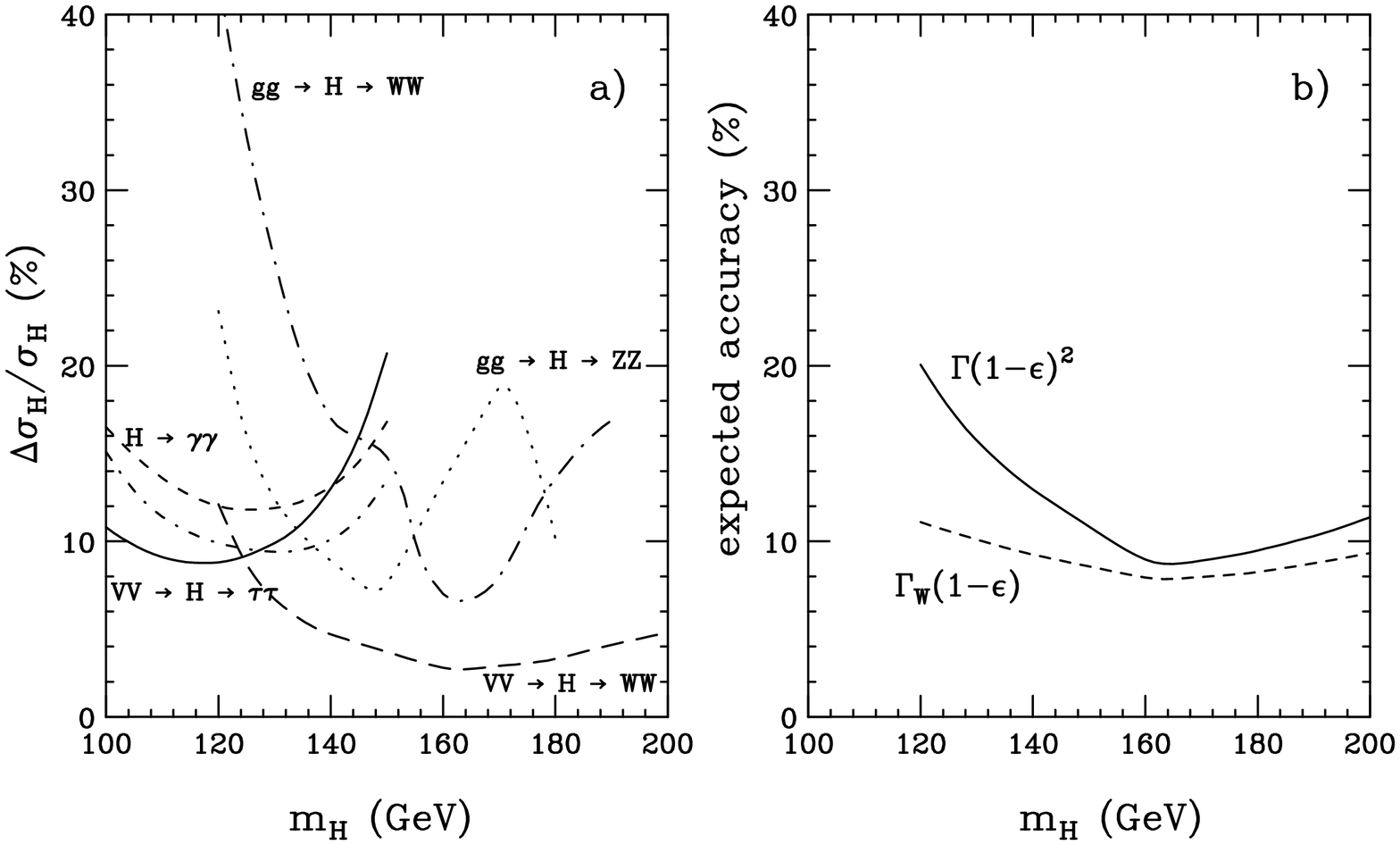,width=8cm,angle=90}}
\caption[0]{Relative accuracy expected at the LHC with 200~fb$^{-1}$ of data.
(a) Cross section times branching fraction for several inclusive modes 
(dotted and dash-dotted lines) and vector boson fusion
 channels (dashed and solid lines).
(b) Extracted total width (solid line) and $H\to WW$ partial width 
(dashed line). In the latter, $\epsilon = 
1-[B(H\to b\bar b)+B(H\to \tau\tau)+B(H\to WW^{(*)})+
B(H\to ZZ^{(*)})+B(H\to gg)+B(H\to \gamma\gamma)]$. To the
extent that $\eps$ is small, the indicated accuracies can be achieved.
}
  \label{f:zeppenfeld}
\end{figure}

Precision measurements for a certain number of quantities will be
possible, depending upon the exact value of $m_{h_{\rm SM}}$.  For instance, in
\cite{atlashsmref} it is estimated that $m_{h_{\rm SM}}$ can be measured to
$<0.1\%$ for $m_{h_{\rm SM}}<400\gev$ and to $0.1$--$1\%$ for
$400<m_{h_{\rm SM}}<700\gev$. Using the $4\ell$ final state, $\Gamma^T_{h_{\rm SM}}$ can
determined for $m_{h_{\rm SM}}>250\gev$ from the shape of the $4\ell$ mass
peak.  Various ratios of branching ratios and a selection of partial
widths times branching ratios can be measured in any given mass
region.  Some early estimates of possibilities and achievable
accuracies appear in \cite{Gunion:1996cn}.  A more recent, but
probably rather optimistic parton-level theoretical study
\cite{Zeppenfeld:2000de} finds that if $m_{h_{\rm SM}}\lsim 200\gev$ then good
accuracies can be achieved for many absolute partial widths and for
the total width provided: (a) $WW$ fusion production can be reliably
separated from $gg$ fusion; (b) the $WW/ZZ$ coupling ratio is as
expected in the SM from the SU(2)$\times$U(1) symmetry; (c) the $WW^*$
final state can be observed in both $gg$ and $WW$ fusion; and (d)
there are no unexpected decays of the $h_{\rm SM}$.  
Invisible Higgs decays may also be addressed by this technique
\cite{higgs-Eboli}; CMS simulations show some promise for
this channel.   The resulting errors
estimated for $L=200\fbi$ of accumulated data are given in
Fig.~\ref{f:zeppenfeld}.

\section{Higgs bosons in low-energy supersymmetry}
\label{sece}

The simplest realistic model of low-energy supersymmetry is the
minimal supersymmetric Standard Model (MSSM),
which consists of the two-Higgs-doublet extension of the 
Standard Model plus the corresponding superpartners~\cite{Haber85}. 
Two Higgs doublets, one  
with $Y=+1$ and one with $Y=-1$, are needed in order
that gauge anomalies due to the higgsino superpartners 
are exactly canceled.   The supersymmetric structure also constrains the 
Higgs-fermion interactions.  In particular, it is
the $Y=-1$ Higgs doublet that generates mass for ``up''-type quarks and 
the $Y=+1$ Higgs doublet that generates mass for ``down''-type
quarks (and charged leptons) \cite{Inoue82,Gunion86}. 

After electroweak symmetry breaking, one finds
five physical Higgs particles: a charged Higgs
pair ($\hpm$), two CP-even neutral Higgs bosons (denoted by $\hl$
and $\hh$ where $\mhl \leq \mhh$) and one CP-odd neutral
Higgs boson ($\ha$).\footnote{The tree-level MSSM Higgs sector
automatically conserves CP.  Hence, the two neutral Higgs vacuum
expectation values can be chosen to be real and positive, and the
neutral Higgs eigenstates possess definite CP quantum numbers.}   
Two other relevant parameters are
the ratio of neutral Higgs vacuum expectation values,
$\tan\beta$, and
an angle $\alpha$ that measures the component of the original
$Y=\pm 1$ Higgs doublet states in the physical CP-even
neutral scalars.

\subsection{MSSM Higgs sector at tree-level}
\label{secea}

The supersymmetric structure of the theory imposes constraints on
the Higgs sector of the model \cite{hhgchap4}. 
As a result, all Higgs sector
parameters at tree-level are determined by two free parameters:
$\tanb$ and one Higgs mass, conveniently chosen to be $\mha$. 
There is an upper bound
to the tree-level mass of the light CP-even Higgs boson:
$\mhl^2\leq\mz^2\cos 2\beta\leq\mz^2$.  However, radiative corrections
can significantly alter this upper bound as described in Section~\ref{seceb}.

The limit of $\mha\gg\mz$ is of particular interest,
with two key consequences.
First, $\mha\simeq\mhh \simeq\mhpm$, up to corrections of ${\cal
O}(\mz^2/\mha)$.  Second, $\cos(\beta-\alpha)=0$ up to corrections
of ${\cal O}(\mz^2/\mha^2)$. This limit is known as the {\it
decoupling} limit \cite{decoupling} because when $\mha$ is large,
the effective low-energy theory below the scale of $\mha$ contains
a single CP-even Higgs boson, $\hl$, whose properties are nearly
identical to those of the Standard Model Higgs boson, $h_{\rm SM}$.

The phenomenology of the Higgs sector is determined by the various
couplings of the Higgs bosons to gauge bosons, Higgs bosons and
fermions. The couplings of the two CP-even Higgs bosons to $W$ and
$Z$ pairs are given in terms of the angles $\alpha$ and $\beta$ 
by
\begin{eqnarray}
            g_{\hl VV}&=& g_{V} m_{V}\sinbma \nonumber \\[3pt]
           g_{\hh VV}&=& g_{V} m_{V}\cosbma\,,
\label{vvcoup}
\end{eqnarray}
 where 
\begin{equation} 
g_ V\equiv
\begin{array}{cl} 
                     g, & V=W,\\
g/\cos\theta_W, & V=Z. 
\end{array} 
\label{hix} 
\end{equation} 
There are
no tree-level couplings of $\ha$ or $\hpm$ to $VV$.  The couplings
of one gauge boson to two neutral Higgs bosons are given by:
\begin{eqnarray}
g_{\hl\ha Z}&=&{g\cosbma\over 2\cos\theta_W}\,,\nonumber \\[3pt]
           g_{\hh\ha Z}&=&{-g\sinbma\over 2\cos\theta_W}\,.
           \label{hvcoup}
\end{eqnarray}

In the MSSM, the Higgs tree-level couplings to fermions obey the
following property: the neutral member of the $Y=-1$ [$Y=+1$] Higgs 
doublet couples exclusively to down-type [up-type]
fermion pairs. This pattern of Higgs-fermion couplings defines the
Type-II two-Higgs-doublet model \cite{wise,hhg}.
Consequently, the couplings of the neutral Higgs bosons to $f\bar
f$ relative to the Standard Model value, $gm_f/2\mw$, are given by
(using third family notation):
\begin{eqnarray}
 \label{qqcouplings}
\hl b\bar b \;\;\; ({\rm or}~ \hl \tau^+ \tau^-)&:&~~ -
{\sin\alpha\over\cos\beta}=\sin(\beta-\alpha)
-\tan\beta\cos(\beta-\alpha)\,,\nonumber\\[3pt]
\hl t\bar t&:&~~ \phm{\cos\alpha\over\sin\beta}=\sin(\beta-\alpha)
+\cot\beta\cos(\beta-\alpha)\,,\nonumber\\[3pt]
\hh b\bar b \;\;\; ({\rm or}~ \hh \tau^+ \tau^-)&:&~~
\phm{\cos\alpha\over\cos\beta}=
\cos(\beta-\alpha)
+\tan\beta\sin(\beta-\alpha)\,,\nonumber\\[3pt]
\hh t\bar t&:&~~ \phm{\sin\alpha\over\sin\beta}=\cos(\beta-\alpha)
-\cot\beta\sin(\beta-\alpha)\,,\nonumber\\[3pt]
\ha b \bar b \;\;\; ({\rm or}~ \ha \tau^+
\tau^-)&:&~~\phm\gamma_5\,{\tan\beta}\,,
\nonumber\\[3pt]
\ha t \bar t&:&~~\phm\gamma_5\,{\cot\beta}\, .
\end{eqnarray}
In these expressions,  $\gamma_5$ indicates a pseudoscalar coupling.

The neutral Higgs boson couplings to fermion pairs
(\ref{qqcouplings}) have been written in such a way that their
behavior can be immediately ascertained in the decoupling limit
($\mha\gg\mz$) by setting $\cosbma=0$. In
particular, in the decoupling limit, the couplings of $\hl$ to
vector bosons and fermion pairs are equal to the corresponding
couplings of the \SM\ Higgs boson.

\begin{figure}[htb]
\centering
\centerline{\psfig{file=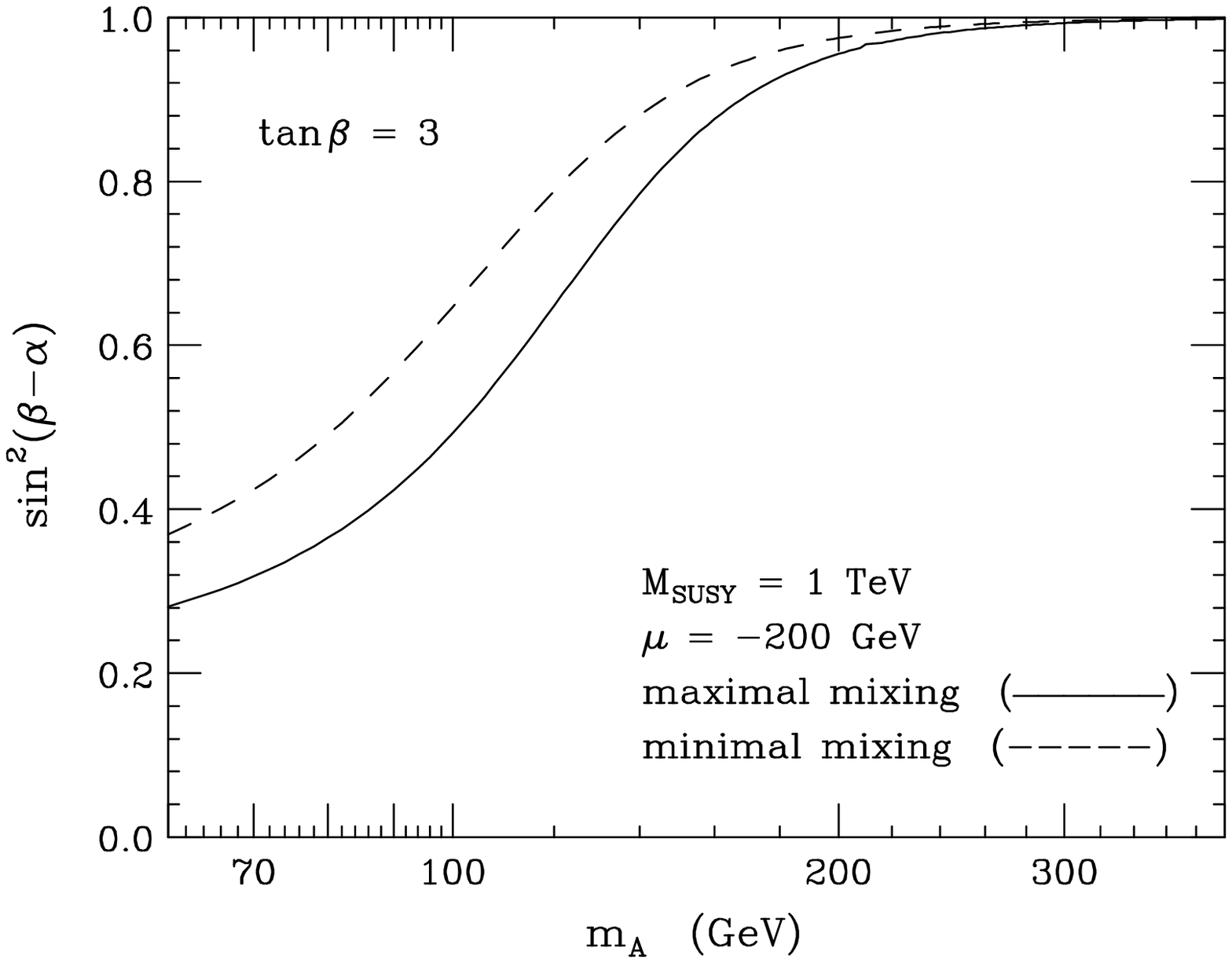,width=8cm}
\hfill
\psfig{file=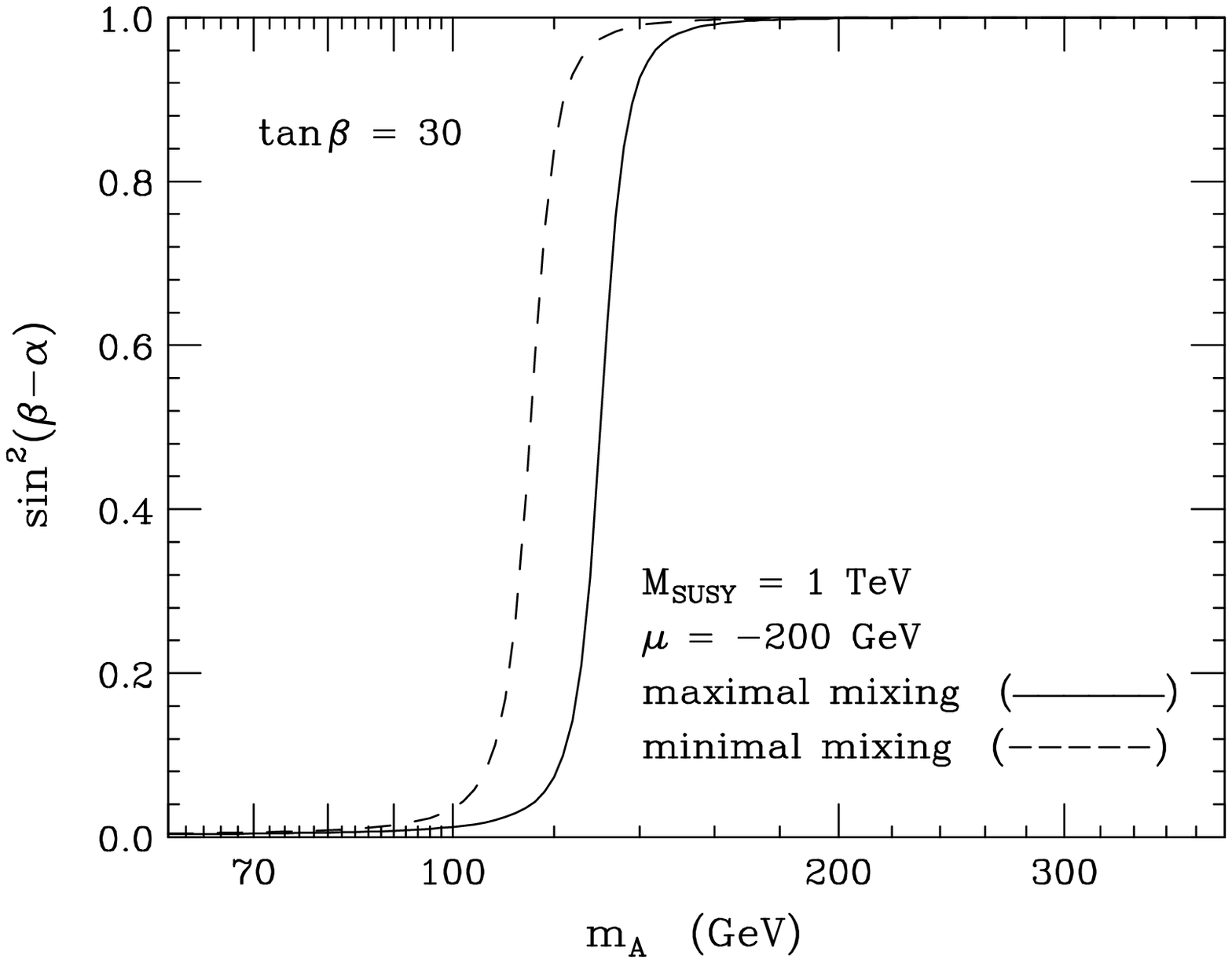,width=8cm}}
\caption[0]{\label{cosgraph} The value of $\sin^2(\beta-\alpha)$
is shown as a function of
$\mha$ for two choices of $\tan\beta = 3$ and $\tan\beta = 30$.
When radiative corrections are included, one can define an approximate
loop-corrected angle $\alpha$ as a function of $\mha$, $\tan\beta$ and
the MSSM parameters.  In the figures above, we
have incorporated radiative corrections, assuming that
$M_{\rm SUSY}=1$~TeV.  In addition,
two extreme cases for the squark mixing parameters
are shown (see Section~\ref{seceb} for further discussion of the
radiative corrections and their dependence on the supersymmetric
parameters). The decoupling effect expected when
$\sin^2(\beta-\alpha)\simeq 1$ for $\mha\gg m_Z$,
continues to hold even when radiative corrections are included.
}
\end{figure}

The region of MSSM Higgs sector parameter space in which the
decoupling limit applies is large, because $\sin(\beta-\alpha)$
approaches 1 quite rapidly once $\mha$ is larger than about
200~GeV, as shown in Fig.~\ref{cosgraph}. As a result, over a
significant region of the MSSM parameter space, the search for the
lightest CP-even Higgs boson of the MSSM is equivalent to the
search for the \SM\ Higgs boson.  This result is more general; in
many theories of non-minimal Higgs sectors, there is a significant
portion of the parameter space that approximates the decoupling
limit. Consequently, simulations of the \SM\ Higgs signal are also
relevant for exploring the more general Higgs sector.

\subsection{The radiatively corrected MSSM Higgs sector}
\label{seceb}

When one-loop radiative corrections are incorporated, the Higgs
masses and couplings depend on additional parameters of the
supersymmetric model that enter via the virtual loops. One of the most
striking effects of the radiative corrections to the MSSM Higgs
sector is the modification of the upper bound of the light CP-even
Higgs mass, as first noted in \cite{hhprl}. When $\tanb\gg 1$ and
$\mha\gg\mz$, the {\it tree-level} prediction for $\mhl$
corresponds to its theoretical upper bound, $\mhmax=\mz$.
Including radiative corrections, the theoretical upper bound is
increased, primarily because of an incomplete cancellation of the
top-quark and top-squark (stop) loops.  (These contributions would
cancel if supersymmetry  were exact.) The relevant parameters
that govern the stop sector are the average of the two stop
squared-masses: $M^2_{\rm SUSY}\equiv \half(\mstopa^2+\mstopb^2)$, and
the off-diagonal element of the stop squared-mass matrix: $m_t
X_t\equiv m_t(A_t-\mu\cot\beta)$, where $A_t$ is a soft supersymmetry-breaking 
trilinear scalar interaction term, and $\mu$ is the supersymmetric Higgs mass 
parameter. 
The qualitative behavior of the
radiative corrections can be most easily seen in the large top
squark mass limit, where, in addition, the splitting of the two
diagonal entries and the off-diagonal entry of the stop
squared-mass matrix are both small in comparison to $M^2_{\rm SUSY}$.  In
this case, the upper bound on the lightest CP-even Higgs mass is
approximately given by 
\begin{equation} \label{deltamh} 
\mhl^2  \lsim 
\mz^2+{3g^2\mt^4\over
8\pi^2\mw^2}\left[\ln\left({M^2_{\rm SUSY}\over\mt^2}\right)
+{X_t^2\over M^2_{\rm SUSY}} \left(1-{X_t^2\over
12M^2_{\rm SUSY}}\right)\right]\,. 
\end{equation} 
More complete treatments of the
radiative corrections include the effects of stop mixing,
renormalization group improvement, and the leading two-loop
contributions, and imply that these corrections somewhat overestimate
the true upper bound of $\mhl$ (see \cite{higgsrad} for the most
recent results). 
Nevertheless, Eq.~(\ref{deltamh}) correctly illustrates
some noteworthy features of the more precise result. 
First, the
increase of the light CP-even Higgs mass bound beyond $\mz$ can be
significant.  This is a consequence of the $m_t^4$ enhancement of
the one-loop radiative correction. Second, the dependence of the
light Higgs mass on the stop mixing parameter $X_t$ implies that
(for a given value of $M_{\rm SUSY}$) the upper bound of the light Higgs
mass initially increases with $X_t$ and reaches its {\it maximal}
value at $X_t\simeq\sqrt{6}M_{\rm SUSY}$.  This point is referred to as the
{\it maximal mixing} case (whereas $X_t=0$ corresponds to the {\it
minimal mixing} case).

\begin{figure}[htb]
\centering
\centerline{\psfig{file=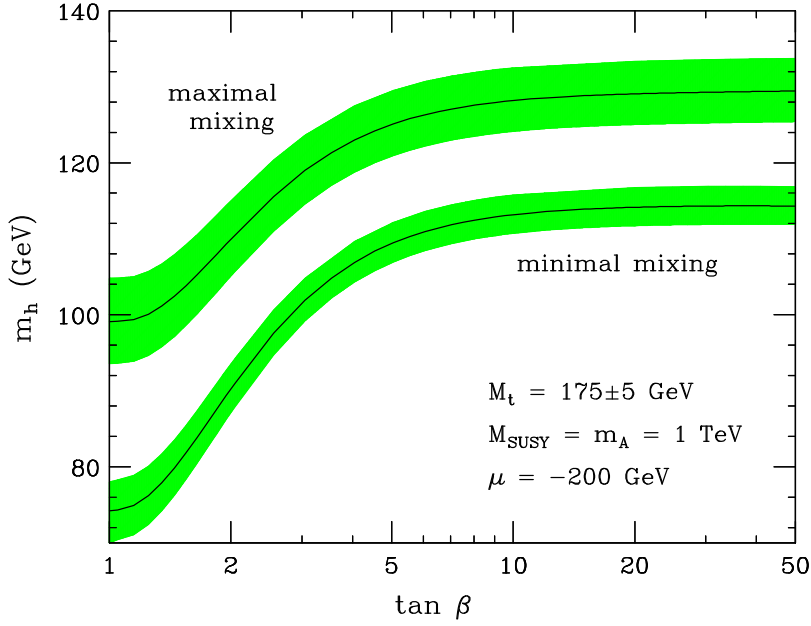,width=0.7\textwidth}}
\caption[0]{The radiatively corrected light CP-even Higgs mass is
plotted as a function of $\tanb$, for the maximal mixing [upper
band] and minimal mixing cases. The impact of the top quark mass
is exhibited by the shaded bands; the central value corresponds to
$m_t=175$~GeV, while the upper [lower] edge of the bands
correspond to increasing [decreasing] $m_t$ by 5~GeV.}
\label{mhtanb}
\end{figure}
Taking $\mha$ large, Fig.~\ref{mhtanb} illustrates that the maximal
value of the lightest CP-even Higgs mass bound is realized at
large $\tanb$ in the case of maximal mixing. Allowing for the
uncertainty in the measured value of $\mt$ and the uncertainty
inherent in the theoretical analysis, one finds for $M_{\rm SUSY}\lsim
2$~TeV that $\mhl\lsim \mhmax$, where
\begin{eqnarray} \label{mhmaxvalue}
\mhmax&\simeq&  122~{\rm GeV}, \quad
\mbox{minimal stop mixing,} \nonumber \\
\mhmax&\simeq&  135~{\rm GeV}, \quad \mbox{maximal stop mixing.}
\end{eqnarray}

The $\hl$ mass bound in the MSSM quoted above does not 
apply to non-minimal supersymmetric extensions of the Standard
Model. If additional Higgs singlet and/or triplet fields are
introduced, then new Higgs self-coupling parameters appear, which
are not significantly constrained by present data.  For example,
in the simplest non-minimal supersymmetric extension of the
Standard Model (NMSSM), the addition of a complex Higgs singlet 
field $S$ adds a new
Higgs self-coupling parameter, $\lambda_S$ \cite{singlets}. The mass
of the lightest neutral Higgs boson can be raised arbitrarily by
increasing the value of $\lambda_S$, analogous to the behavior of
the Higgs mass in the Standard Model. Under the assumption that
all couplings stay perturbative up to the Planck scale, one finds
in essentially all cases that $\mhl\lsim 200$~GeV, independent of
the details of the low-energy supersymmetric model~\cite{Espinosa:1998re}.

\begin{figure}[htb]
\centering
\psfig{file=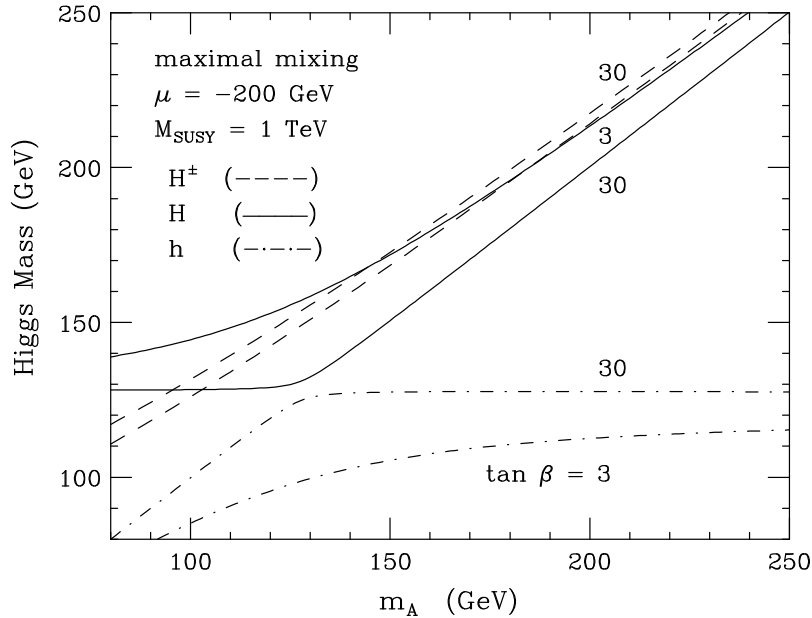,width=0.7\textwidth}
\caption[0]{\label{massvsma} Lightest CP-even Higgs mass ($\mhl$),
heaviest CP-even Higgs mass ($\mhh$) and charged Higgs mass ($\mhpm$) as
a function of $\mha$ for two choices of $\tan\beta=3$ and
$\tan\beta=30$. 
The slight increase in the charged Higgs mass as
$\tan\beta$ is increased from 3 to 30 is a consequence of the
radiative corrections.
}
\end{figure}
In Fig.~\ref{massvsma}, we exhibit the masses of
the CP-even neutral and the charged Higgs masses as a function of
$\mha$.  
Note that $\mhh\geq\mhmax$ for all values of
$\mha$ and $\tan\beta$, where $\mhmax$ is to be evaluated depending on
the top-squark mixing, as indicated in Eq.~(\ref{mhmaxvalue}).

Radiative corrections also significantly modify the tree-level
values of the Higgs boson couplings to fermion pairs and to vector
boson pairs. As discussed above, the tree-level Higgs couplings
depend crucially on the value of $\sinbma$.  In the first
approximation, when radiative corrections of the Higgs
squared-mass matrix are computed, the diagonalizing angle $\alpha$
is modified.  This provides one important source of the
radiative corrections of the Higgs couplings.   In Fig.~\ref{cosgraph},
we show the effect of radiative corrections on the value of
$\sinbma$ as a function of $\mha$  for different values of the
squark mixing parameters and $\tanb$.  One can then simply insert
the radiatively corrected value of $\alpha$ into
eqs.~(\ref{vvcoup}), (\ref{hvcoup}),  and (\ref{qqcouplings}) to
obtain radiatively improved couplings of Higgs bosons to vector
bosons and to fermions.

At large $\tanb$, there is another potentially important class of radiative
corrections in addition to those that enter through the modified $\alpha$.
These corrections arise in the relation between $m_b$ and $\tanb$ and 
depend on the details of the MSSM spectrum (which
enter via loop-effects).   
At tree-level, the Higgs couplings to
$b\bar b$ are proportional to the Higgs--bottom-quark Yukawa
coupling. Deviations from the tree-level relation
due to radiative corrections are calculable and finite
\cite{hffsusyqcd,deltamb,deltamb1,deltamb2,hffsusyprop}.  One of the
fascinating properties of such corrections is that in certain
cases the corrections do {\it not} vanish in the limit of large
supersymmetric mass parameters.  These corrections grow with
$\tanb$ and therefore can be significant in the large $\tanb$
limit. In the supersymmetric limit, $b\bar b$ couples only to
the neutral component of the $Y=-1$ Higgs doublet.
However, when supersymmetry is broken there will be 
a small coupling of $b\bar b$ to the neutral component of the $Y=+1$
Higgs doublet resulting from radiative corrections.  From this result, one can
compute the couplings of the physical Higgs bosons to $b\bar b$ pairs.
A useful approximation at large $\tanb$ yields the following
corrections to Eq.~(\ref{qqcouplings}):
\begin{eqnarray}
 \label{bbcouplings}
\hl b\bar b& :&~~~ -
{\sin\alpha\over\cos\beta}{1\over 1+\Delta_b}
\left[ 1 - \frac{\Delta_b\cot\alpha}{\tan\beta} \right]
\,,\nonumber\\[3pt]
\hh b\bar b& :&~~~
\phm{\cos\alpha\over\cos\beta}{1\over 1+\Delta_b}
\left[ 1 +\frac{\Delta_b\tan\alpha}{\tan\beta} \right]
\,,\nonumber\\[3pt]
\ha b \bar b& :&~~~\phm\gamma_5\,\frac{\tan\beta}{1+\Delta_b}\,,
\end{eqnarray}
where $\Delta_b\propto\tan\beta$.
The explicit form of $\Delta_b$ at one--loop in the limit of
$M_{\rm SUSY} \gg m_b$ is given in \cite{deltamb,deltamb1,deltamb2}.
The correction $\Delta_b$ arises from a bottom-squark--gluino loop,
which depends on the gluino mass and the supersymmetric Higgs mass
parameter $\mu$, and the top-squark--chargino loop, which depends on
the top-squark masses and the top-squark mixing parameters $\mu$ and $A_t$. 
Contributions proportional to the
electroweak gauge couplings have been neglected.

Similarly, the neutral Higgs couplings to $\tau^+\tau^-$ are modified
by replacing $\Delta_b$ in Eq.~(\ref{bbcouplings}) with $\Delta_\tau$
\cite{deltamb1,deltamb2}.
One can also derive radiatively corrected couplings
of the charged Higgs boson to fermion pairs \cite{chhiggstotop2,eberl}.
The tree-level couplings
of the charged Higgs boson to fermion pairs
are modified accordingly by replacing
$m_b \rightarrow m_b/(1 + \Delta_b)$ and
$m_{\tau} \rightarrow m_{\tau}/(1 + \Delta_{\tau})$,
respectively.

One consequence of the above results is that the neutral Higgs
coupling to $b\bar b$ (which is expected to be the dominant decay
mode over nearly all of the MSSM Higgs parameter space), can be
significantly suppressed at large $\tan\beta$~\cite{CMW,Wells,bdhty}
if $\Delta_b\simeq {\cal O}(1)$.  Typically $|\Delta_\tau|\ll
|\Delta_b|$, since the correction proportional to $\alpha_s$ in the latter is absent 
in the former.  For this reason,
the $\tau^+\tau^-$ decay mode can be the dominant Higgs decay
channel for the CP-even Higgs boson with SM-like couplings to gauge
bosons.  

In the decoupling limit, one can show that
$\cot\alpha\cot\beta=-1+{\cal O}(m_Z^2/\mha^2)$.  Inserting this
result into Eq.~(\ref{bbcouplings}), one can check that the 
$\hl b\bar b$ coupling does indeed approach its Standard Model value.
However, because $\Delta_b\propto\tan\beta$, the deviation of
the $\hl b\bar b$ coupling from the corresponding SM result is of 
${\cal O}(m_Z^2\tan\beta/\mha^2)$.  That is, at large $\tan\beta$,
the approach to decoupling may be ``delayed'' 
\cite{loganetal}, depending on the
values of other MSSM parameters that enter the radiative
corrections.

\subsection{MSSM Higgs boson decay modes}
\label{secec}

In this section, we consider the decay properties of
the three neutral Higgs bosons
($\hl$, $\hh$ and $\ha$) and of the charged Higgs pair ($\hpm$).
Let us start with the lightest state, $\hl$.
When $\mha\gg m_Z$, the decoupling limit
applies, and the couplings of $\hl$ to SM particles are nearly
indistinguishable from those of $h_{\rm SM}$.
If some superpartners are light, there may be some additional decay
modes, and hence the $\hl$ branching ratios would be different
from the corresponding Standard Model values, even though
the partial widths to Standard Model particles are the same.
Furthermore, loops of light charged or colored superpartners could modify
the $\hl$ coupling to photons and/or gluons, in which case the one-loop
$gg$ and $\gamma\gamma$ decay rates would also be different.
On the other hand, if all superpartners are heavy, all the decay
properties of $\hl$ are essentially those of the SM Higgs boson, and the
discussion of Section~\ref{secca} applies.

The heavier Higgs states, $\hh$, $\ha$ and $\hpm$, are roughly 
mass-degenerate and have negligible couplings to vector boson pairs.
In particular, $\Gamma(\hh\to VV)\ll\Gamma(h_{\rm SM}\to VV)$, while the
couplings of $\ha$ and $\hpm$ to the gauge bosons are loop-suppressed.
The couplings of $\hh$, $\ha$ and $\hpm$
to down-type (up-type) fermions are significantly
enhanced (suppressed) relative to those of $h_{\rm SM}$ if $\tanb\gg 1$.
Consequently, the decay modes $\hh,\ha \to b\bar b$,
$\tau^+\tau^-$ dominate the neutral Higgs decay modes for 
moderate-to-large values of
$\tanb$ below the $t\bar t$ threshold,
while $H^+\to\tau^+\nu$  dominates the charged Higgs decay below the
$t\bar b$ threshold.

For values of $\mha$ of order $\mz$, all Higgs boson states lie
below 200~GeV in mass, and would all be accessible at the LC.
In this parameter
regime, there is a significant area of the parameter space in which
none of the neutral Higgs boson decay properties approximates those of
$h_{\rm SM}$.  For example, when
$\tan\beta$ is large, supersymmetry-breaking effects can significantly
modify the $b\bar b$ and/or the $\tau^+\tau^-$ decay rates with
respect to those of $h_{\rm SM}$.
Additionally, the heavier Higgs bosons can decay into lighter
Higgs bosons.  Examples of such decay modes are: $\hh\to \hl\hl$,
$\ha\ha$, and $Z\ha$, and $\hpm\to W^\pm\hl$, $W^\pm\ha$ (although in
the MSSM, the Higgs branching ratio into vector boson--Higgs boson
final states, if kinematically allowed, rarely exceeds a few percent).  
The decay of the heavier Higgs boson into two lighter Higgs bosons can
provide information about Higgs self-couplings.  
For values of $\tanb\lsim 5$, the branching ratio of
$\hh\to \hl\hl$ is dominant for a Higgs mass range of $200~{\rm
GeV}\lsim\mhh\lsim 2m_t$. The dominant radiative corrections to
this decay arise from the corrections to the self-interaction
$\lambda_{\hh\hl\hl}$ in the MSSM and are large \cite{7}.

The phenomenology of charged Higgs bosons is less model-dependent,
and is governed by the values of $\tanb$ and $\mhpm$. Because
charged Higgs couplings are proportional to fermion masses, 
the decays to third-generation quarks and leptons are
dominant.  In particular, for $\mhpm<m_t+m_b$ (so that the channel
$H^+\to t\bar b$ is closed), $H^+\to\tau^+\nu_\tau$ is favored if
$\tan\beta\gsim 1$, while $H^+\to c\bar s$ is favored only if
$\tan\beta$ is small. Indeed, ${\rm BR}(H^+\to\tau^+\nu_\tau)
\simeq 1$ if $\tan\beta\gsim 5$.  These results apply
generally to Type-II two-Higgs doublet models. For $\mhpm\gsim
180$~GeV, the decay $H^+\to t\bar b\to W^+b \bar b$ is the
dominant decay mode.

In addition to the above decay modes, there exist new Higgs decay
channels that involve supersymmetric final states. Higgs decays
into charginos, neutralinos and third-generation squarks and
sleptons can become important, once they are kinematically allowed
\cite{13a}. For Higgs masses below 130~GeV, the range of
supersymmetric parameter space in which supersymmetric decays are
dominant is rather narrow when the current bounds on
supersymmetric particle masses are taken into account.  One
interesting possibility is a significant branching ratio of
$\hl\to\widetilde\chi^0 \widetilde\chi^0$, which could arise  for
values of $\mhl$ near its upper theoretical limit.  
Such an invisible decay mode
could be detected at the LC by searching for the missing mass
recoiling against the $Z$ in $e^+e^-\to\hl Z$.

\subsection{MSSM Higgs boson production at the LC}
\label{seced}

For $\mha\gsim 150$~GeV, Fig.~\ref{cosgraph} shows that the MSSM Higgs
sector quickly approaches the decoupling limit, where
the properties of $\hl$ approximately coincide with
those of $h_{\rm SM}$.  Thus, the Higgsstrahlung and 
vector-boson-fusion cross-sections for $h_{\rm SM}$ production also apply to $\hl$
production.  In contrast, the $\hh VV$ and $\ha VV$ 
couplings are highly suppressed, 
since $|\cos(\beta-\alpha)|\ll 1$.  Equation~(\ref{vvcoup})
illustrates this for the $H^0W$ coupling.  Thus, 
these mechanisms are no longer useful for $\hh$ and $\ha$
production.  The most robust production mechanism is $e^+e^-\to 
Z^*\to \hh\ha$, which is not suppressed since the $Z\hh\ha$ coupling
is proportional to $\sin(\beta-\alpha)$, as indicated in Eq.~(\ref{hvcoup}).  
Radiatively corrected cross-sections for $Z\hl$, $Z\hh$, $\hh\ha$, and
$\hl\ha$ have been recently obtained in \cite{rosiek}.  The
charged Higgs boson is also produced in pairs via $s$-channel photon
and $Z$ exchange.
However, since $\mhh\simeq\mha\simeq\mhpm$ in the decoupling limit, $\hh\ha$
and $\hp\hm$ production are kinematically allowed only when 
$\mha\lsim\sqrt{s}/2$.\footnote{The pair production of scalars is
P-wave suppressed near threshold, so in practice the corresponding
Higgs mass reach is likely to be somewhat lower than $\sqrt{s}/2$.}
In $\gamma\gamma$ collisions, one can extend the Higgs mass reach for
the neutral Higgs bosons.  As described
in Section~\ref{secj}, the $s$-channel resonant production of
$\hh$ and $\ha$ (due primarily to the top and bottom-quark loops in
the one-loop Higgs--$\gamma\gamma$ triangle)
can be detected for some choices of $\mha$ and $\tanb$ 
if the heavy Higgs masses are less than
about 80\% of the initial $\sqrt{s}$ of the primary $e^+e^-$ system.
The corresponding cross sections are a 
few~fb \cite{Gunion:1993ce,Muhlleitner:2001kw}.

If $\mha\lsim 150$~GeV, deviations from the decoupling limit become
more apparent, and $\hh$ can now be produced via Higgsstrahlung and
vector boson fusion at an observable rate.  In addition, 
the factor of $\cos(\beta-\alpha)$ in the $Z\hl\ha$ coupling
no longer significantly suppresses $\hl\ha$ production.
Finally, if $\mhpm\lsim
170$~GeV, the charged Higgs boson will also be produced in $t\to H^+ b$.
In the non-decoupling regime, all non-minimal
Higgs states can be directly produced and studied at the LC.

The associated production of a single Higgs boson and a
fermion-antifermion pair can also be considered.  Here, the new
feature is the possibility of enhanced Higgs--fermion Yukawa couplings.
Consider the behavior of the Higgs couplings at large
$\tan\beta$, where some of the Higgs couplings to down type
fermion pairs (denoted generically by $b\bar b$) 
can be significantly enhanced.\footnote{We do not consider
the possibility of $\tan\beta\ll 1$, which would lead to enhanced Higgs
couplings to up-type fermions.  In models of low-energy
supersymmetry, there is some theoretical prejudice that suggests that
$1 \lsim\tanb \lsim m_t/m_b$, with the fermion masses evaluated
at the electroweak scale.
For example, $\tanb\lsim 1$ is disfavored
since in this case, the Higgs--top quark 
Yukawa coupling blows up at an
energy scale significantly below the Planck scale.
The Higgs-bottom quark Yukawa coupling has a similar problem if
$\tanb \gsim m_t/m_b$.
As noted in Section~\ref{secfa}, some of the low $\tan\beta$ region is already
ruled out by the MSSM Higgs search.}
Let us examine two
particular large $\tan\beta$ regions of interest.  In the
decoupling limit (where $\mha\gg\mz$ and $|\cos(\beta-\alpha)|\ll
1$), it follows from
Eq.~(\ref{qqcouplings}) that the $b\bar b\hh$
and $b\bar b\ha$ couplings have equal strength and are significantly
enhanced by a factor of
$\tanb$ relative to the $b\bar bh_{\rm SM}$ coupling, while the
$b\bar b\hl$ coupling is given by the corresponding Standard Model value.
If $\mha\lsim \mz$ and $\tanb\gg 1$, then $|\sin(\beta-\alpha)|\ll
1$, as shown in Fig.~\ref{cosgraph}, and $\mhl\simeq\mha$.
In this case, the $b\bar b\hl$
and $b\bar b\ha$ couplings have equal strength and are significantly
enhanced (by a factor of
$\tanb$) relative to the $b\bar bh_{\rm SM}$ coupling.\footnote{%
However in this case, the value of the $b\bar b\hh$ coupling can differ
from the corresponding $b\bar bh_{\rm SM}$ coupling when $\tanb\gg 1$,
since in case (ii), where $|\sin(\beta-\alpha)|\ll 1$,
the product $\tan\beta\sin(\beta-\alpha)$ need not
be particularly small.}  Note that in both cases above,
only two of the three neutral Higgs bosons have enhanced
couplings to $b\bar b$.   If  $\phi$ is one of the two neutral Higgs 
bosons with enhanced $b\bar b\phi$ couplings,
then the cross-section for $e^+e^-\to f\bar f\phi$ 
($f=b$ or $\tau$) will be
significantly enhanced relative to the corresponding Standard Model
cross-section by a factor of $\tan^2\beta$.  The phase-space
suppression is not as severe as in $e^+e^-\to t\bar t\phi$ (see
Fig.~\ref{ttbarhiggs}), so this process could extend the mass reach of the
heavier neutral Higgs states at the LC given sufficient luminosity. 
The production of the charged Higgs boson via  $e^+e^-\to t\bar b H^-$
is also enhanced by $\tan^2\beta$, although this process has a more
significant phase-space suppression because of the final state top quark.
If any of these processes can be observed, it would provide a direct
measurement of the corresponding Higgs--fermion Yukawa coupling. 

\section{MSSM Higgs boson searches before the LC}
\label{secf}

\subsection{Review of direct search limits}
\label{secfa}

Although no direct experimental evidence for the Higgs boson yet exists,
there are both experimental as well as theoretical constraints on
the parameters of the MSSM Higgs sector. Experimental limits on
the charged and neutral Higgs masses have been obtained at LEP.
For the charged Higgs boson, $\mhpm>78.7$~GeV \cite{LEPHIGGS}.
This is the most model-independent bound.  It is valid for more
general non-supersymmetric two-Higgs doublet models and assumes
only that the $H^+$ decays dominantly into $\tau^+\nu_\tau$ and/or
$c \bar s$. The LEP limits on the masses of $\hl$ and $\ha$ are
obtained by searching simultaneously for $e^+e^- \to Z \to Z\hl$
and $e^+e^- \to Z \to\hl\ha$.  
Radiative corrections can be significant, as shown in Section~\ref{seceb},
so the final limits depend on the choice of MSSM parameters that
govern the radiative corrections.  The third generation
squark parameters are the most important of these.  
The LEP Higgs working group \cite{LEPHIGGSgroup} quotes limits for
the case of $M_{\rm SUSY}=1$~TeV in the maximal-mixing scenario, which
corresponds to the choice of third generation squark parameters
that yields the largest corrections to $\mhl$.  The present LEP
95\%~CL lower limits are $\mha>91.9$~GeV and $\mhl>91.0$~GeV.  The
theoretical upper bound on $\mhl$ as a function of $\tanb$,
exhibited in Fig.~\ref{mhtanb}, can then be used to exclude a region of
$\tanb$ in which the predicted value of $\mhl$ lies below the
experimental bound.  Under the same MSSM Higgs parameter
assumptions stated above, the LEP Higgs search excludes the region
$0.5<\tanb<2.4$ at 95\%~CL.  

In discussing Higgs discovery 
prospects at the Tevatron and LHC, we shall quote
limits based on the assumption of $M_{\rm SUSY}=1$~TeV and maximal squark
mixing.  This tends to be a conservative assumption;  that is,
other choices give sensitivity to {\it more} of the $\mha$ versus $\tan\beta$ 
plane.  However, there are a number of other parameter regimes  
in which certain Higgs search strategies become more difficult.
While these issues are of vital importance to the Tevatron and LHC
Higgs searches, they are much less important at the LC.

\subsection{MSSM Higgs searches at the Tevatron}
\label{secfb}

\begin{figure}[tbh]
\centering
\centerline{\psfig{file=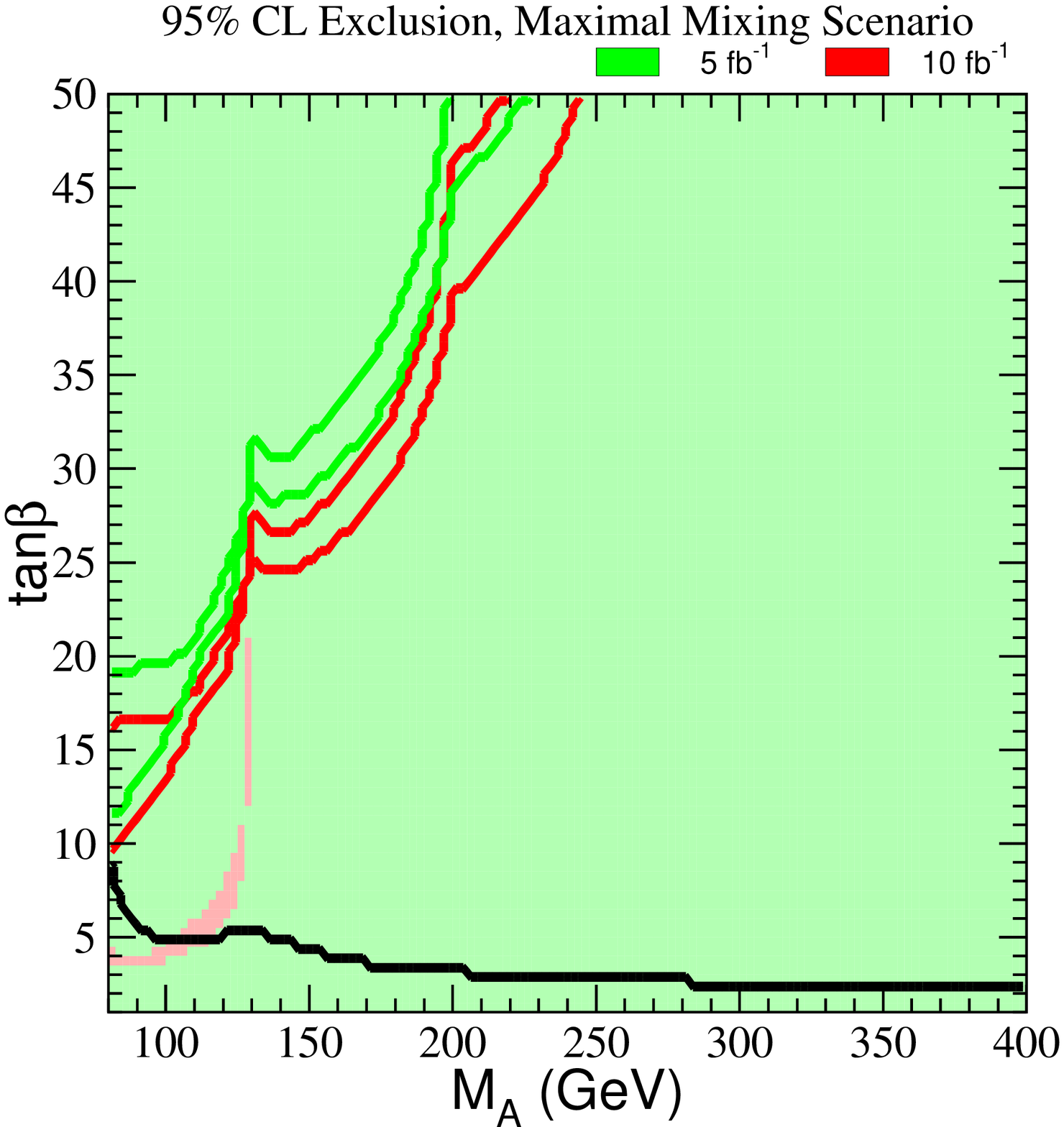,width=8cm}
\hfill
\psfig{file=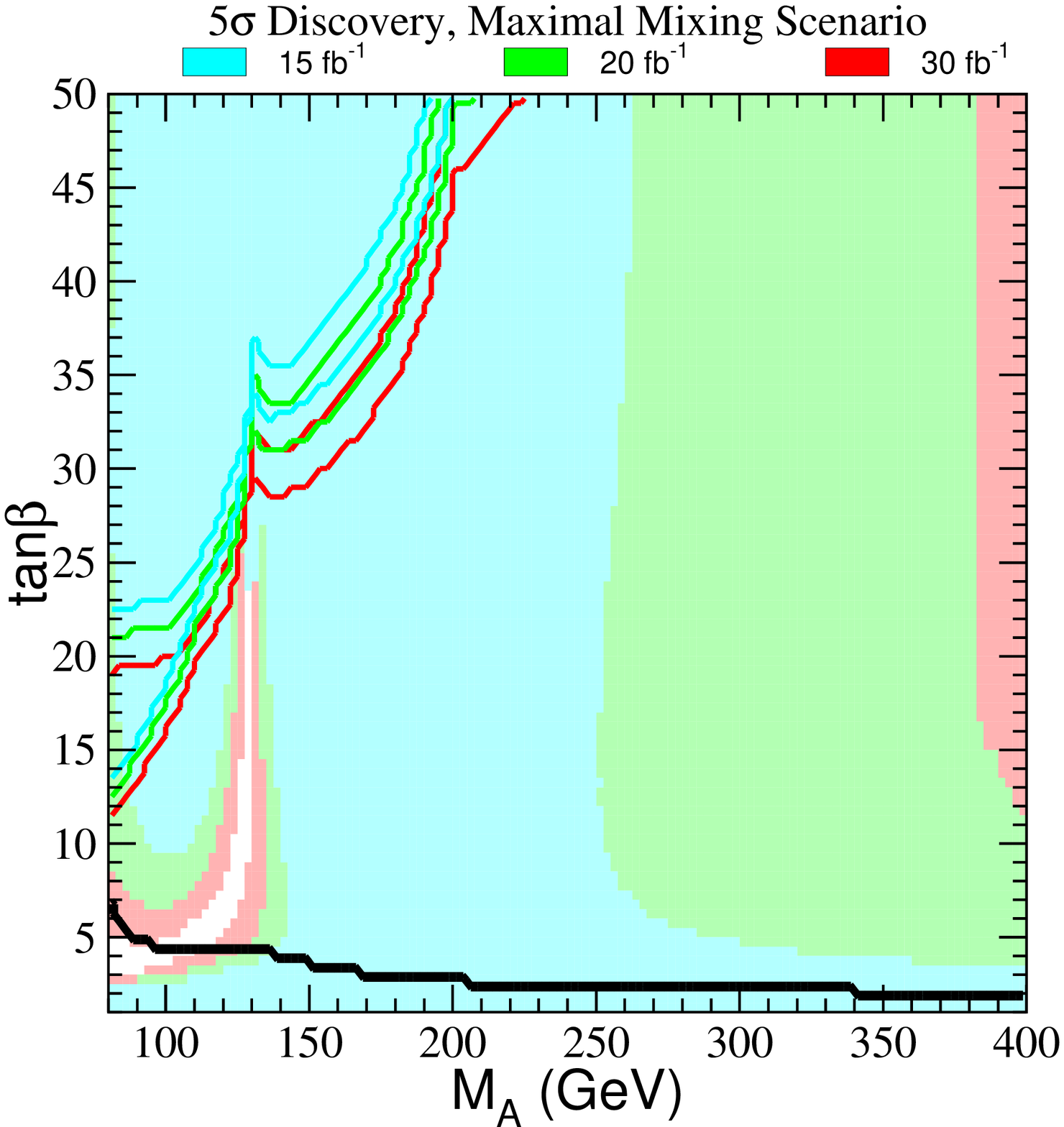,width=8cm}}
\vskip2pc
\caption[0]{\label{fullmhmax95} (a) 95$\%$ CL exclusion region
and (b) $5\sigma$ discovery region on the $\mha$--$\tan \beta$
plane, for the maximal mixing scenario
and two different search channels:
$q\bar q\to V\phi$ ($\phi=\hl$, $\hh$), $\phi\to b\bar b$
(shaded regions) and 
$gg$, $q\bar q\to b\bar b\phi$ ($\phi=\hl$, $\hh$, $\ha$),
$\phi\to b\bar b$ (region in the upper left-hand corner bounded by the
solid lines).  Different integrated 
luminosities are explicitly shown by the color coding.
The two sets of lines (for a given color) 
correspond to the CDF and D\O\ simulations, 
respectively.  The region below the solid black line near the bottom
of the plot is excluded by the absence of observed $e^+e^-\to Z\phi$
events at LEP2.
}
\end{figure}

At the Tevatron, the SM Higgs search can be reinterpreted in terms of
the search for the CP-even Higgs boson of the MSSM. Since the theoretical
upper bound was found to be $\mhl\lsim 135$~GeV  (for $M_{\rm SUSY}<2$~TeV),
only the Higgs search of the low-mass region, 100~GeV $\lsim\mhl\lsim
135$~GeV,  applies.  In the MSSM at large $\tanb$, the enhancement of
the $\ha b\bar b$ coupling (and a similar enhancement of either the
$\hl b\bar b$ or $\hh b\bar b$ coupling)  provides a new search
channel: $q\bar q$, $gg\to b\bar b\phi$, where $\phi$ is a neutral
Higgs boson with enhanced couplings to $b\bar b$.  Combining both
sets of analyses, the Tevatron Higgs Working Group obtained the
anticipated 95\% CL exclusion and 5$\sigma$ Higgs discovery contours for 
the maximal mixing scenario as a function of total integrated
luminosity per detector (combining both CDF and D0 data sets) shown in
Fig.~\ref{fullmhmax95} \cite{tevreport}.

From these results, one sees that 5~fb$^{-1}$ of 
integrated luminosity per experiment will allow one to
test nearly all of the MSSM Higgs parameter space at 95\% CL. 
To assure discovery of a CP-even Higgs boson at the 5$\sigma$ level,
the luminosity requirement becomes very important.
Figure~\ref{fullmhmax95}(b) shows that 
a total integrated luminosity of about 20~fb$^{-1}$ per experiment is 
necessary in order to assure a significant, although not exhaustive,
coverage of the MSSM parameter space.  If the anticipated 15~fb$^{-1}$
integrated luminosity is achieved, the discovery reach will
significantly extend beyond that of LEP.  A Higgs discovery would
be assured if the Higgs interpretation of the Higgs-like
LEP events is correct.
Nevertheless, the MSSM Higgs boson could still evade capture at the
Tevatron.  We would then turn to the LHC to try to obtain a definitive
Higgs boson discovery.

\subsection{MSSM Higgs searches at the LHC}
\label{secfc}

The potential of the LHC to discover one or more of the MSSM Higgs
bosons has been exhaustively studied for the minimal and maximal
mixing scenarios described above. One of the primary goals of these
studies has been to demonstrate that at least one of the MSSM Higgs
bosons will be observed by ATLAS and CMS
for any possible choice of $\tanb$ and $\mha$
consistent with bounds coming from current LEP data. In order to establish
such a `no-lose' theorem, 
an important issue is whether or not the Higgs bosons have substantial
decays to supersymmetric particle pairs.  It is reasonable to suppose
that these decays will be absent or relatively insignificant
for the light $\hl$.  Current mass
limits on SUSY particles are such that 
only $\hl\to\widetilde\chi^0_1\widetilde\chi^0_1$ might possibly be
kinematically allowed and this possibility arises only
in a very limited class of models. For $\mha\gsim 200\gev$,
decays of the $\ha,\hh,\hpm$
to SUSY pair states (especially pairs of light charginos/neutralinos)
are certainly a possibility, but the branching
ratios are generally not all that large.  The discovery
limits we discuss below would be weakened, but not dramatically.
Further, at high $\tanb$ the enhancement
of the $b\anti b$ and $\tau^+\tau^-$ couplings of the heavy $\ha$
and $\hh$ imply that SUSY decay
modes will not be important even for quite high $\mha\sim\mhh\sim\mhpm$.
We will summarize the LHC discovery prospects for the MSSM
Higgs bosons assuming that SUSY decays are not significant.

\begin{figure}[b!]
\centering \psfig{file=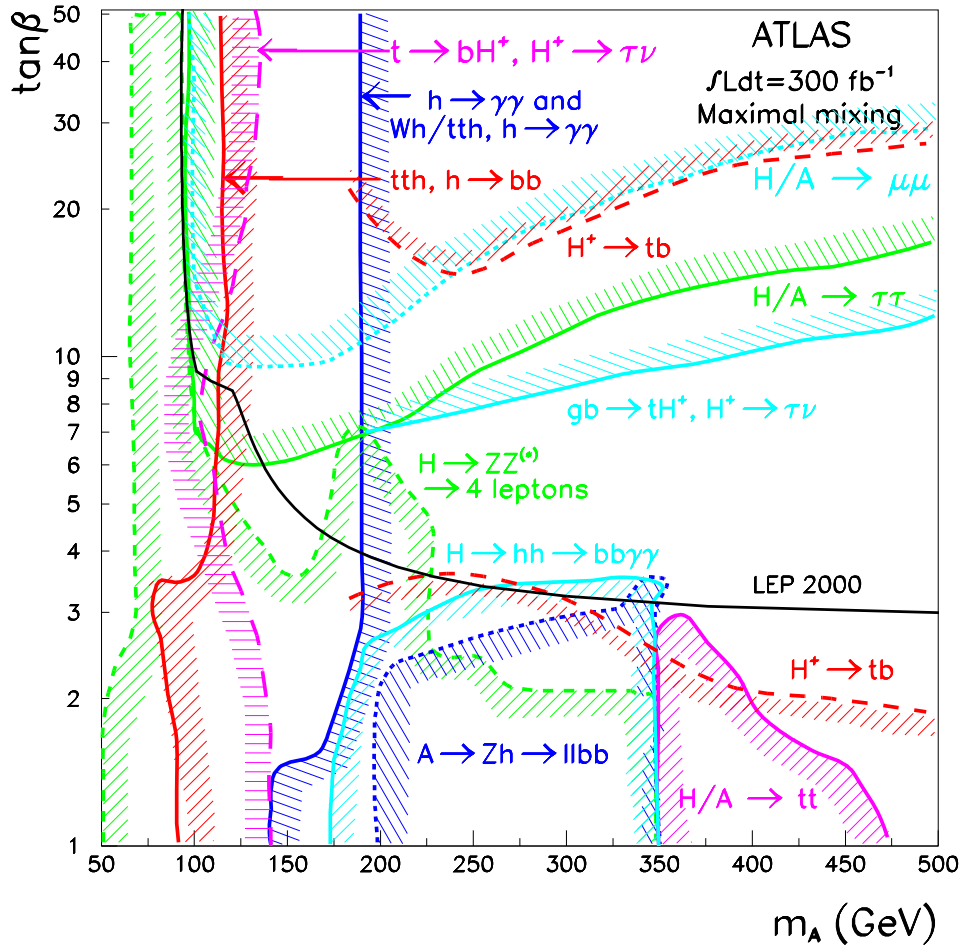,width=13cm}
\caption[0]{$5\sigma$ discovery contours for MSSM Higgs boson detection
in various channels are shown in the $[\mha,\tanb]$ parameter space, 
assuming maximal mixing and an integrated luminosity of $L=300\fbi$
for the ATLAS detector. This figure is preliminary \cite{atlasmaxmix}.}
  \label{f:atlasmssm}
\end{figure}

One of the primary Higgs discovery modes is detection of the relatively
SM-like $\hl$ using the same
modes as employed for a light $h_{\rm SM}$. Based on Fig.~\ref{f:atlasmssm} 
(which assumes $L=300\fbi$)
\cite{atlasmaxmix}, we see that for $\mha\gsim 180\gev$,
the $\hl$ will be detected via  $gg,WW\to\hl$ and $W\hl,t\anti t\hl$ with
$\hl\to \gam\gam$, while 
the $t\anti t\hl$ with $\hl\to b\anti b$ mode is viable
down to $\mha\gsim 100-120\gev$, depending on $\tanb$. 
There are also many possibilities for detecting the other MSSM Higgs
bosons.
We give a descriptive list. First, there is a small domain
in which $\mha\lsim 130\gev$, but yet $\mha$ is still large enough
for consistency with LEP limits, in which $t\to b\hpm$ discovery will
be possible.  However, the most interesting alternative
detection modes are based on $gg\to\ha,\hh$ and $gb\to \hpm t$
production. We focus first on the former. For low-to-moderate $\tanb$ values,
the channels $\hh\to ZZ^{(*)}\to 4\ell$, $\hh\to \hl\hl\to
b\anti b\gam\gam$ and $\ha\to Z\hl\to\ell\ell b\anti b$ 
are viable when $\mha\lsim 2m_t$, whereas the 
$\ha,\hh\to t\anti t$ modes are viable for $\mha>2m_t$.
For large enough $\tanb$ the
$gg\to \ha,\hh\to \tau^+\tau^-,\mu^+\mu^-$ discovery modes become viable.  
For the $gb\to \hpm t$ process, the
$\hpm\to t b$ decays provide a $5\sigma$ signal
both for low-to-moderate $\tanb\lsim 2$--3 and for high $\tanb\gsim
15$--25, depending upon mass. In addition, the $\hpm\to \tau^\pm \nu$
decay mode yields a viable signal for $\tanb\gsim 7$--12. Of course,
if the plot were extended to higher $\mha$, the minimum $\tanb$
value required for $\hh,\ha$ or $\hpm$ detection would gradually
increase.

It is important to notice that 
current LEP constraints exclude all of the low-to-moderate $\tanb$
regime in the case of maximal mixing (and, of course, even
more in the case of minimal mixing). 
Thus, it is very likely that $\tanb$ and $\mha$
will be in one of two regions: (a) the increasingly large (as
$\mha$ increases) wedge of moderate $\tanb>3$ in which only the $\hl$
will be detected; or, (b) the high $\tanb$ region for which
the $gg\to\hh,\ha\to \tau^+\tau^-,\mu^+\mu^-$ and 
$gb\to \hpm t\to \tau^\pm \nu t,tbt$ modes are viable as well.
If the $\hh,\ha,\hpm$ are heavy and cannot be detected 
either at the LHC (because $\tanb$ is not large enough) or at the LC
(because they are too heavy to be pair-produced),  precision measurements
of the $\hl$ branching ratios and other properties will 
be particularly crucial. The precision measurements might provide the
only means for constraining or approximately determining the value
of $\mha$ aside from possible direct detection in $\gam\gam\to \hh,\ha$
production. Expected LC precisions are such that deviations of
$\hl$ branching ratios from the predicted SM values can be detected 
for $\mha\lsim 700\gev$ \cite{Gunion:1996cn,deviations}.

At the LHC there is another important possibility for $\hl$ detection.
Provided that the mass of the second-lightest neutralino exceeds that of
the lightest neutralino (the LSP) by at least $\mhl$, gluino
and squark production will lead to chain decays in which 
$\widetilde\chi_2^0\to\hl\widetilde\chi_1^0$
occurs with substantial probability.  In this way, an enormous number
of $\hl$'s can be produced, and the $\hl\to b\anti b$
decay mode will produce a dramatic signal.  

\section{Non-exotic extended Higgs sectors}
\label{secg}

In this section, we consider the possibility of extending
only the Higgs sector of the SM, leaving unchanged the gauge
and fermionic sectors of the SM. We will also consider
extensions of the two-doublet Higgs sector of the MSSM.

The simplest extensions of the minimal one-doublet Higgs sector
of the SM contain additional doublet and/or singlet Higgs fields.
Such extended Higgs sectors will be called non-exotic (to distinguish
them from exotic Higgs sectors with higher representations, which will be
considered briefly in Section~\ref{seck}).
Singlet-only extensions have the advantage of not introducing
the possibility of charge violation, since there are no charged Higgs bosons.
In models with more than one Higgs doublet, tree-level Higgs-mediated 
flavor-changing neutral currents are present unless additional symmetries 
(discrete symmetries or supersymmetry) are introduced to restrict the 
form of the tree-level Higgs-fermion interactions \cite{gwp}.  
Extensions containing additional doublet fields allow for spontaneous
and explicit CP violation within the Higgs sector.  These could be
the source of observed CP-violating phenomena. 
Such models require that the mass-squared of the charged Higgs boson(s)
that are introduced be chosen positive in order to avoid
spontaneous breaking of electric charge conservation.

Extensions of the 
SM Higgs sector containing doublets and singlets can certainly be considered
on a purely {\it ad hoc} basis. But there are also many dynamical 
models in which the effective low-energy sector
below some scale $\Lambda$ of order 1 to 10 TeV, or higher, 
consists of the SM fermions and gauge bosons plus an extended Higgs
sector.  Models with an extra doublet of Higgs fields 
include those related to technicolor,  in which the effective Higgs
doublet fields are composites containing new heavier fermions.  See Chapter 5,
Section 3 for further discussion of this case. The heavy fermions
should be vector-like to minimize extra contributions to precision
electroweak observables. In many of these models,
the top quark mixes with the right-handed component of a new 
vector-like fermion. The top quark could also mix with the right-handed
component of a Kaluza-Klein (KK) excitation of a fermion field, so
that Higgs bosons would be composites of the top quark and fermionic
KK excitations. (For a review and references to the literature,
see \cite{Dobrescu:1999cs}.)
Although none of these
(non-perturbative) models have been fully developed, 
they do provide significant motivation for
studying the Standard Model with a Higgs sector containing
extra doublets and/or singlets if only as 
the effective low-energy theory below a scale $\Lambda$ in the TeV range.

When considering Higgs representations in the context
of a dynamical model with strong couplings
at scale $\Lambda$, restrictions on Higgs self-couplings and Yukawa
couplings that would arise by requiring perturbativity for such couplings
up to some large GUT scale do not apply.  At most, one should only demand 
perturbativity up to the scale $\Lambda$ at which the new (non-perturbative)
dynamics enters and the effective theory breaks down.

The minimal Higgs sector of the MSSM
is a Type-II two-doublet model, where one Higgs doublet ($H_d$)
couples at tree-level
only to down quarks and leptons while the other ($H_u$) couples
only to up quarks.  Non-minimal extended Higgs sectors 
are also possible in low-energy supersymmetric models.
Indeed, string theory realizations of low-energy supersymmetry often
contain many extra singlet, doublet and even higher representations, some of
which can yield light Higgs bosons  
(see, {\it e.g.}, \cite{Cvetic:2000nc}).
However, non-singlet Higgs representations spoil gauge coupling unification,
unless additional intermediate-scale matter fields 
are added to restore it.  
A particularly well-motivated extension is the inclusion of 
a single extra complex singlet Higgs field, often denoted $S$.
Including $S$, the superpotential
for the theory can contain the term $\lambda_S H_u H_d S$, which can then
provide a natural source of a weak scale value for the $\mu$
parameter appearing in the bilinear superpotential form $\mu H_u H_d$
required in the MSSM. A weak-scale value for
$s\equiv \VEV{S^0}$, where $S^0$ is the scalar component of the superfield $S$,
is natural and yields an effective $\mu=\lambda_S s$.
This extension of the MSSM is referred to as the next-to-minimal supersymmetric
model, or NMSSM, and has received considerable attention. For
an early review and references, see \cite{hhg}.

\subsection{The decoupling limit}
\label{secga}

In many extended Higgs sector models, the most natural parameter
possibilities correspond to a decoupling limit in which there is only
one light Higgs boson, with Yukawa and vector boson couplings close to
those of the SM Higgs boson.  In contrast, all the other Higgs bosons
are substantially heavier (than the $Z$) with negligibly small relative mass
differences, and with suppressed vector boson couplings (which vanish in
the exact limit of decoupling).  By assumption, the decoupling limit
assumes that all Higgs self-couplings are kept fixed and perturbative
in size.~\footnote{In the decoupling limit, the heavier Higgs bosons
may have enhanced couplings to fermions ({\it e.g.}, at large
$\tan\beta$ in the 2HDM).  We assume that these couplings also remain
perturbative.}  In the MSSM, such a decoupling limit arises for
large $\mha$, and quickly becomes a very good approximation for
$\mha\gsim 150$~GeV.

The decoupling limit can be evaded in special cases, in which the
scalar potential exhibits a special form ({\it e.g.}, a discrete
symmetry can forbid certain terms).  In such models, there could exist
regions of parameter space in which all but one Higgs boson are
significantly heavier than the $Z$, but the light scalar state does
{\it not} possess SM-like properties~\cite{Chankowski:2000an}. A
complete exposition regarding the decoupling limit in the 2HDM, and
special cases that evade the limit can be found in \cite{gunhabdecoup}.

\subsection{Constraints from precision electroweak data and LC implications}
\label{secgb}

In the minimal SM, precision electroweak
constraints require $m_{h_{\rm SM}}\lsim 230\gev$ at 90\% CL.
This is  precisely
the mass region preferred in the MSSM and its extensions.
However, in the context of general doublets + singlets
extensions of the Higgs sector
there are many more complicated possibilities.  First, it could
be that there are several, or even many, 
Higgs bosons that couple to vector bosons and it is only 
their average mass weighted by the square of their $VV$ coupling
strength (relative to the SM strength) that must obey this limit.
Second, there can be weak isospin violations either within the
Higgs sector itself or involving extra dynamics (for example related
to the composite Higgs approach) that can compensate for 
the excessive deviations predicted if there is a SM-like
Higgs with mass substantially above $\sim 230\gev$.  

\begin{figure}[htb]
\centerline{\psfig{file=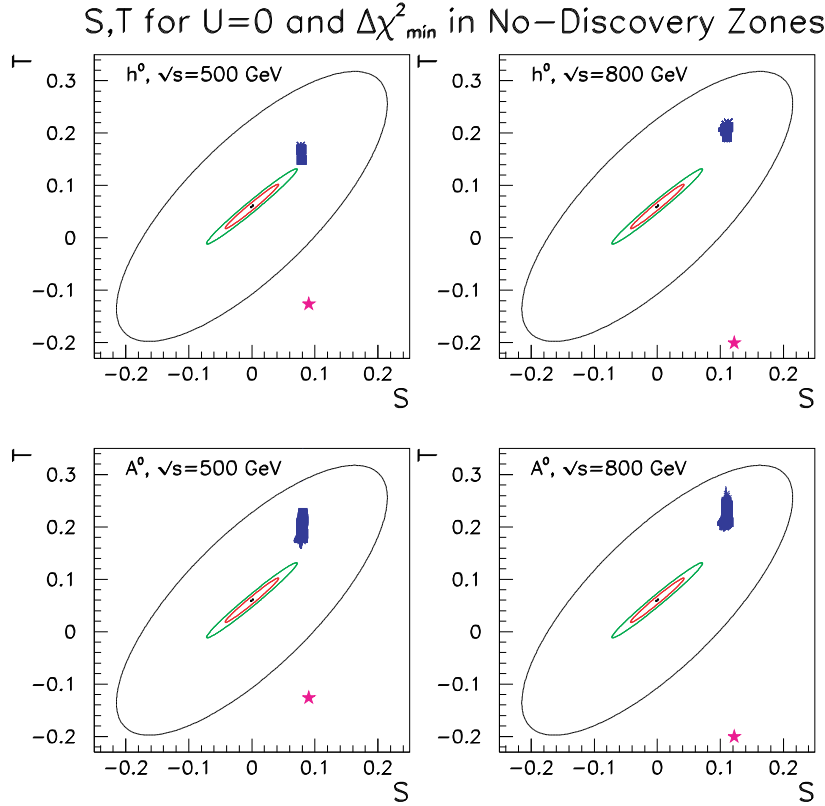,width=12cm}}
\caption[0]{\label{dsdt}
The outer ellipse gives the current 90\% CL region
for $U=0$ and SM Higgs mass of 115 GeV. The blobs show the
$S,T$ predictions for the 2HDM  models described
in the text that have minimum $\Delta\chi^2$
relative to this SM benchmark and for which no Higgs boson of the 2HDM
will be detected at the LC.
The innermost (middle) ellipse gives the 90\% (99.9\%) CL
region for $m_{h_{\rm SM}}=115\gev$ obtained after
Giga-Z precision measurements 
and a $\Delta m_W\lsim 6$ MeV threshold
scan measurement of $m_W$. The stars indicate the minimal 
SM $S,T$ prediction if $m_{h_{\rm SM}}=\sqrt s$.
}
\end{figure}

A particularly simple
example of this latter situation arises in the context of the 
2HDM \cite{Chankowski:2000an}.
Consider a 2HDM in which one of the CP-even neutral
Higgs bosons has SM-like couplings but has mass
just above a particular presumed value of $\sqrt s$ ($500$ or $800\gev$)
for the linear collider. In addition, focus on cases in which
there is a lighter $\ha$ or $\hl$ 
with no $VV$ coupling (for either, we use the notation $\what h$) 
and in which all other Higgs bosons have
mass larger than $\sqrt s$. Next, isolate mass and $\tanb$
choices for which detection of the $\what h$ 
will also be impossible at the LC. Finally, scan over masses of the
heavy Higgs bosons so as to achieve the smallest
precision electroweak $\Delta\chi^2$ relative to that found
in the minimal SM for $m_{h_{\rm SM}}=115\gev$. The blobs 
of overlapping points in Fig.~\ref{dsdt} 
indicate the $S,T$ values for the optimal choices
and lie well within the current 90\% CL ellipse.
The heavy Higgs boson with SM couplings gives a large positive contribution
to $S$ and large negative contribution to $T$, and in the absence
of the other Higgs bosons would give the $S,T$ location indicated
by the star.  However, there is an additional positive contribution
to $T$ arising from a slight mass non-degeneracy among the
heavier Higgs bosons.  For instance, for the case of a light $\what h=\ha$,
the $\hl$ is heavy and SM-like and 
\begin{equation}
   \Delta \rho\equiv\alpha\Delta T
=\frac{\alpha}{16 \pi m_W^2 c_W^2}\left\{\frac{c_W^2}{s_W^2}
   \frac{m_{H^\pm}^2-m_{H}^2}{2}-3m_W^2\left[\log\frac{m_{h}^2}{m_W^2}
   +\frac{1}{6}+\frac{1}{s_W^2}\log\frac{m_W^2}{m_Z^2}\right]\right\}
\label{drhonew}
\end{equation}
can be adjusted to place the $S,T$ prediction at the location of the 
blob in Fig.~\ref{dsdt}
by an appropriate choice of $\mhpm^2-\mhh^2$. 
Indeed, even if the ``light'' decoupled Higgs boson is not so light,
but rather has mass equal to $\rts$ (and is therefore unobservable),
one can still obtain entirely adequate agreement with current
precision electroweak data. Fortunately, one can only
push this scenario so far. To avoid moving beyond the current 90\%
ellipse (and also
to maintain perturbativity for the Higgs self-couplings), 
the Higgs with SM-like $VV$ coupling must have mass $\lsim 1\tev$.
 
In composite Higgs models with extra fermions, there are similar
non-degeneracies of the fermions that can yield a similar positive contribution
to $\Delta\rho$ and thence $T$.  As reviewed in \cite{peskinwells},
consistency with current precision electroweak data inevitably constrains
parameters so that some type of new physics (including a possible heavy
scalar sector) would again have to lie below a TeV or so.  
Future Giga-Z data could provide much stronger constraints
on these types of models, as discussed in Section~\ref{seci}.

\subsection{Constraints on Higgs bosons with $VV$ coupling}
\label{secgc}

In the MSSM, we know that the Higgs boson(s) that carry the $VV$ coupling must
be light: if $\mha$ is large (the decoupling limit) then it is
the mass-bounded $\hl$ that has all the $VV$ coupling strength; if
$\mha\lsim 2\mz$, then the $\hh$ can share the $VV$ coupling with
the $\hl$, but then $\mhh$ cannot be larger than  about $2\mz$.
In the NMSSM, assuming Higgs-sector
CP conservation, there are 3 neutral CP-even Higgs 
bosons, $h_{1,2,3}$ ($m_1<m_2<m_3$), 
which can share the $VV$ coupling strength. One can show 
(see \cite{Ellwanger:1999ji} for a recent update)
that the masses of the $h_i$ with substantial $VV$
coupling are strongly bounded from above.
This result generalizes to the most general supersymmetric Higgs 
sector as follows.  Labeling the neutral Higgs bosons by $i$
with masses $m_{h_i}$ and denoting the $ZZ$ squared-coupling relative to the SM
by $K_i$, it can be shown that
\begin{equation} 
\sum_i K_i \geq 1\,,\quad \sum_i K_i m_{h_i}^2\leq (200\gev)^2\,.
\label{hsumrules}
\end{equation}
That is, the aggregate strength of the $VV$ coupling-squared 
of all the neutral Higgs bosons is at least that of the SM, and the
masses-squared of the neutral $h_i$ weighted by the coupling-squared
must lie below a certain bound.  The upper bound
of $(200\gev)^2$ in Eq.~(\ref{hsumrules}) is obtained \cite{Espinosa:1998re} by 
assuming that the MSSM remains
perturbative up to the the GUT scale of order $10^{19}\gev$.
This bound applies for the most general
possible Higgs representations (including triplets) in
the supersymmetric Higgs sector and for arbitrary numbers
of representations. If only doublet and singlet representations are
allowed for, the bound would be lower.  The $(200\gev)^2$ 
bound also applies to general Higgs-sector-only extensions
of the SM by requiring consistency with precision electroweak
constraints {\it and} assuming the absence of a large contribution
to $T$ from the Higgs sector itself or from new physics,
such as discussed in Section~\ref{secgb}.

\subsection{Detection of non-exotic extended Higgs sector scalars
at the Tevatron and LHC}
\label{secgd}

In the case of extended Higgs sectors, all of
the same processes as discussed for the SM and MSSM will again
be relevant. However, we can no longer guarantee Higgs discovery
at the Tevatron and/or LHC.
In particular, if there are many Higgs bosons sharing
the $WW,ZZ$ coupling, Higgs boson discovery 
based on processes that rely on the $VV$ coupling could be much more difficult 
than in models with just a few light Higgs bosons with substantial $VV$
coupling.  This is true even if the sum rule of Eq.~(\ref{hsumrules}) applies.
For example, at the LHC even the NMSSM addition
of a single singlet to the minimal two-doublet structure
in the perturbative supersymmetric context 
allows for parameter choices such that no Higgs boson can be
discovered \cite{Gunion:1996fb} using any of the processes
considered for SM Higgs and MSSM Higgs detection. 
The $\gam\gam$ decay channel 
signals are all weak (because of decreased $W$-loop
contribution to the coupling). Further, if a moderate value
of $\tanb$ is chosen then
$t\anti t+$Higgs processes are small and $b\anti b+$Higgs
processes are insufficiently enhanced. In short, the 
equivalent to the wedge of Fig.~\ref{f:atlasmssm}
enlarges.  The $\hl$ signal is divided among the three
light neutral CP-even Higgs bosons and diluted to too low a statistical significance.

However, in other cases, the Tevatron and LHC could observe
signals not expected in an approximate decoupling limit.  For example,
in the 2HDM model discussed earlier the light $\what h$
with no $VV$ couplings  decays via $\what h \to b\anti b,\tau^+\tau^-$
and discovery in  $t\anti t \what h$, $b\anti b \what h$
and even $gg\to \what h$ \cite{gungrnew} is  possible, 
though certainly not guaranteed. Further, in these models
there is a heavy neutral Higgs boson having the bulk of the $VV$ coupling
and (for consistency with current precision
electroweak constraints or with perturbativity) mass $\lsim 1\tev$.  
This latter
Higgs boson would be detected at the LHC 
using $gg,WW$ fusion production and $ZZ\to 4\ell,WW\to 2j\ell\nu,\ldots$
decay modes, just like a heavy minimal SM Higgs boson.

\subsection{LC production mechanisms for non-exotic extended 
Higgs sector scalars} 
\label{secge}

Any physical Higgs eigenstate  with substantial $WW$ and $ZZ$ coupling
will be produced in Higgsstrahlung and $WW$ fusion at the LC.
Although there could be considerable cross section dilution and/or 
resonance peak overlap, the LC will nonetheless
always detect a signal.  This has been discussed for the MSSM
in Section~\ref{seced}.  In the NMSSM,
if one of the heavier CP-even $h_i$ has most
of the $VV$ coupling, the strong bound on its mass \cite{Ellwanger:1999ji}
noted earlier implies that it will be detected at any LC with 
$\sqrt s>350\gev$ within a small fraction of a 
year when running at planned luminosities. 
The worst possible case is that
in which there are many Higgs bosons 
with $VV$ coupling with masses spread
out over a large interval with separation smaller than the mass
resolution. In this case, the Higgs signal becomes a kind of continuum
distribution. Still, in \cite{Espinosa:1999xj} it is shown
that the sum rule of Eq.~(\ref{hsumrules}) guarantees
that the Higgs continuum signal will still be detectable for sufficient
integrated luminosity, $L\gsim 200\fbi$, as a broad excess in the recoil
mass spectrum of the $e^+e^-\to ZX$ process.  (In this case, $WW$ fusion events
do {\it not} allow for the reconstruction of Higgs events independently
of the final state Higgs decay channel.)
As already noted, the value of $200\gev$ appearing in Eq.~(\ref{hsumrules}) 
can be derived from perturbative RGE constraints
for the most general Higgs sector
in supersymmetric theories and is also required 
by precision electroweak data for general SM Higgs sector
extensions, at least in theories that do not have a large positive contribution
to $T$ from a non-decoupling structure in the Higgs sector
or from new physics not associated with the Higgs sector.

Other production modes of relevance include 
Higgs pair production, $t\anti t+$Higgs, and $b\anti b+$Higgs.
In multi-doublet models, $t\anti b H^-$ and $b\anti t H^+$
reactions are present.
However, none of these are guaranteed to be either kinematically accessible 
or, if accessible, to have a sufficiently high event rate to be observed.

Regardless of the production process,
relevant decay channels could include cases where 
heavier Higgs bosons decay to lighter ones.
If observed, such decays would provide vital information regarding
Higgs self-couplings.

\begin{figure}[htb]
\centerline{\psfig{file=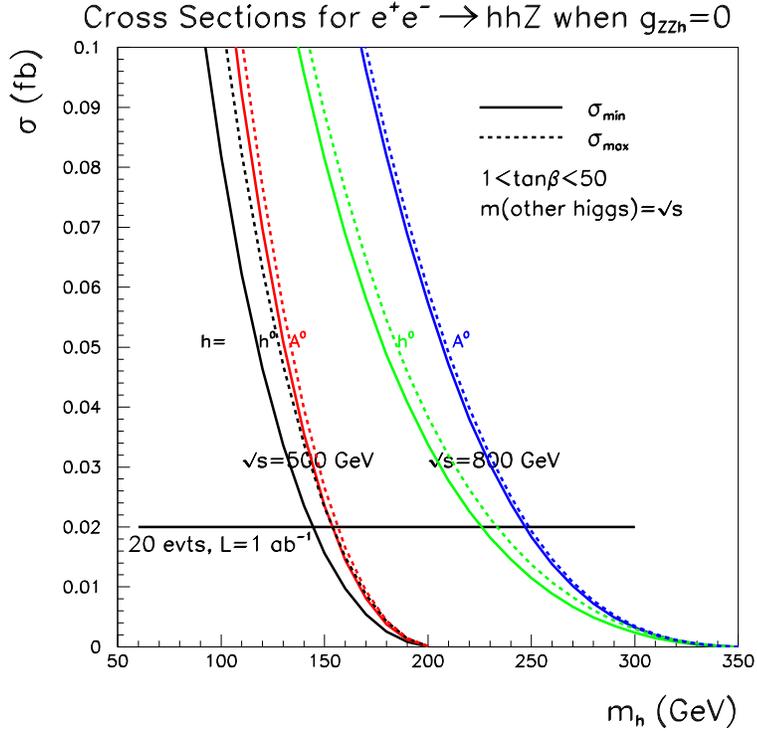,width=11cm}}
\caption{
For $\protect\sqrt s=500\gev$  and $800\gev$ and for $\what h=\hl$
and $\what h=\ha$,
we plot as a function of $m_{\what h}$ the maximum and minimum values of
$\sigma(e^+e^-\to \what h\what h Z)$  found after scanning $1<\tanb<50$
taking all other higgs masses equal to $\sqrt s$. For $\what h=\hl$, we
require $\sin(\beta-\alpha)=0$ during the scan. The 20 event level
for $L=1000$  fb$^{-1}$ is indicated.}
\label{f:hhz}
\end{figure}

We should particularly consider what production processes
are most relevant for those Higgs bosons (denoted $\what h$)
 that do not have substantial $VV$ coupling.
Such processes have particular relevance in the non-decoupling scenario
for the general 2HDM model discussed earlier. There,
such a $\what h$ is the only Higgs boson light enough to be produced at an
LC with $\rts\lsim 1\tev$ and it
cannot be produced and detected in $WW$ fusion or Higgsstrahlung.
Since the other Higgs bosons are heavy, the $\what h$ also cannot
be produced in association with another Higgs boson.
As shown in \cite{Grzadkowski:2000wj,Chankowski:2000an}, the
$b\anti b\what h$ and $t\anti t\what h$ processes will also
not be detectable at the LC if $\tanb$ is moderate
in value. The most interesting tree-level processes
are then those based on the quartic couplings $WW\what h\what h$ and $ZZ\what h
\what h$ required 
by gauge invariance \cite{Haber:1993jr,Djouadi:1996ah}.
These couplings allow for $WW\to \what h\what h$ fusion and 
$Z^*\to Z\what h\what h$
production, respectively.  The exact cross sections
for these processes are only mildly sensitive to the 
masses of the other heavier Higgs bosons via 2HDM Higgs 
self-couplings. Of course, phase space restrictions
imply an upper limit on the $\what h$ masses that can be probed in this way.
Cross sections in the case of $Z^*\to Z{\what h}{\what h}$ are plotted in
Fig.~\ref{f:hhz} for both $\what h=\ha$ and $\what h=\hl$ taking $\sqrt s=500$ 
 \cite{gunfarris}. Assuming optimistically
that 20 events in $L=1000\fbi$ could be detected, 
$Z^*\to Z\what h\what h$ could be detected for $m_{\what h}$ as large as
$150\gev$.  At $\sqrt s = 800\gev$, this limit increases to $250\gev$.
Similar results are obtained for $WW\to \what h\what h$ fusion production.


\section{Measurements of Higgs boson properties at the LC}
\label{sech}

The strength of the LC physics program is that it cannot only observe
one or more Higgs boson(s),
but also precisely determine the Higgs boson 
mass, width, couplings, and quantum numbers, and parameters of the Higgs potential.
These measurements are crucial to establish the nature of the Higgs and thus to
illuminate the mechanism of electroweak symmetry breaking. 
Measurements of the Higgs couplings can demonstrate that
a Higgs boson generates the masses of vector bosons, charged leptons, and
up- and down-type quarks.  If the measured couplings are not simply
proportional to mass, this will require a Higgs sector more complex
than a single complex Higgs doublet.  Accurate measurements are needed
to distinguish the SM Higgs and $\hl$ of the MSSM near the decoupling
limit. Couplings are determined
through measurements of Higgs branching ratios and cross sections.
Higgs bosons are also expected to couple to themselves, and this self-coupling
$\lambda$ can only be explored through the direct production of two or
more Higgs bosons.  The measurement of {\it direct} and {\it model independent}
absolute Higgs couplings is a major cornerstone of the LC program.

Details of some of the studies of Higgs coupling measurements can be found in
\cite{Battaglia:2000jb}.  A comprehensive description of
European studies using the simulated TESLA
detector can be found in \cite{tesla_report}.
North American studies consider simulations of detectors
with capabilities described in Chapter 15.
The program of measurements of Higgs boson properties strongly
impacts detector design. Measurement of branching ratios into fermions
requires sophisticated vertex detectors to separate $b$ from $c$
(and gluon) jets.  Precise recoil mass measurements 
need excellent momentum resolution (particularly for $\mu^+\mu^-$)
from charged particle tracking. 
The performance of the combined tracking and calorimetry systems
needs to result in precise jet-jet invariant masses, missing mass
measurements, and the ability to separate hadronic $W$ from hadronic
$Z$ decays.

The specific measurements used to determine the Higgs couplings to
vector bosons, fermions and scalars are significantly different
depending on the mass of the Higgs boson.  A generic neutral CP-even 
Higgs boson will be denoted by $h$  in this section.  We 
treat three cases separately: a light Higgs boson ($m_h < 2m_W$), an
intermediate mass Higgs boson ($2 m_W \leq m_h < 2m_t$), and a heavy
Higgs boson ($m_h \geq 2m_t$).

\subsection{Mass}
\label{secha}

In the Standard Model, the Higgs mass determines all its other
properties.  Thus, the precision of the mass 
measurement affects the comparison of
theory and experiment, for example, in a global fit of cross sections,
branching ratios, and precision electroweak data.  Similarly, in the
MSSM or other models with extended Higgs sectors, the masses of all the
Higgs bosons are an important input in determining the underlying
model parameters.

For this fundamental mass measurement, 
a LC can reconstruct the system recoiling against a $Z$ (independent of Higgs decay).
 Full event reconstruction, plus kinematic constraints, can improve resolution
and clean up mass tails.
For a light or intermediate mass Higgs boson, the 
optimal running conditions would have a smaller
center-of-mass energy such as $\sqrt{s} = 350$~GeV,  to allow  better 
momentum resolution and to minimize the beamstrahlung. 
Under such conditions, one can precisely measure the
recoil mass in $e^+e^- \rightarrow Z h$ events opposite to the
reconstructed leptonic decay $Z \rightarrow e^+e^-$ or 
$\mu^+\mu^-$.  This measurement is independent of the 
Higgs decay mode.
Accuracy can be improved by reconstructing specific decay modes,
leading, for example, to a four-jet topology where effective (5-C)
kinematic constrained fits can be employed.

\begin{figure}[htb]
\centering
\psfig{file=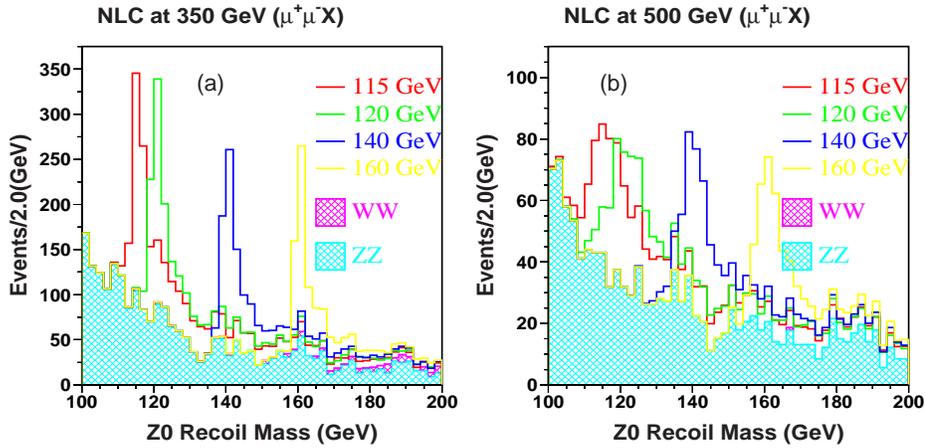,height=2.5in}
\caption{Recoil mass from a pair of leptons for different Higgs masses
at (a) $\sqrt{s} = 350$~GeV and (b) 500 GeV simulated in the 
L detector described in Chapter 15.}
\label{fig:haijun_mass}
\end{figure}

Figure~\ref{fig:haijun_mass} shows the distribution of the recoil mass,
\begin{equation}
M_{\rm recoil} = \sqrt{s - 2 \sqrt{s} \cdot E_{\ell^+ \ell^-} + 
M^2_{\ell^+ \ell^-}}\,,
\end{equation}
in a simulation of the L  linear collider detector~\cite{LCD} described in Chapter 15 for 
Higgs masses between 115 and 160 GeV~\cite{haijun_mass}. 
Using Monte Carlo
shape templates and an integrated luminosity of 500 fb$^{-1}$, precisions 
of $\Delta m_{h_{\rm SM}} \simeq 80$~MeV at $\sqrt{s} = 350$~GeV and
$\Delta m_{h_{\rm SM}} \simeq 140$~MeV at $\sqrt{s} = 500$~GeV have been estimated 
for either the $e^+e^-$ or $\mu^+ \mu^-$ mode.

\begin{figure}[htb]
\centering
\psfig{file=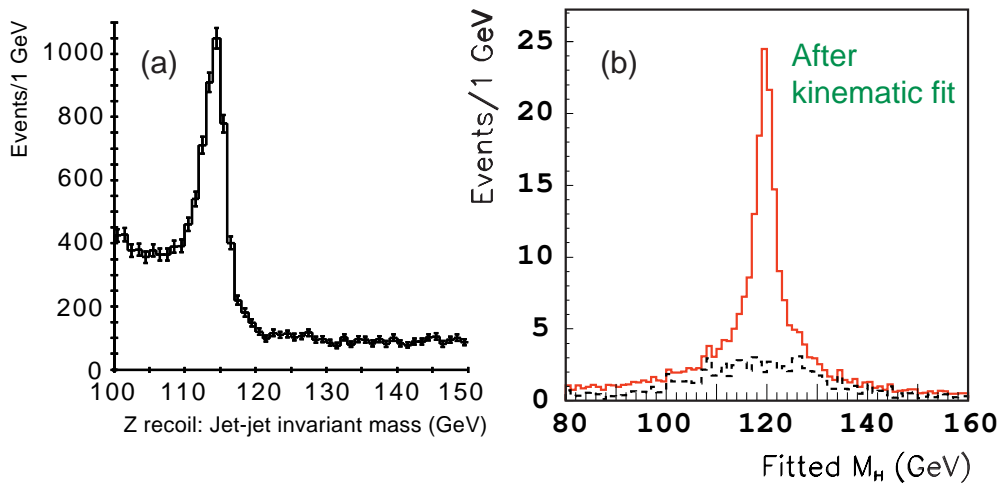,height=2.5in}
\caption{(a) Jet-jet invariant mass of the jets recoiling
from a $Z$ reconstructed hadronically simulated in the
LCD Large detector, $m_{h_{\rm SM}} = 115$~GeV.
(b) Direct reconstruction of the four-jet $q\bar{q}h_{\rm SM}$ state
simulated in the L detector after fitting with full kinematic
constraints, $m_{h_{\rm SM}} = 120$~GeV.}
\label{fig:juste_direct}
\end{figure}

Realistic simulations have also been made with the 
L detector for the process $Z h \rightarrow q\bar{q} h$ resulting 
in four jets.  
Figure~\ref{fig:juste_direct}(a) shows the jet-jet invariant mass
distribution for pairs of jets for Higgs with $m_{h_{\rm SM}} = 115$~GeV recoiling 
against a $Z$ reconstructed from its hadronic decay mode~\cite{Ronan}. 
A clean Higgs
signal with a mass resolution of approximately 2~GeV is observed. The central
Higgs mass is shifted down by the loss of low-energy charged and
neutral particles in the simulated event reconstruction.  A low-mass tail
of the Higgs signal arises from missing neutrinos in semi-leptonic $b$ and
$c$ quark decays.
Using neural net tags and full kinematic fitting~\cite{Juste:1999xv}, the
mass peak shown in Fig.~\ref{fig:juste_direct}(b) 
is obtained for $m_{h_{\rm SM}} = 120$~GeV, 
$\sqrt{s} = 500$~GeV, and 500~fb$^{-1}$ resulting in 
$\Delta m_{h_{\rm SM}} \simeq 50$~MeV.
If a second lower-energy IR is available, it might be attractive
to perform a scan across the $Z h$ threshold.  With a total integrated
luminosity of 100~fb$^{-1}$, $\Delta m_{h_{\rm SM}} \simeq 100$~MeV 
at $m_{h_{\rm SM}} = 150$~GeV
is achievable~\cite{Barger:1997pv}, competitive with the methods above.

Further work is necessary to confirm analogous precisions for heavier Higgs
bosons and MSSM Higgs bosons with different decay modes and possible 
close mass-degenera\-cies.  The number of $Z h$ events with 
$Z \rightarrow \ell^+ \ell^-$ for an intermediate-mass 
($m_h > 2m_W$) or heavy Higgs ($m_h > 2m_t$) with SM 
coupling falls quickly~\cite{fnal_report}.
In this case, and for the decays $ h \rightarrow ZZ$, hadronic decays
of the $Z$ would have to be considered to gain sufficient statistics.
For the heavier MSSM Higgs boson states,
European studies~\cite{kuskinen}
have shown typical mass precisions of
$\Delta m_{H^{\pm}}$ and $\Delta m_{\ha,\hh}$ of around 1~GeV for
500~fb$^{-1}$, but at $\sqrt{s} = 800$~GeV.
The MSSM $H^0$ and $A$ may be studied separately using $\gamma \gamma \rightarrow H/A$
with different states of $\gamma$ linear polarization, thus helping to refine 
mass determinations in the nearly degenerate case.

\subsection{Coupling determinations---light Higgs bosons}
\label{sechb}

\subsubsection{Cross sections}
\label{sechba}

For Higgs masses below $2 m_W$, the couplings $g_{h ZZ}$ 
and $g_{h WW}$ are best measured through measurements of the Higgsstrahlung
and $WW$ fusion cross sections, respectively.  These cross sections are
also critical in the extraction of branching ratios since the experimental
measurement will be a product of cross section and branching ratio.

\begin{figure}[htb]
\centering
\psfig{file=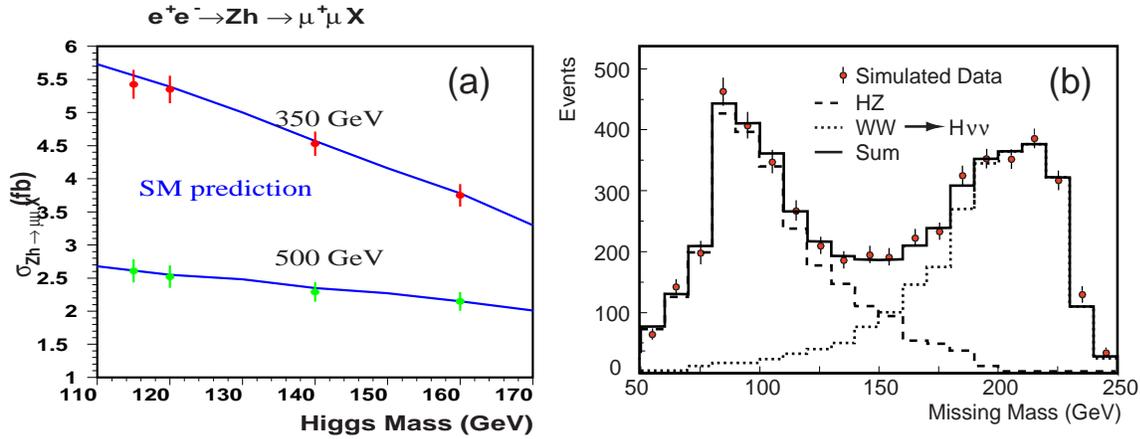,height=2.5in}
\caption{(a) Cross section measurement for 500~fb$^{-1}$ and
(b) separation of Higgsstrahlung and $WW$ fusion ($\sqrt{s} = 350$~GeV)
through a fit (after background subtraction), both simulated in the 
L detector.}
\label{fig:haijun_cross}
\end{figure}

Measurement of the cross section $\sigma(Z^* \rightarrow Zh)$ is best 
addressed via 
the recoil mass method
outlined above~\cite{haijun_mass}.  
Again, in this case, to reduce the contribution from
the $WW$ fusion process, it may be preferrable to run at a lower energy, {\it i.e.}, 
$\sqrt{s} = 350$~GeV, and to examine recoil against $\mu^+\mu^-$ to avoid
large Bhabha backgrounds. The study with the L detector described
above finds $\Delta\sigma / \sigma \simeq 4$\% at $\sqrt{s} = 350$~GeV 
and $\simeq$6.5\% at 500 GeV with 500~fb$^{-1}$ as shown in 
Fig.~\ref{fig:haijun_cross}(a).  These agree roughly with estimates from
European studies~\cite{Garcia-Abia:1999kv}.

With efficient and pure $b$-jet tagging, events due to
$e^+e^- \rightarrow W^+W^- \nu \bar{\nu} \rightarrow \nu \bar{\nu}h
\rightarrow \nu \bar{\nu}b\bar{b}$ can be separated from those
due to Higgsstrahlung, 
$Zh  \rightarrow  \nu \bar{\nu}h \rightarrow \nu\bar{\nu}b\bar{b}$ by 
examining the missing mass distribution and
fitting to the expected shapes of a peak at $m_Z$ from Higgsstrahlung
and the higher missing masses from $WW$ fusion.  This technique
has been confirmed with simulations of the L detector as shown
in Fig.~\ref{fig:haijun_cross}(b)~\cite{rvk_separate}.  
With 500~fb$^{-1}$ and a precision 
$\BR(h_{\rm SM} \rightarrow b \bar{b}) \simeq 3$\% (see below), the fusion-process
cross section with this analysis can be found with a precision
$\Delta\sigma / \sigma$ = 3.5\% for $m_{h_{\rm SM}} = 120$~GeV.

\subsubsection{Branching ratios}
\label{sechbb}

A key advantage of the linear collider in Higgs studies is the 
identification of Higgsstrahlung $Zh$ events through the tag
of the $Z$ decays. This selection is essentially independent of the decay
mode of the $h$ and simplifies the measurement of Higgs boson branching
ratios.

Small beam sizes, the possibility of a first track measurement as close as
1~cm from the beam axis, and sophisticated pixel vertex detectors allow
for efficient and clean separation of quark flavors.  Separate tagging of
$b$, $c$ and $g$ jets is possible.

\begin{figure}[htbp]
\centerline{
\hskip -0.15in
\psfig{file=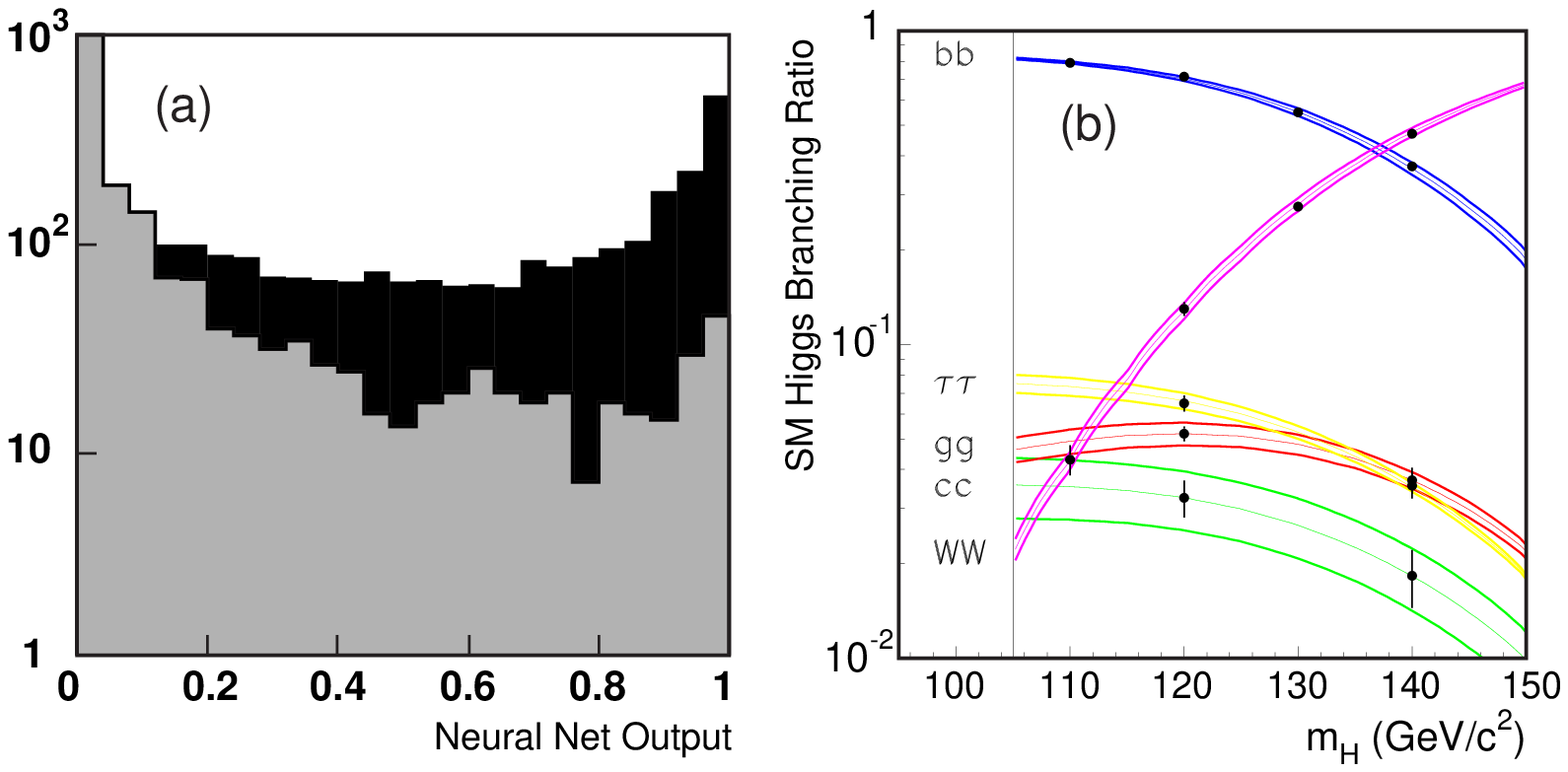,height=3in}}
\caption{(a) For the simulated L detector with CCD vertex 
detector, neural
net $h_{\rm SM} \rightarrow c\bar{c}$ output for $h_{\rm SM}\rightarrow c\bar{c}$ events
(dark) compared to output for $h_{\rm SM} \rightarrow b\bar{b}$ events (gray).
(b) Variation of branching ratios with SM Higgs mass (bands are 1$\sigma$
uncertainties on the theoretical predictions) and measurement 
precisions in the TESLA detector (points with error bars).}
\label{fig:nnet_brs}
\vskip.1in
\centering
\psfig{file=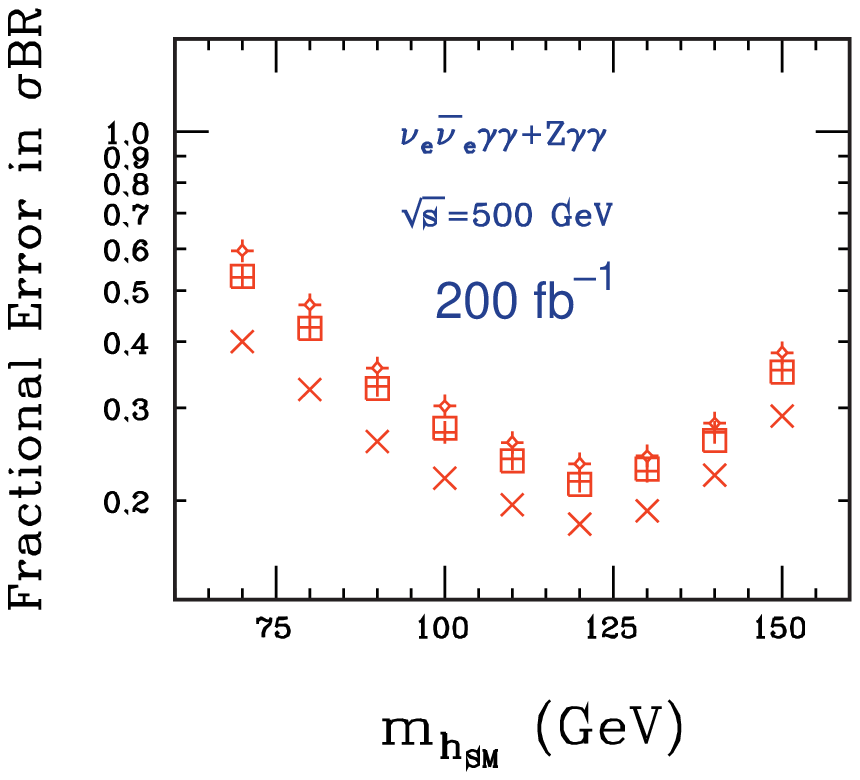,height=2.5in}
\caption{Fractional error on the branching ratio 
$\BR(h_{\rm SM} \rightarrow \gamma \gamma)$. The open squares are for a typical
LC detector electromagnetic energy resolution of 
$\Delta E/E = 10\% / \sqrt{E} \oplus 1.0\%$.}
\label{fig:hgamgam}
\end{figure}

In a study~\cite{brau_vtx} of vertexing using a CCD vertex detector in a standard 
LC detector configuration (C1 in \cite{brau_def}), 
topological vertexing~\cite{Jackson:1997sy} with neural net
selection was used for flavor (or anti-flavor, {\it i.e.},  $WW^*$) tagging.  
The separation of $b\bar b$ and $c\bar c$ events by this method is
illustrated in Fig.~\ref{fig:nnet_brs}(a).
Assuming 500~fb$^{-1}$ and 80\% polarization, the
results shown in Table~\ref{tab:higgs_br} were obtained.

\begin{table}[t!]
\centering
\begin{tabular}{l||c|c||c|c}  \hline\hline

    &  \multicolumn{2}{c||}{$m_{h_{\rm SM}} = 120$~GeV} & 
       \multicolumn{2}{c}{$m_{h_{\rm SM}} = 140$~GeV} \\ \hline
       
    & $\BR$ & $\delta\BR / \BR$ & $\BR$ & $\delta \BR /\BR$ \\ \hline

$h_{\rm SM} \rightarrow b\bar{b}$ &  $(69 \pm 2.0)$\% & 2.9\% & 
                            $(34 \pm 1.3)$\% & 4.1\% \\
                           
$h_{\rm SM}\rightarrow WW^*$     &  $(14 \pm 1.3)$\% & 9.3\% & 
                            $(51 \pm 1.8)$\% & 3.7\% \\

$h_{\rm SM}\rightarrow c \bar{c}$ & $(2.8 \pm 1.1)$\% & 39\% & 
                            $(1.4 \pm 0.64)$\% & 45\% \\
                            
$h_{\rm SM}\rightarrow gg$        & $(5.2 \pm 0.93)$\% & 18\% & 
                            $(3.5 \pm 0.79)$\% & 23\% \\  
                            
$h_{\rm SM}\rightarrow \tau^+ \tau^-$   & $(7.1 \pm 0.56)$\% & 7.9\% & 
                            $(3.6 \pm 0.38)$\% & 10\% \\  \hline\hline
                                                    
\end{tabular}
\caption{Predicted branching ratio precisions in the L detector
and typical vertex detector
configuration for 500~fb$^{-1}$ and $\sqrt{s} = 500$~GeV.}
\label{tab:higgs_br}
\end{table}

These results scale approximately as 
$( \sigma \int \L dt)^{-1/2}$ when taken together with other 
studies~\cite{hildreth,Borisov:1999mu,battaglia}, 
but the results of~\cite{battaglia} (shown in  Fig.~\ref{fig:nnet_brs}(b))
are noticeably more precise
for the $c\bar{c}$ and $gg$ modes.   These  branching ratio
measurements can then be used to either distinguish a SM Higgs boson
from an MSSM Higgs boson, or 
to probe
higher-mass states and
extract MSSM parameters such as $\mha$ even if the CP-odd $\ha$ is not 
accessible.  That analysis is described in more detail below.

An accessible decay mode for lighter Higgs bosons is 
$h \rightarrow \gamma \gamma$, which requires excellent electromagnetic
calorimetry.  As shown in Fig.~\ref{fig:hgamgam}, for a SM
Higgs boson in a typical LC detector, this is a difficult measurement 
requiring a large 
luminosity, which is best optimized for masses around 120 GeV~\cite{Gunion:1996qg}. A
higher-luminosity study~\cite{Boos:2000bz} with 1000~fb$^{-1}$ and 
$m_{h_{\rm SM}} = 120$ GeV for the TESLA detector finds $\delta\BR/\BR = 14$\%.
A $\gamma\gamma$ collider, discussed in Section~\ref{secj}, 
would be a more powerful tool for
determining the Higgs coupling to photons.

For light Higgs bosons, the coupling
to top quarks is still accessible via the radiative process $t\bar{t}h$
described below, or indirectly through $\BR(h \rightarrow gg)$.

A set of difficult decay channels for the LHC is invisible decays of the
Higgs boson into, {\it e.g.}, neutralinos, majorans or heavy neutrinos.
The LC can close this loophole
and measure the branching ratio easily,  even for branching ratios as small
as 5\% for a relatively narrow Higgs state, by using
the recoil mass method and demanding no detector activity opposite the
$Z$, or by comparing the number of events tagged with 
$Z \rightarrow \ell^+ \ell^-$ with the total number of observed Higgs decays
into known states.

\subsubsection{Radiative production and $t\bar{t}h$ coupling}
\label{sechbc}

For a light Higgs boson,  production through radiation
off a top quark is feasible, resulting in a final state of $t\bar{t}h$ .  This allows
a determination of the Yukawa top quark coupling 
$g_{h tt}$ \cite{Djouadi:1992tk,Dittmaier:1998dz}.
For a SM-like Higgs boson with $m_h = 120$~GeV, the $t\bar{t}h$ 
cross section is roughly
10 times larger at $\sqrt{s} = 700$--800 GeV than at 500~GeV.
At $\sqrt{s} = 800$~GeV, a statistical error of 
$\delta g_{h tt} / g_{h tt} \sim 5\%$ was 
estimated~\cite{Gunion:1996vv} for $L=500\fbi$ 
on the basis of an optimal observable analysis.
At $\sqrt{s} = 500$~GeV, a statistical error of 
$\delta g_{h tt} / g_{h tt} \simeq 21$\% is estimated~\cite{dawsontth}
using 1000~fb$^{-1}$.
A more sophisticated analysis using
neural net selections, full simulation, and the same integrated luminosity
at $\sqrt{s} = 800$~GeV finds a total error of 6\% on the 
coupling~\cite{Juste:1999af}. More details on this process can
be found in Chapter 6, Section 3.1.

\subsubsection{Higgs self-coupling}
\label{sechbd}

To delineate the Higgs sector fully, it is essential to  measure
the shape of the Higgs potential.  The cross section for double Higgs
production ({\it e.g.}, $Zhh$) is related to the triple Higgs coupling 
$g_{hhh}$, which in turn is related to the spontaneous
symmetry breaking shape of the Higgs potential.
The Higgs mass, $m_h^2=4\lambda v^2$, also measures the potential shape 
parameter $\lambda$, so independent determinations through $hh$ production
give a cross-check.
In the MSSM, a variety of double Higgs production processes would
be required to determine $g_{\hl\hl\hl}$,
$g_{\ha\hl\hl}$, {\it etc.} \cite{Djouadi:1996ah}.

These cross sections are low, and 
high integrated luminosity is needed, bolstered by polarization
and neural net selections. Experimental 
studies~\cite{Miller:1999ji,Castanier:2001sf} indicate that
for a SM-like Higgs boson with $m_h = 120$~GeV at $\sqrt{s} = 500$~GeV and
1000~fb$^{-1}$, a precision of $\delta g_{hhh} /g_{hhh} = 23$\% is
possible.  Regions of accessibility in MSSM parameters for 
MSSM Higgs self-couplings have also been 
determined~\cite{Lafaye:2000ec,Djouadi:1999ei}.

The cross section for SM triple Higgs production is very  low,
$\sigma(Zhh) < 10^{-3}$~fb, so measurement of the quartic coupling
$g_{hhhh}$ is hopeless with currently envisioned luminosities.

\subsubsection{Implications for the MSSM Higgs sector}
\label{sechbe}

The discussion of light Higgs coupling determinations has been based
on the assumption that the actual Higgs couplings to fermions, vector
bosons and scalars are close to the corresponding Standard Model
expectations.  In Section~\ref{secga}, it was argued that such an expectation
is rather generic, and applies to the decoupling limit of models of
Higgs physics beyond the Standard Model.  In particular, the
decoupling limit of the MSSM Higgs sector sets in rather rapidly once
$\mha\gsim 150$~GeV [see Section~\ref{secea}].  Since $\mhl\lsim 135$~GeV in
the MSSM [Eq.~(\ref{mhmaxvalue})], the precision study of $\hl$ using the
techniques discussed above can distinguish between $\hl$ and $h_{\rm SM}$
with a significance that depends on how close the model is to the
decoupling limit.  Said another way, the detection of deviations in
the Higgs couplings from their Standard Model predictions
would yield evidence for the existence of the
non-minimal Higgs sector, and in the context of the MSSM would provide
constraints on the value of $\mha$ (with some dependence on
$\tan\beta$ and other MSSM parameters that enter in the Higgs
radiative corrections).

In \cite{chlm}, the potential
impact of precision Higgs measurements at the LC on
distinguishing $\hl$ from $h_{\rm SM}$ was examined.  
The fractional deviation of the $\hl$ branching ratios into a
given final state from the corresponding result for $h_{\rm SM}$ (assuming
the same Higgs mass in both cases) is defined as:
\begin{equation}
        \delta \BR = \frac{\BR_{\rm MSSM} - \BR_{\rm SM}}
        {\BR_{\rm SM}}.
        \label{eq:deltaBR}
\end{equation}
For the MSSM Higgs boson decay, both $\mhl$ and the corresponding
branching ratios were computed including the radiative corrections due
to the virtual exchange of Standard Model and supersymmetric
particles, as described in Section~\ref{seceb}.  Thus, the $\hl$ branching
ratios depend on $\mha$ and $\tan\beta$ (which fix the tree-level MSSM
Higgs sector properties) and a variety of MSSM parameters that govern
the loop corrections.  Four scenarios were considered: the minimal and
maximal top-squark mixing cases [see Eq.~(\ref{mhmaxvalue}) and surrounding text],
and two additional cases with large $|\mu|=|A_t|$ (for $\mu A_t<0$
and two possible sign choices of $\mu$), where $\mu$ and $A_t$ control
the top-squark mixing.  In the latter two scenarios, significant
renormalization of the CP-even Higgs mixing angle $\alpha$ and 
$\Delta_b$ [see Eq.~(\ref{bbcouplings})] can arise.

\begin{figure}[t!]
\centerline{\psfig{file=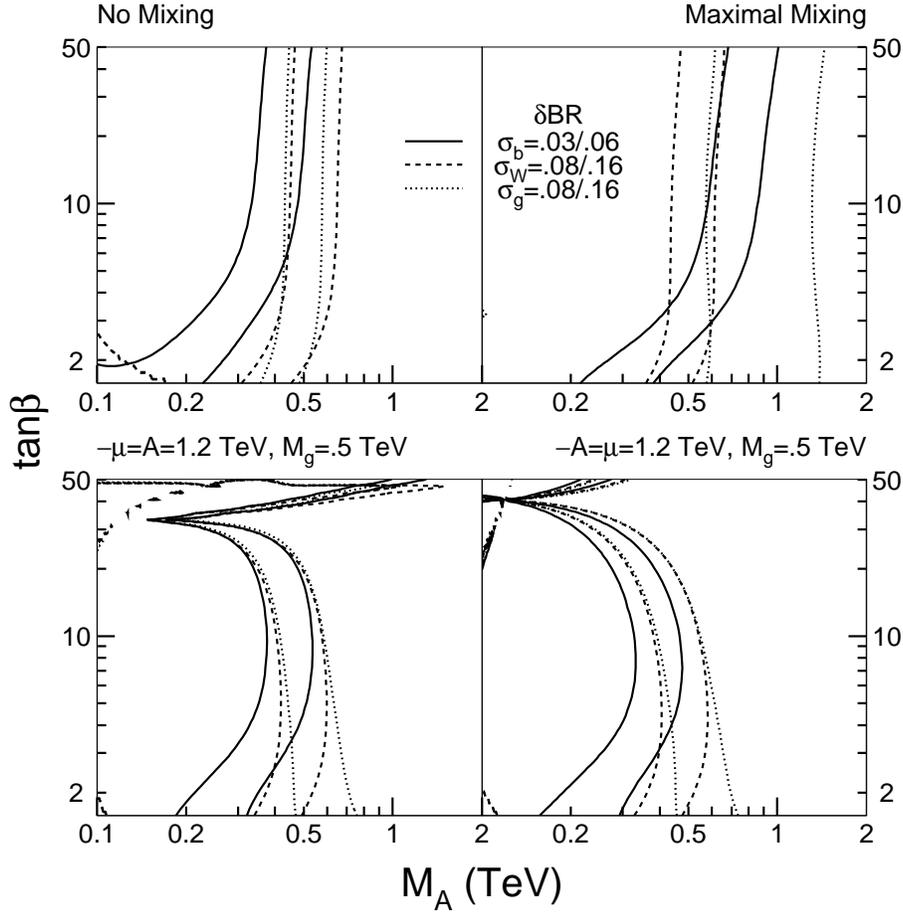,width=0.8\textwidth}}
\caption[0]{Contours of 
$\delta \BR(b\bar b) = 3$ and 6\% (solid), 
$\delta \BR(WW^*) = 8$ and 16\% 
(dashed) and 
$\delta \BR(gg) = 8$ and 16\% (dotted) [BR deviations defined in
Eq.~(\ref{eq:deltaBR})]
in the no ({\it i.e.} minimal) mixing scenario (top left),
the maximal mixing scenario (top right),
and the large $\mu$ and $A_t$ scenario with  
$\mu = -A_t = 1.2$ TeV (bottom left) and 
$\mu = -A_t = -1.2$ TeV (bottom right).  Taken from \cite{chlm}.}  
\label{mssmdbrs}
\end{figure}

In Fig.~\ref{mssmdbrs}, contours of $\delta \BR$ are plotted for three
$\hl$ decay modes:  $b\bar b$, $WW^*$ and $gg$.  The contours shown
correspond roughly to the $1\sigma$ and $2\sigma$ measurements claimed
by \cite{battaglia}, rescaled for the LC at
$\sqrt{s}=500$~GeV (see also the
$b\bar b$ and $WW^*$ branching ratio precisions given in 
Table~\ref{tab:higgs_br}).  In the minimal and maximal scenarios, the
dependence on $\mha$ is nearly independent of $\tan\beta$, and
demonstrates that one can achieve sensitivity to values of $\mha$ that
lie significantly beyond $\sqrt{s}/2$ where direct production at the
LC via $e^+e^-\to \hh\ha$ is kinematically forbidden.  However, the
cases with large $|\mu|=|A_t|$ exhibit the possibility of
``premature'' decoupling, that is, relatively low values of $\mha$ (at
a particular large value of $\tan\beta$) at which the properties of
$\hl$ and $h_{\rm SM}$ cannot be distinguished by the decay modes
considered above.\footnote{The premature decoupling is a consequence of the
renormalization of the mixing angle $\alpha$ which just happens to
yield $\cos(\beta-\alpha)=0$, in which case the $\hl$ couplings reduce
to those of $h_{\rm SM}$ as shown in Section~\ref{secea}.}
Thus, a measured deviation of Higgs branching ratios that
distinguishes $\hl$ from $h_{\rm SM}$ can place significant constraints on
the heavier non-minimal Higgs states, although the resulting constraints can
depend in a nontrivial way on the value of the MSSM parameters that
control the Higgs radiative corrections.

\subsection{Coupling determinations---intermediate mass Higgs bosons}
\label{sechc}

For $m_h < 2 m_W$, the measurement of branching ratios is extremely
rich, yielding couplings to both many of the fermions and bosons.
For larger masses, decays to $f\bar{f}$ become rarer until the threshold
for decays into top is crossed.  
In this intermediate mass range, a LC can measure
the $W$ and $Z$ couplings more precisely than the LHC
both through Higgs production rates and via branching ratios
for decays into these bosons. 
Whether the observed Higgs boson fully generates the $W$ 
and $Z$ mass can then be checked.

Precision electroweak measurements in the framework of the 
Standard Model indirectly predict~\cite{erler,degrassi} 
$m_{h_{\rm SM}}\lsim 205$--230~GeV at 95\% CL, and a Higgs observed
with mass much greater than this would imply new physics. At this point,
measurements from a Giga-Z dataset would be particularly useful to
probe this new sector.

\subsubsection{Cross sections}
\label{sechca}

Techniques described earlier~\cite{haijun_mass,rvk_separate} for 
cross section measurements of both the Higgsstrahlung and $W$-fusion
processes, with subsequent Higgs decays into $b\bar{b}$,
can still be used for the lower portion of the intermediate
mass range, {\it i.e.}, $m_h \sim 160$~GeV.  Even in this intermediate mass
range, it is beneficial to run at the peak of the cross section
at roughly $m_h + m_Z + 50$~GeV. 
The typical precisions that can be obtained are 
$\Delta\sigma(Zh_{\rm SM}) / \sigma(Zh_{\rm SM}) \simeq 5$\% and
$\Delta\sigma(\nu \bar{\nu}h_{\rm SM}) / \sigma(\nu \bar{\nu}h_{\rm SM}) \simeq 17$\%
for $m_{h_{\rm SM}} = 160$~GeV,
at $\sqrt{s} = 350$~GeV 
with 500~fb$^{-1}$.

For heavier Higgs bosons in this mass range, cross sections for
both Higgsstrahlung and $W$-fusion will
need to be extracted from using the decay
$h\rightarrow WW^*$,  for example, as described in~\cite{Borisov:1999mu}. 
Couplings determined from
$t\bar{t}h$ and $Zhh$ production would clearly need higher $\sqrt{s}$.

\subsubsection{Branching ratios}  
\label{sechcb}

Using Higgsstrahlung events at an optimal $\sqrt{s}$, the statistical
error on $\BR(h_{\rm SM}\rightarrow b\bar{b})$ is still only 6.5\% at $m_{h_{\rm SM}} =
160$~GeV~\cite{battaglia}.  At $\rts=500\gev$, with leptonic decays of
the $Z$ only, the statistical error on this branching ratio reaches
25\% at $m_{h_{\rm SM}} \simeq 165$~GeV with 250~fb$^{-1}$ and remains below 30\%
for $m_{h_{\rm SM}} < 200$~GeV with 2000~fb$^{-1}$~\cite{fnal_report}.  However,
in addition to the leptonic decays of the $Z$, hadronic decays can
also be used to  tag the associated $Z$.  Extrapolating
from full LCD detector simulations, it is conservatively estimated
that including the hadronic decays of the $Z$ results in an increase
in signal statistics above background by a factor of four.  With these
assumptions and 500~fb$^{-1}$, again with the optimal $\sqrt{s} \simeq
350$~GeV, the error on the $b\anti b$ branching ratio can then be
estimated to reach 25\% at $m_{h_{\rm SM}} \simeq 200$~GeV.  Measurement of
branching ratios to $c\bar{c}$, $\tau^+\tau^-$, $gg$, and $\gamma
\gamma$ does not seem feasible in this mass range.

Branching ratios into vector bosons can be measured with good
precision in the intermediate mass range.
For $m_{h_{\rm SM}} = 160$~GeV and 500~fb$^{-1}$, a predicted excellent
precision of 2.1\% on $\BR(h_{\rm SM} \rightarrow WW)$, has been 
reported~\cite{Borisov:1999mu}, with
extrapolated estimated precision of better than 7\% over the mass range of 150 
to 200~GeV~\cite{fnal_report}.  

To measure $\BR(h \rightarrow ZZ)$, it will be necessary to 
distinguish hadronic $Z$ decays from hadronic $W$ decays.  This serves 
as an important benchmark for electromagnetic and hadronic calorimetry.
With 500~fb$^{-1}$, and
assuming that this separation allows one to identify
one of the two $Z$'s in the Higgs decays (through leptons or
$b\bar{b}$) 40\% of the time, the statistical uncertainty of this
branching ratio would be approximately 8\% for  
$m_{h_{\rm SM}} \simeq 210$~GeV~\cite{fnal_report}, degrading to
17\% for $m_{h_{\rm SM}} = 160$~GeV~\cite{tesla_report} where the branching
ratio into $Z$'s is still small.

\subsection{Coupling determinations---heavy Higgs bosons}
\label{sechd}

If the Higgs boson is heavy, {\it i.e.}, $\mh>2m_t$, and if this
Higgs boson possesses couplings close to those expected in the SM,
then consistency with the precision electroweak data
(which implies $\mhsm\lsim 230$~GeV at 95\% CL) would require the
existence of new physics beyond the SM.
A high statistics measurements at the $Z$ peak
could be useful to elucidate the non-SM effects. In addition, with high
center of mass energy and large integrated luminosity, an experiment at
the LC could directly observe heavy Higgs decay and make measurements
of the Higgs couplings.  These measurements could reveal
departures from the SM Higgs properties and provide indirect evidence for
the nature of the new physics, which would modify the SM Higgs couplings
through loop effects.

\subsubsection{Cross sections}
\label{sechda}

As a specific case, for $m_h = 500$~GeV,
a SM-like Higgs boson would have a width of 70~GeV
and dominant decay modes into $W^+W^-$ (55\%), $ZZ$ (25\%), 
and $t\bar{t}$ (20\%). The production cross section at $\sqrt{s} = 800$~GeV
for $Zh$ would
be 6~fb, but Higgs production would be dominated by  the
$W$-fusion process, whose cross section would be 10~fb.  With 
1000~fb$^{-1}$, one would expect 
400 $Zh$ events where the $Z$ decays to electrons or muons.
With reasonable selection and acceptance cuts, a measurement
of $\sigma(Zh)$ to better than 7\% should be feasible.

\subsubsection{Branching ratios}
\label{sechdb}

The LHC will have great difficulty distinguishing $h\rightarrow t\bar{t}$
decays from the huge QCD $t\bar{t}$ backgrounds.
On the other hand,
this mode should be observable at a LC.  
In the SM, the important coupling $g^2_{tth_{\rm SM}} \simeq 0.5$ can be
compared to $g^2_{bbh_{\rm SM}} \simeq 4 \times 10^{-4}$.
If the Higgs boson is heavier than 350~GeV, it will be possible
obtain a good determination of the top-Higgs Yukawa coupling.
Full simulations are needed
for heavy Higgs decays into top, but with reasonable assumptions,
one can expect a statistical error of $\delta \BR /\BR \simeq 14$\% with
500~fb$^{-1}$~\cite{fnal_report}.
Simulations using the TESLA detector of the 
$W^+W^- \rightarrow h_{\rm SM} \rightarrow t\bar{t}$ process with 1000~fb$^{-1}$
and 6-jet final states
show impressive signal significance for $\sqrt{s} = 1000$~GeV and 
reasonably good significance at $\sqrt{s} = 800$ GeV~\cite{Alcaraz:2000xr}.
These studies find that
a relative error of better than 10\% in the top quark Yukawa coupling measurement
can be achieved for Higgs masses in the
350--500~GeV and 350--650~GeV ranges at $\sqrt{s} = 800$~GeV
and 1000~GeV, respectively.

Assuming that detector performance allows separation of
hadronic $W$ and $Z$ decays, and using production through 
$W$-fusion, 
the $WW$ and $ZZ$ coupling of the Higgs boson can be studied by using
methods similar to those for $t\bar t$.  This gives the
estimates on $\BR(h_{\rm SM} \rightarrow W^+W^-)$ and 
$\BR(h_{\rm SM} \rightarrow ZZ)$ shown in Table~\ref{tab:summary}.

\subsection{Summary of couplings}
\label{seche}

The relative measurement errors for a SM Higgs at various masses are
summarized in Table~\ref{tab:summary}.
As much as possible, the entries have been collected from 
simulations with the L detector  described in Chapter 15. 
For uniformity, the entries have been scaled to 500~fb$^{-1}$, except
where  otherwise noted.
The significant measurements of many branching ratios
and couplings demonstrate the strength of the LC Higgs program.

\begin{table}[b!]
\centering
\begin{tabular}{c||c|c|c|c|c}  \hline\hline

 $\Delta m_h$ &  \multicolumn{5}{l}{$\simeq 140$~MeV (recoil against leptons
                                          from $Z$)} \\ 
      
                 &  \multicolumn{5}{l}{$\simeq 50$~MeV (direct reconstruction)}
                                                       \\ \hline
 $m_h$~(GeV)     & 120 & 140 & 160 & 200 
                 & 400--500  \\ \hline 
 $\sqrt{s}$~(GeV)    &  \multicolumn{4}{c|}{500}
                 &  800 \\ \hline \hline
 $\Delta \sigma(Zh) / \sigma(Zh)$
     &  $6.5$\% & $6.5$\% & $6$\% &  $7$\%   & $10$\% \\ \hline
     
 $\Delta \sigma(\nu\bar{\nu}h)\BR(b\bar{b})/ \sigma \BR$ 
     & $3.5$\%  & $6$\% & $17$\% 
     &  --             & --            \\ \hline \hline
       
 $\delta g_{h xx} / g_{h xx}$ (from $\BR$'s)&  &  &  &  &  
\\ \hline \hline

 $t\bar{t} $    & 7 -- 20\% \dag\ & -- & -- & --  & 10\%   \\

 $b\bar{b}$      & $1.5$\% & $2$\% & $3.5$\% 
                 & $12.5$\% & -- \\ 

 $c\bar{c}$      & $20$\%  & $22.5$\%  & -- & -- & -- \\

 $\tau^+ \tau^-$ & $4$\%   & $5$\%     & -- & -- & -- \\ \hline
 
 $WW^{(*)}$      & $4.5$\% & $2$\% & $1.5$\% 
                 & $3.5$\% & $8.5$\%    \\ 

 $ZZ^{(*)}$      & --             & --           & $8.5$\% 
                 & $4$\%   & $10$\%    \\
 
 $gg$            & $10$\%  & $12.5$\%  & -- & -- & -- \\ 
 
 $\gamma \gamma $ & $7$\%  & $10$\%    & -- & -- & -- \\ \hline
  
 $g_{hhh}$        & $23$\% \S\ & --  & -- & -- & -- \\ \hline\hline

\end{tabular}
\caption{Summary of measurement precisions for 
the properties of a SM-like Higgs boson, $h$, 
and couplings for a range of Higgs boson masses for
500~fb$^{-1}$,
unless otherwise indicated.
\dag\ radiative $t\bar{t}h$ production, 
1000~fb$^{-1}$, $\sqrt{s} = 800$ -- 1000 GeV;
\S\ 1000~fb$^{-1}$. 
}
\label{tab:summary}
\end{table}

Just as the computer program {\tt ZFITTER}~\cite{Bardin:2001yd} is used
with $Z$ mass, widths, asymmetries and branching ratios to make
global fits for $Z$ couplings, a program {\tt HFITTER}~\cite{hfitter} is
now available that performs a global fit taking into account correlations
between measurements of Higgs boson properties. 
Individual couplings of the Higgs boson can then be extracted optimally,
for example through the correct combination of cross section and branching
ratio measurements for such couplings as $g_{h WW}$ and $g_{h ZZ}$.
Such precision fits can
be used to probe  for indirect evidence of higher-mass states.

\subsection{Total width}
\label{sechf}

Determination that a Higgs boson total width is anomalously large
would indicate new non-SM effects.
For light Higgs bosons, the predicted SM width is too small to be
measured directly, but a combination of branching ratios and coupling measurements
allows the indirect and {\it model-independent} measurement of the total 
width through 
\begin{equation}
\Gamma_{tot} = \Gamma(h \rightarrow X)/\BR(h \rightarrow X)\,.
\end{equation}
For $m_{h_{\rm SM}}< 115$~GeV, the total width measurement would very likely require a 
$\gamma \gamma$ collider, an $e^+e^-$ LC, and input from the 
LHC~\cite{Gunion:1996cn}.
However, limits
from LEP2 indicate $m_{h_{\rm SM}} \gsim 115$~GeV and therefore a significant
branching ratio to $WW^*$.  This gives  the attractive prospect of 
a  model-independent measurement of the total width using LC
measurements alone.

First, measurements of $\sigma(h\nu\nu) \cdot \BR(h\rightarrow b\bar{b})$ and
$\BR(h\rightarrow b\bar{b})$,  through recoil Higgsstrahlung measurements, 
give $\Gamma(h \rightarrow WW^*)$.  Then, a similar independent measurement
of $\BR(h \rightarrow WW^*)$ gives the total width,
through the relation
$\Gamma_{tot} = \Gamma(h \rightarrow WW^*)/\BR(h \rightarrow WW^*)$.
For example, from Table~\ref{tab:summary}, even with as little as
200~fb$^{-1}$, $\Gamma_{tot}$ can be found to approximately 10\% for
$m_{h_{\rm SM}}= 120$~GeV, improving to a few percent for $m_{h_{\rm SM}} = 150$~GeV.
Even better precision can be attained with the introduction of some
model assumptions in the value used for $\Gamma(h_{\rm SM} \rightarrow WW^*)$, 
{\it e.g.}, assuming the SU(2) relation between $W$ and $Z$ couplings
along with $\sigma_{meas}(Zh_{\rm SM})$.

For $m_{h_{\rm SM}} \gsim 205$~GeV, $\Gamma_{tot}(h_{\rm SM})$
exceeds 2~GeV, and the
physical width would be directly resolvable with typical LC detector
resolutions.  References~\cite{Gunion:1996cn,Drollinger:2001bc} track these
variations of precision for indirect and direct measurements for
different values of $m_{h_{\rm SM}}$ and inputs from different machines.  The
jet-jet mass resolution assumed in~\cite{Gunion:1996cn} has been
verified by full simulations~\cite{Ronan} in the L detector with
200~fb$^{-1}$ of data,
resulting in estimated direct measurements of the total width whose
accuracy reaches a 
minimum value of  6\% in the mass range of 240--280~GeV.
 The indirect determination described above can
also be pursued, and the combination would allow even better
precision.

\subsection{Quantum numbers}
\label{sechg}

\begin{figure}[htb]
\centering
\psfig{file=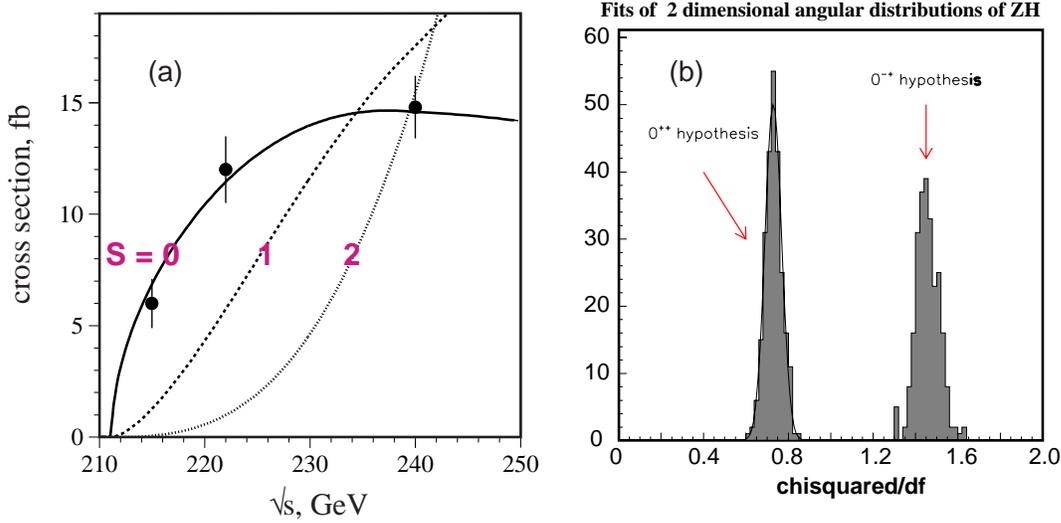,height=2.8in}
\caption{(a) Behavior of Higgsstrahlung threshold for various spin states
along with typical measurement precisions on the cross section.
(b) Fit to the double-differential angular distribution in
$Z\phi$ events (see text) to distinguish CP-even and CP-odd states.
}
\label{fig:thresh}
\end{figure}

The spin, parity, and charge conjugation quantum numbers $J^{PC}$ of
a Higgs boson, generically denoted by $\phi$
in this subsection, can potentially be determined in a model-independent
way. Useful ingredients include the following:
\bit
\item A Higgs boson produced in $\gam\gam$ collisions 
cannot have $J=1$ and must have positive $C$~\cite{yangsak}.
\item 
The behavior of the $Z\phi$ Higgsstrahlung cross section at threshold 
constrains the possible values of $J^{PC}$ of the state. If the
spin of the $\phi$ is 2 or less, a cross
section growing as $\beta$ indicates a CP-even object, whereas
a cross section growing as $\beta^3$ signals a CP-odd 
state~\cite{Miller:2001bi}, as shown in 
Fig.~\ref{fig:thresh}(a).
\eit

\bit
\item
The angular dependence of the $e^+e^-\to Z\phi$ cross section depends
upon whether the $\phi$ is CP-even, CP-odd, or a
mixture~\cite{Miller:2001bi,Hagiwara:1994sw,Barger:1994wt,Han:2001mi}.
Following~\cite{Han:2001mi} we parameterize the $ZZ\phi$ vertex as
\begin{equation}
    \Gamma_{\mu\nu}(k_1,k_2) = 
        a g_{\mu\nu} +
        b \frac{k_{1\mu} k_{2\nu} - g_{\mu\nu}k_1\cdot k_2}{\mz^2}  +
   \tilde{b}\frac{\eps_{\mu\nu\alpha\beta} k_1^\alpha k_2^\beta}{\mz^2}\,,
    \label{eq:zzh}
\end{equation}
where $k_1$ and $k_2$ are the momenta of the two $Z$s.
The first term arises from a Standard-Model-like $ZZ\phi$ coupling, and the
last two from effective interactions that could be induced by high-mass
virtual particles.
With this vertex the Higgsstrahlung cross section becomes
\begin{equation}
\frac{d\sigma}{d\cos\theta_Z} \propto 1
    + \frac{p_Z^2}{\mz^2} \sin^2\theta_Z
    - 4\, {\rm Im}\left[\tilde{b}\over\tilde{a}\right]
      \frac{v_ea_e}{v_e^2+a_e^2}
      \frac{p_z\sqrt{s}}{\mz^2} \cos\theta_Z
    +\left|\tilde{b}\over \tilde{a}\right|^2
      \frac{p^2_zs}{2\mz^4} (1+\cos^2\theta_Z)\,,
    \label{eq:zzhX}
\end{equation}
where $\theta_Z$, $p_Z$, and $E_Z$
are the scattering angle, momentum, and energy of the final-state $Z$
boson; $v_e$ and $a_e$ are the vector and axial-vector couplings
at the $e^+e^-Z$ vertex; and $\tilde{a}\equiv a-bE_Z\sqrt{s}/\mz^2$.
The term in Eq.~(\ref{eq:zzhX}) proportional to $\cos\theta_Z$ arises
from interference between the CP-even and CP-odd couplings in
Eq.~(\ref{eq:zzh}).
If the CP-odd coupling~$\tilde{b}$ is large enough, it can be extracted
from the forward-backward asymmetry.
Even upper limits on this asymmetry would be interesting.
Note that the CP-even component
of a Higgs boson will typically couple at tree-level whereas the
CP-odd component will only couple via one-loop diagrams (typically
dominated by the $t$ quark loop). As a result the coupling
strength $\tilde b$ is typically proportional to $\mz^2/s$ 
times a loop suppression factor.  Thus, an asymmetry measurement may be
able to provide a crude determination of 
the $\tilde b/a$ term.
If $\phi$ is a purely CP-odd state with one-loop coupling, 
the resulting $Z\ha$ cross section will simply be too small to
provide a useful measurement of the asymmetry.

\item
The angular distribution of the fermions in the $Z\to f\anti f$
decays in $Z\phi$ production also reflects the CP nature of
the state $\phi$ \cite{Hagiwara:1994sw,Barger:1994wt}.  
For the decay $Z \rightarrow e^+ e^-$ or $\mu^+\mu^-$, the following 
angles can be defined: the angle between the initial $e^-$ and the $Z$; 
the angle between the final state $e^-$ or $\mu^-$ and the direction of motion
of the $Z$, in the rest frame of the $Z$; and the angle between the $Z$
production plane and $Z$ decay plane.  Correlations between these angles
can be exploited, {\it e.g.}, a fit to the double-differential angular distribution
of the first two of these angles results in a 14$\sigma$ separation between
the $0^{++}$ (CP-even, scalar) and the 
$0^{-+}$ (CP-odd, pseudoscalar)~\cite{fnal_report},
assuming that the $Z\phi$ cross section is independent
of the CP nature of $\phi$ (see Fig.~\ref{fig:thresh}(b)). Even more powerful are
fits to the triple-differential angular distribution, 
where sufficient luminosity 
can uncover non-standard $ZZ\phi$ couplings.
However, this technique again suffers from the 
difficulty described in the previous item; 
namely, the CP-odd part of the state $\phi$
is typically so weakly coupled to $ZZ$ that there is little
sensitivity to the CP-odd component 
if there is any significant CP-even component
in $\phi$), or a very small cross section,  if $\phi$ is almost purely CP-odd.

\item
If $\phi$ has significant branching ratios to either $\tau^+\tau^-$
or $t\anti t$, the polarization of the decay fermions can be measured.
This can provide a direct
determination of the ratio $b_f/a_f$ in the
$y_f\anti f (a_f+ib_f\gamma_5)f\phi$ ($f=\tau$ or $t$) Yukawa
coupling structure of $\phi$ 
\cite{Grzadkowski:1995rx,Grzadkowski:1994kv,Kramer:1994jn}.
\item The angular distributions in the $t\anti t \phi$ final state,
which has adequate cross section for $\rts\gsim 800\gev$
for modest values of $m_\phi\lsim 200\gev$, assuming Yukawa
coupling $y_t\anti t (a_t+ib_t\gamma_5)t\phi$ comparable to SM values, 
appear to provide an excellent
means for determining the CP nature of $\phi$ by allowing one
to probe the ratio $b_t/a_t$ \cite{Gunion:1996bk,Gunion:1996vv}.

\item It is likely that the CP properties of the 
$\phi$ can  be well determined using photon polarization asymmetries in
$\gam\gam\to \phi$ 
collisions \cite{Grzadkowski:1992sa,Gunion:1994wy,Kramer:1994jn}.
This is discussed in Section 10.

\item
If the $\phi$ has substantial $ZZ$ coupling, then
$e^-e^-\to ZZ e^-e^-\to \phi e^-e^-$ 
can be used to probe its CP nature~\cite{Boe:1999kp} 
via the energy distributions of the $\phi$
and the final electrons, which are much harder in the
case of a CP-odd state than for a CP-even state.
Certain correlations are also useful probes of the CP properties of
the $\phi$. However, if the CP-odd portion of $\phi$ couples at one-loop
(as expected for a Higgs boson), there will be either little sensitivity
to this component or little cross section.

\eit

\subsection{Precision studies of non-SM-like Higgs bosons}
\label{sechh}

We confine our remarks to a two-doublet Higgs model 
(either the MSSM Higgs sector or a more general 2HDM).
In the MSSM, we noted in Section~\ref{seced} that for $\mha\lsim\sqrt{s}/2$, 
as long as one is not too close to threshold, it is possible to
observe all Higgs scalars of the non-minimal Higgs sector.  In
particular, in parameter regions away from the decoupling limit, none
of the CP-even Higgs scalars may resemble the SM Higgs boson.
Precision studies of all the Higgs bosons will provide a detailed
profile of the non-minimal Higgs sector.
Once $\mha\gsim\sqrt{s}/2$, only the $\hl$ will be visible at the LC.
  There may still be some possibilities for observing the heavier
Higgs states produced singly, either in association with a $b\bar b$
pair at large $\tan\beta$ where the coupling to $b\bar b$ is enhanced,
or by $s$-channel resonance production at a $\gamma\gamma$ collider.

Masses $\mha$ and $\mhh$  
in excess of 500 GeV to 1 TeV are certainly possible.  In such cases,
very substantial energy for the LC will be required to observe these
states directly, either in association with $b\bar b$ (at large
$\tan\beta$) or via $\hh\ha$ production.
Measuring the former will provide a 
crucial determination of the $b\anti b$ couplings,
which in the given model context will provide a determination
of $\tanb$, with accuracy determined by the production rates. 
Moreover, if the $\hh$ and $\ha$
can be produced at a high rate (by whatever process),
a detailed study of their branching ratios has the potential
for providing very vital information regarding model parameters.
In the supersymmetric context, 
the heavy $\hh$, $\ha$ and $\hpm$ would generally
decay to various pairs of supersymmetric particles 
as well as to $b$'s and $t$'s.
A study of the relative branching ratios would provide powerful
determinations of $\tanb$ and many of the soft-SUSY-breaking
parameters \cite{Gunion:1997cc,Gunion:1996qd,Feng:1997xv}.

\section{The Giga-Z option---implications for the Higgs sector}
\label{seci}

Measurements of the effective leptonic mixing angle 
and the $W$ boson mass to
precisions of $\delta\sweff \simeq 10^{-5}$ and 
$\delta\mw \approx 6 \mev$ at Giga-Z can be exploited in many ways.
The size of the Giga-Z $90\%$ CL ellipses is illustrated in Fig.~\ref{dsdt}.
Potential implications include the following.
\bit
\item 
Within the SM context, the Higgs boson mass can be determined indirectly to a
precision of about 7\%. Deviation between the directly observed
value and the value implied by Giga-Z data would require new
physics beyond the SM.
\item In the MSSM context it will be possible
to obtain information about new high mass scales beyond the
direct reach of the collider. 
This would be of particular importance if 
the heavier scalar top quark, ${\tilde t_2}$, and the heavy Higgs
bosons $\ha$, $\hh$ and $\hpm$ were beyond the kinematical reach of the
LC and background problems precluded their observation at the LHC.
\item
In the context of a non-minimal Higgs sector, such as the general 2HDM
extension of the minimal SM, constraints on the Higgs sector
and/or new physics can be obtained. These would be particularly
important in those cases where none of the Higgs bosons or new 
particles could be observed at the LC without higher $\sqrt s$
or at the LHC because of backgrounds.
\eit

\subsection{The MSSM context} 

In the case of the MSSM,
the relation between $\mw$ and $\sweff$ is affected by the parameters
of the supersymmetric sector, especially the
${\tilde t}$ sector. At a 
LC, the mass of the light ${\tilde t}$, $m_{\tilde t_1}$, and the
${\tilde t}$ mixing angle, $\theta_{\tilde t}$, 
should be measurable very well if the process $e^+\,e^- \to {\tilde t}_1 
\bar{{\tilde t}_1}$ is accessible~\cite{lcstop}.

\begin{figure}[tbh]
\vspace{1em}
\centering
\psfig{file=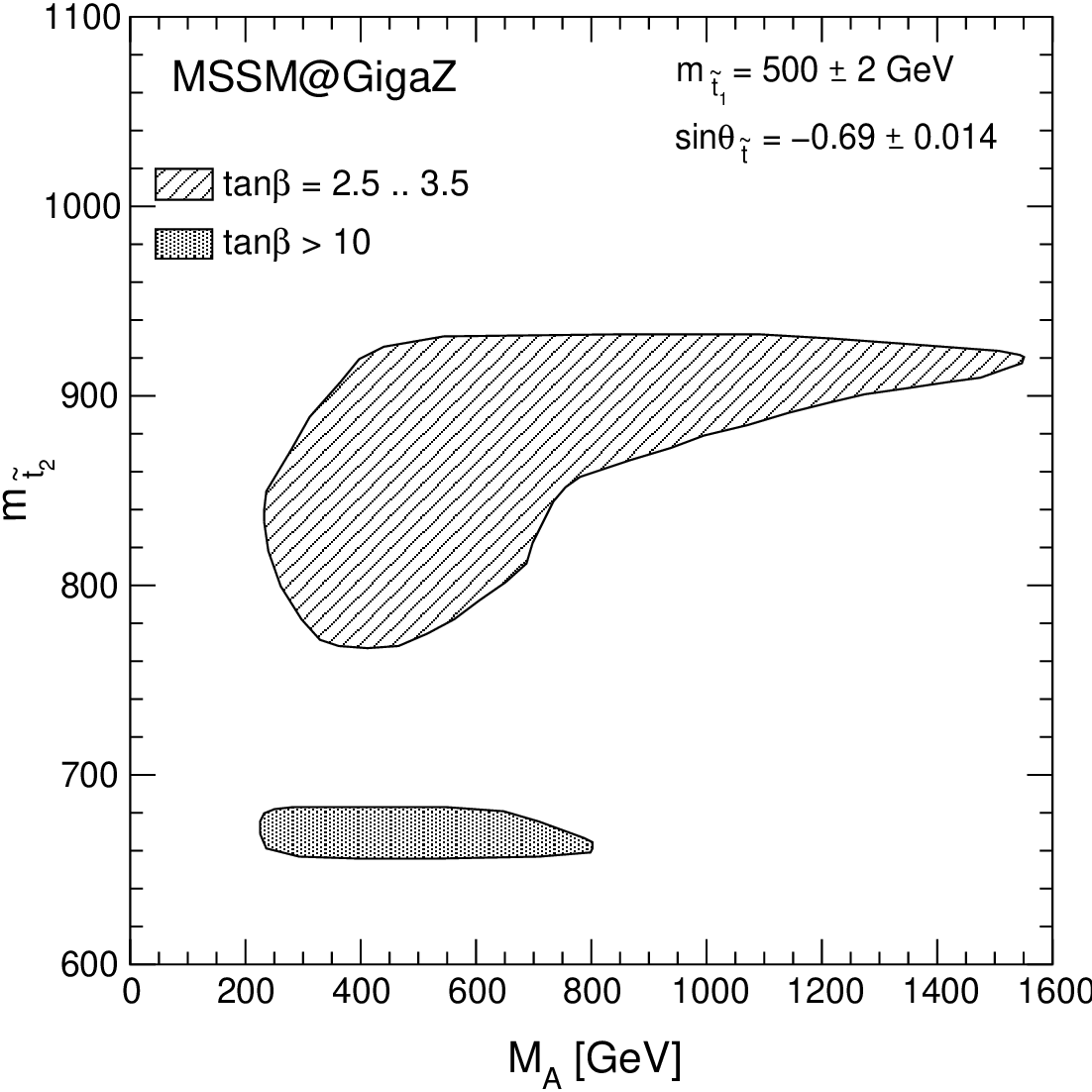,width=10cm,height=8cm}
\caption{
 The region in the $\mha- m_{\tilde t_2}$ plane, allowed by
$1\,\sigma$ errors obtained from the Giga-Z measurements of $\mw$ and 
$\sweff$: 
$\mw = 80.400 \pm 0.006 \gev$, 
$\sweff = 0.23140 \pm 0.00001$, 
and from the LC measurement of $\mhl$:
$\mhl = 115 \pm 0.05~({\rm exp.}) \pm 0.5~({\rm theo.}) \gev $. 
$\tanb$ is assumed to be $\tanb = 3 \pm 0.5$ or $\tanb > 10$. 
The other parameters are given by 
$ m_{\tilde t_1} = 500 \pm 2 \gev$,
$\sin\theta_{\tilde t} = -0.69 \pm 0.014$,
$ A_b =  A_t \pm 10\%$,
$\mgl = 500 \pm 10 \gev$,
$\mu = -200 \pm 1 \gev$ and
$M_2 = 400 \pm 2 \gev$.
}
\label{fig:MSt2MA}
\end{figure}

In Fig.~\ref{fig:MSt2MA} (from~\cite{gigaz}), it is demonstrated how 
{\em upper} bounds on $\mha$ and 
$ m_{\tilde t_2}$ can be derived from measurements of $\mhl$, $\mw$
and $\sweff$,
supplemented by precise determinations of 
$ m_{\tilde t_1}$ and $ \theta_{\tilde t}$.
The analysis assumes a lower bound, $\tanb \geq 10$, which
can be expected from  
measurements in the gaugino sector (see, {\it e.g.}, \cite{tbmeasurement}).
The other parameters values 
are assumed to have the uncertainties as expected from LHC~\cite{lhctdr} and
a LC~\cite{tesla_report}.

For low $\tanb$ (where the prediction for $\mhl$ depends sensitively on
$\tanb$) the heavier ${\tilde t}$ mass, $ m_{\tilde t_2}$, can be restricted to
$ 760 \gev \lsim  m_{\tilde t_2} \lsim 930 \gev$ from the $\mhl$, $\mw$ and 
$\sweff$ 
precision measurements. 
The mass $\mha$ varies between $200 \gev$ and $1600 \gev$. 
If $\tanb \ge 10$ (where $\mhl$ has only a mild dependence on $\tanb$),
the allowed region for the ${\tilde t_2}$ turns out to be
much smaller, $ 660 \gev \lsim  m_{\tilde t_2} \lsim 680 \gev$, and 
the mass $\mha$ is restricted to $\mha \lsim 800 \gev$.

In deriving the bounds on the heavier ${\tilde t}$~mass, $ m_{\tilde t_2}$, 
the constraints from $\mhl$ and from $\sweff$ and $\mw$ 
play an important role. For the bounds
on $\mha$, the main effect comes from $\sweff$.
The assumed value of $\sweff = 0.23140$ differs slightly from the 
corresponding value obtained in the SM limit.
For this value the (logarithmic) dependence on $\mha$ is 
still large enough (see \cite{gigaz4}) so that from 
the high precision in 
$\sweff$ at Giga-Z an {\em upper limit} on $\mha$ can be set.
For the error of $\sweff$ that could be  obtained at an LC without the Giga-Z mode
(which is at least ten times larger), 
no bound on $\mha$ could be inferred.

\subsection{Non-exotic extended Higgs sector context} 

Building on the discussion of the general 2HDM given earlier,
one can imagine many situations for which 
the very small Giga-Z $90\%$ CL ellipses illustrated in Fig.~\ref{dsdt}
would provide crucial (perhaps the only) constraints.  
For example, suppose the LHC observes a $1\tev$ Higgs boson with 
very SM-like properties and no other new
physics below the few-TeV scale.  
We have seen that this is possible in the 2HDM scenarios consistent with
current precision electroweak constraints. Suppose
further that it is not immediately possible to increase
$\sqrt s$ sufficiently so that $\hl \ha$ production is allowed (typically
requiring $\sqrt s> 1.5\tev$ in these models). Giga-Z
measurements would provide strong guidance as to the probable masses
of the non-SM-like Higgs bosons of any given non-minimal
Higgs sector. However, it must be accepted that a particular Giga-Z
result for $S,T$ might have other non-Higgs interpretations as well.

\section{\boldmath The $\gam\gam$ collider option}
\label{secj}

Higgs production in $\gamma\gamma$ collisions offers a unique 
capability to measure the two-photon width of the Higgs and
to determine its CP composition through control of the photon
polarization. A brief discussion of photon collider technology
can be found in Chapter 13.

The $\gam\gam$ coupling of a SM-like Higgs boson $h_{\rm SM}$ of relatively light 
mass receives contributions from loops containing any particle
whose mass arises in whole or part from the vacuum expectation
value of the corresponding neutral Higgs field.
A measurement of $\Gamma(h_{\rm SM}\to\gam\gam)$ 
provides the possibility of revealing the presence of arbitrarily
heavy particles that acquire mass via the Higgs 
mechanism.\footnote{Loop contributions from particles that acquire
a large mass from some other mechanism will decouple as $({\rm mass})^{-2}$
and $\Gamma(h_{\rm SM}\to\gam\gam)$ will not be sensitive to their presence.}
However, since such masses are basically proportional to
some coupling times $v$, if the coupling is perturbative
the masses of these heavy particles are unlikely to be much 
larger than $0.5-1\tev$. Since
$B(h_{\rm SM}\to X)$ is entirely determined by the spectrum
of light particles, and is thus not affected by heavy states,
$N(\gam\gam\to h_{\rm SM}\to X)\propto \Gamma(h_{\rm SM}\to\gam\gam)B(h_{\rm SM}\to X)$
will provide an extraordinary probe for such heavy states.
Even if there are no new particles that acquire mass via the Higgs
mechanism, a precision measurement of $N(\gam\gam\to\widehat h\to X)$
for specific final states $X$ ($X=b\anti b,WW^*,\ldots$)
can allow one to distinguish between a ${\widehat h}$ that is part
of a larger Higgs sector and the SM $h_{\rm SM}$.  The
deviations from the SM predictions typically exceed 5\% if the other heavier Higgs bosons
have masses below about 400 GeV.

\begin{figure}[htbp]
\centering
\psfig{figure=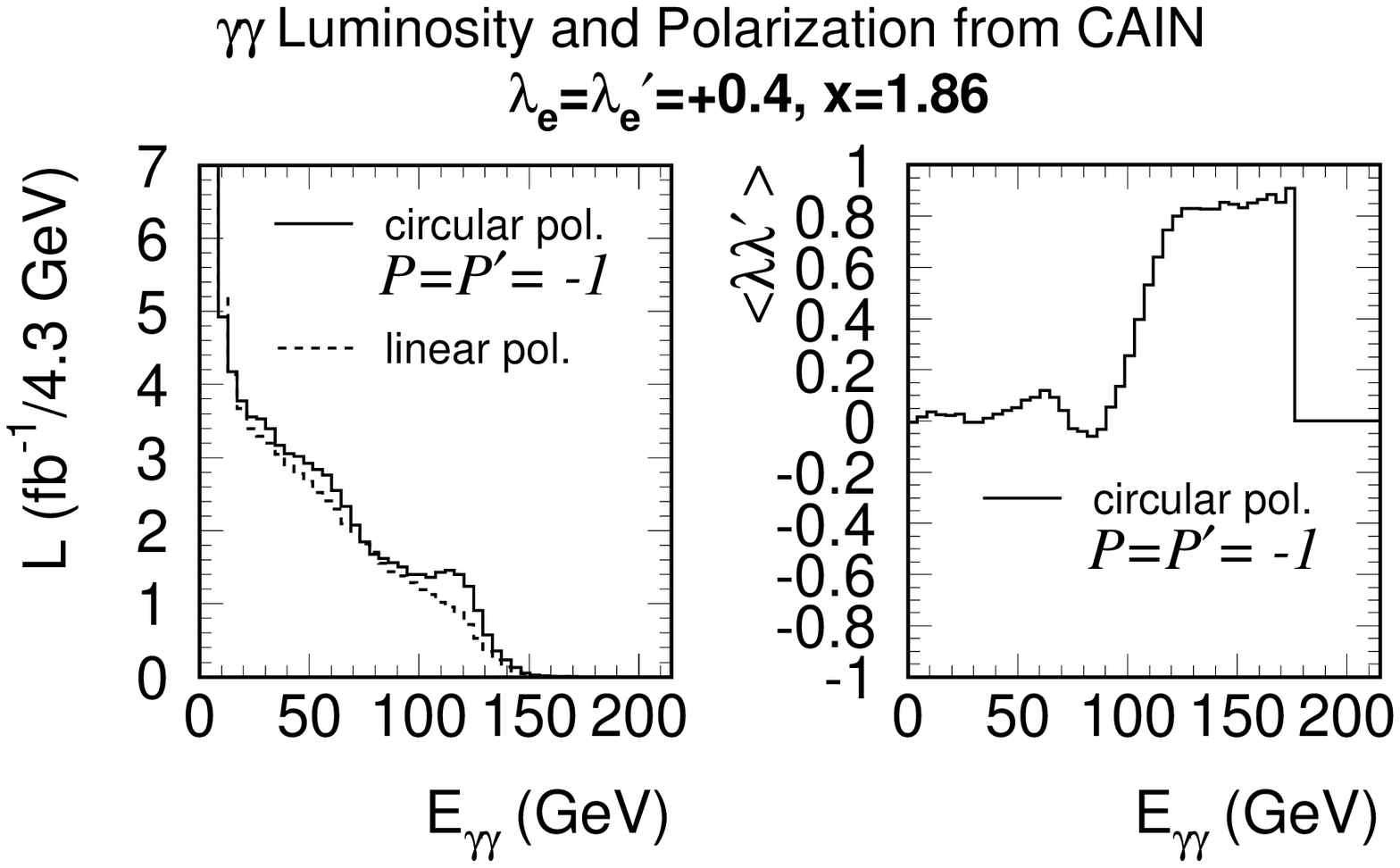,width=12cm,angle=0}
\caption[0]{Left: CAIN \cite{cainref} predictions for the $\gam\gam$ 
luminosity distribution for circularly polarized 
($\lambda_e = \lambda_e' = 0.4$, $P = P' = -1$)
and linearly polarized
photons assuming  $10^7$~sec/year, $\rts=206\gev$, 
80\% electron beam polarization,
and a 1.054 micron laser wavelength, after including beamstrahlung
and other effects, from~\cite{twogamhiggs}. 
Right: The corresponding value of $\VEV{\lambda\lambda'}$,
for circular polarization.} 
\label{fig:higgsspec}
\vskip.1in
\centering
\psfig{figure=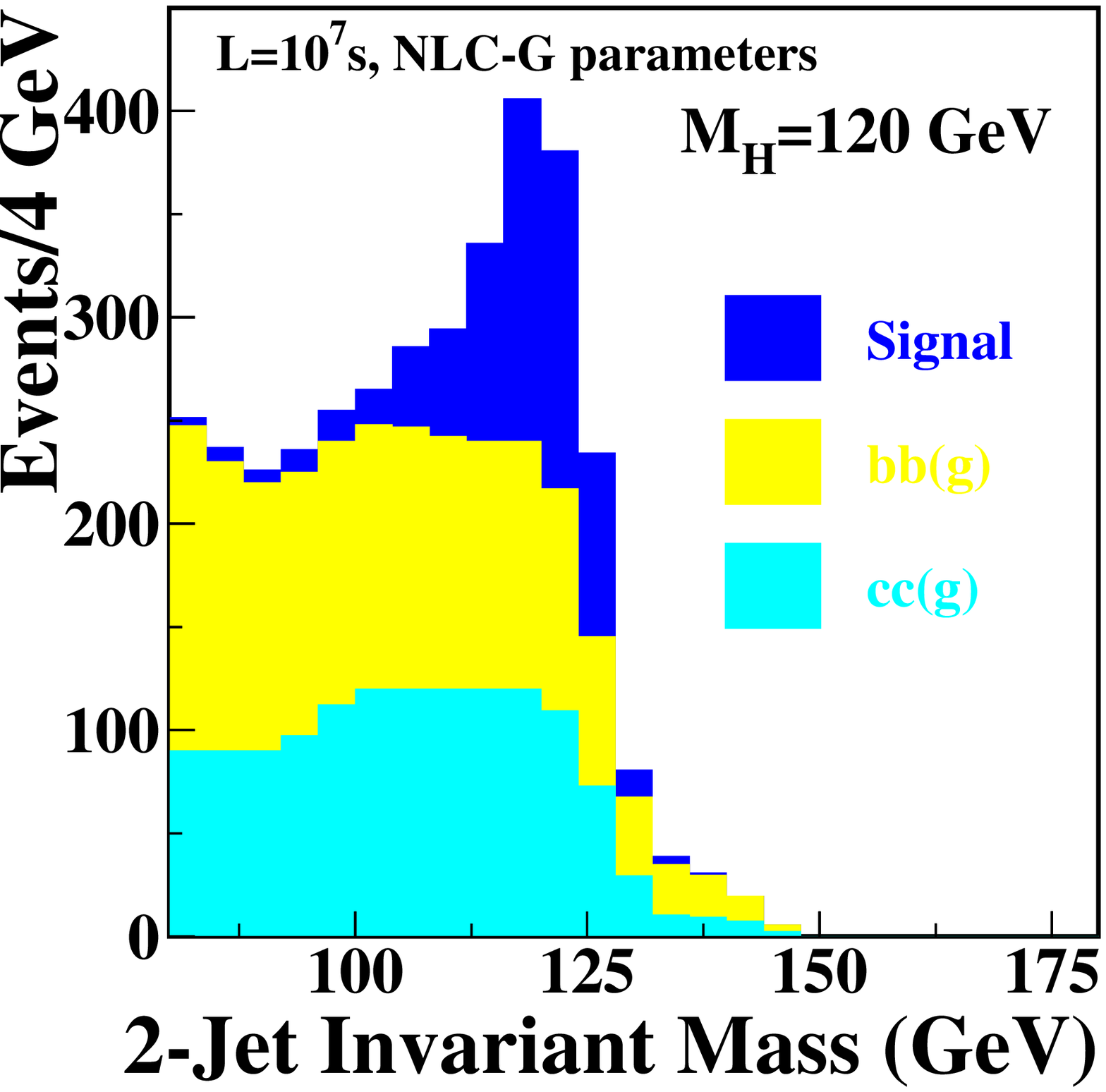,width=9cm,angle=0}
\caption{
Higgs signal and heavy quark background in $\gamma\gamma
\to h$ 
for a Higgs mass of 120~GeV~\cite{twogamhiggs}.
}
\label{fig:higgs}
\end{figure}

The predicted rate for Higgs boson production followed by decay to
final state $X$ can be found in \cite{Gunion:1993ce}.
This rate depends strongly on
$d{\cal L}_{\gam\gam}/dy$, the differential $\gam\gam$ collider luminosity,
where $y=m_{\widehat h}/\rts$
and $\rts$ is the $ee$ collider center-of-mass energy.
An important parameter to maximize peak luminosity is $\VEV{\lambda\lambda'}$, 
the average value of the product of the helicities of the two colliding
photons after integration over their momentum fractions $z$ and $z'$.
Larger values of this parameter also suppress the dominant $J_z = \pm 2$,
$\gamma\gamma\to b\bar b g$ background, which is proportional
to $(1 - \VEV{\lambda\lambda'})$.
The computation of $d{\cal L}_{\gam\gam}/dy$ was first considered in 
\cite{Ginzburg:1983vm,Ginzburg:1984yr}. More realistic 
determinations~\cite{cainref}
including beamstrahlung, secondary collisions between scattered electrons
and photons from the laser beam, and other non-linear effects result in
a substantial enhancement of the luminosity in the low-$E_{\gamma\gamma}$
region as shown in Fig.~\ref{fig:higgsspec}.

The choice of parameters that gives a peaked spectrum is well suited
for light Higgs studies. Using the spectrum of~~Fig.~\ref{fig:higgsspec}
as an example, the di-jet invariant
mass distributions for the Higgs signal and
for the $b\anti b(g)$ background for $m_{h_{\rm SM}}=120\gev$ 
are shown in Fig.~\ref{fig:higgs}~\cite{twogamhiggs}. 
After a year of operation,
$\Gamma(h_{\rm SM}\to\gam\gam)B(h_{\rm SM}\to b\anti b)$ could be measured with
an accuracy of about $5\%$.  (A much more optimistic error of close
to $2\%$ is quoted in \cite{hjikia} for $m_{h_{\rm SM}}=120\gev$,
based upon a substantially higher peak luminosity.)
The error for this measurement increases to about $20\%$
for $m_{h_{\rm SM}}= 160\gev$,
primarily due to the decrease of the Higgs di-jet branching fraction
by a factor of 18.

In many scenarios, it is possible that by combining this 
result with other types of
precision measurements for the SM-like Higgs boson, 
small deviations can be observed indicating the possible 
presence of heavier Higgs
bosons.  For a 2HDM (either the MSSM or a two-Higgs-doublet model
with partial decoupling),
if $\mhh\sim\mha>\rts/2$ then $e^+e^-\to \hh\ha$  is not possible and
$\gam\gam\to \hh,\ha$ may be the only option
allowing their discovery (other than implementing higher $\rts$).
The alternatives of $b\bar{b}H$ and $b\bar{b}A$ production
will only allow $H$ and $A$ detection if $\tan\beta$ is 
large \cite{Grzadkowski:2000wj}.
A LC for which the maximum energy is
$\rts=630\gev$ can potentially probe Higgs masses as
high as $500\gev$. 
If $\mhh$ and $\mha$ are known to within roughly $50\gev$
on the basis of precision $\hl$ data, 
then there is an excellent chance of detecting them 
by scanning, {\it i.e.} stepping in $\rts$, using a peaked $\gam\gam$ spectrum
\cite{Muhlleitner:2001kw,twogamhiggs}.
If no constraints have been placed on the $\hh,\ha$ masses (other
than $\mha \sim\mhh>\rts/2$), it is best to employ a broad $\gam\gam$
spectrum, which would yield a visible signal for $\hh,\ha$ production
for only some  parameter choices of $\mha$ and $\tanb$ \cite{twogamhiggs}. 

In the non-decoupling 2HDM model with a light decoupled $\what h$
and all other Higgs bosons heavier than $\rts$,
$\gam\gam\to \what h\to b\anti b$ 
might allow detection of the $\what h$
for some of the $\tanb$ values in the wedge where the $b\anti b \what h$ 
and $t\anti t\what h$ production processes
both yield fewer than 20 events for  an integrated luminosity of 1000~fb$^{-1}$~\cite{twogamhiggs}.

Once one or several Higgs bosons have been detected, 
precision studies can be
performed including: determination of CP properties; a detailed scan to
separate the $\hh$ and $\ha$ in the decoupling limit of a 2HDM; and
branching ratios measurements.  The branching ratios to supersymmetric final states
are especially important for determining the basic supersymmetry breaking 
parameters \cite{Gunion:1995bh,Gunion:1997cc,Feng:1997xv,Muhlleitner:2001kw}.

The CP properties can be determined for 
any spin-0 Higgs $\widehat h$ produced in $\gam\gam$
collisions.  Since $\gam\gam\to {\widehat h}$ is of one-loop order,
whether ${\widehat h}$ is CP-even, CP-odd or a mixture, 
the CP-even and CP-odd parts of ${\widehat h}$ have
$\gam\gam$ couplings
of similar size.   However, the structure of the couplings is very different:
\begin{equation}
{\cal A}_{{\rm CP}=+}\propto \vec \eps_1\cdot\vec \eps_2\,,\quad
{\cal A}_{{\rm CP}=-}\propto (\vec\eps_1\times\vec \eps_2)\cdot \what p_{\rm beam}\,.
\end{equation}
By adjusting the orientation of the initial laser
photon polarization vectors with
respect to one another, it is possible to determine the relative
amounts of CP-even and CP-odd content in the resonance ${\widehat h}$
\cite{Grzadkowski:1992sa}. 
If ${\widehat h}$ is a mixture, one can use helicity asymmetries for this 
purpose
 \cite{Grzadkowski:1992sa,Kramer:1994jn}.
However, if ${\widehat h}$ is either purely CP-even
or purely CP-odd, then one must employ transverse linear polarizations
\cite{Gunion:1994wy,Kramer:1994jn}. 
Substantial luminosity with transverse polarization can 
be obtained, although the spectrum is not peaked, as shown
in Fig.~\ref{fig:higgsspec}.

One measure of the CP nature of a Higgs is the asymmetry
for parallel {\it vs.} perpendicular orientation 
of the linear polarizations of the initial laser beams,
\begin{equation}
{\cal A}\equiv{N_{\parallel}-N_{\perp}\over N_{\parallel}+N_{\perp}}\,,
\end{equation}
which is positive (negative) for a CP-even (odd) state. Since 
100\% linear polarization for the laser beams translates into
only partial linear polarization for the colliding photons, 
both $N_{\parallel}$
and $N_{\perp}$ will be non-zero for the signal.  In addition,
the heavy quark background contributes to both. The expected value
of ${\cal A}$ must be carefully computed for a given model.  For the
SM Higgs with $m_{h_{\rm SM}}=120\gev$, it is estimated 
\cite{twogamhiggs} that ${\cal A}$ 
can be measured with an accuracy of about 20\% in one year
of operation, assuming the linear polarization spectrum of
Fig.~\ref{fig:higgsspec}, 60\% linear polarization of the colliding 
photons, and $S/B$ comparable to that shown in Fig.~\ref{fig:higgs}. 
This measurement would thus provide a 
moderately strong test of the CP=+ nature of the $h_{\rm SM}$.

We end by noting that the $e^-\gam$ and $e^-e^-$ collider
options are most relevant to exotic Higgs scenarios,
as discussed in Section~\ref{seck}.

\section{Exotic Higgs sectors and other possibilities}
\label{seck}

As we have seen, there are many scenarios and models in which the
Higgs sector is more complicated than the one-Higgs-doublet of the
minimal SM. Supersymmetry requires at least two Higgs doublets. Even
in the absence of supersymmetry, a two-doublet Higgs sector allows for
CP-violating phenomena. Singlets can also be added without altering
the tree-level prediction of $\rho=1$.  However, the possibility of
Higgs representations with still higher weak (left handed, denoted
$L$) isospin should not be ignored. The primary negative is that, for
triplets and most higher representations, if the vacuum expectation
value of the neutral Higgs field member of the representation is
non-zero ($v_L\neq 0$) then $\rho$ becomes infinitely renormalized and
can no longer be computed \cite{gvw}; instead it becomes a parameter
that must be input as part of the renormalization program.  Triplets
have received the most attention, as they arise naturally in
left-right symmetric extensions of the Standard Model gauge group
\cite{leftright}.  (These and other models that utilize Higgs triplets are
reviewed in \cite{hhg}.)  In this section we will also briefly
consider the Higgs-like pseudo-Nambu-Goldstone bosons that arise in
generic technicolor theories.

\subsection{A triplet Higgs sector} 

Including a single complex  SU(2)-triplet Higgs representation, in addition
to some number of doublets and singlets, 
results in six additional physical Higgs eigenstates:  
$H^{--,++}$, $H^{-,+}$, $H^0$ and $H^{0\,\prime}$.
All but the doubly-charged states
can mix with the doublet/singlet Higgs states 
under some circumstances. Even if $v_L\neq 0$
for the neutral field, $\rho=1$ can be preserved
at tree-level if, in addition, a real triplet field
is also included \cite{Georgi:1985nv,Chanowitz:1985ug}. 
However, $\rho$ will still be infinitely renormalized
at one-loop unless $v_L=0$ is chosen.
Left-right symmetric models capable of yielding
the see-saw mechanism for neutrino mass generation
{\it require} two triplet Higgs representations (an L-triplet and an R-triplet).
The large see-saw mass entry, $M$, arises from a 
lepton-number-violating Majorana coupling (which L-R symmetry requires
to be present for both the L-triplet and R-triplet
representations). Again, $\rho$ will not be altered
if $v_L=0$, but $v_R$ must be non-zero and large for large $M$.
We will briefly discuss the phenomenology of an L-triplet.
That for the R-triplet of the L-R symmetric model
is quite different. (See \cite{hhg} for a review.)
 
The resulting Higgs sector phenomenology can be very complex.
We focus on the most unequivocal signal for a triplet 
representation, namely observation of a doubly-charged Higgs boson.
Pair production, $Z^*\to H^{++}H^{--}$, has limited
mass reach, $m_{H^{++}}<\sqrt s/2$. Fortunately, single production is also 
generally possible. Most interestingly, the generically-allowed
lepton-number-violating Majorana coupling 
leads to an $e^-e^-\to H^{--}$ coupling and the possibility of $s$-channel
resonance production of the $H^{--}$ in $e^-e^-$ collisions.
Observation of this process 
would provide a dramatic confirmation of the presence of
the Majorana coupling and, in many cases, the ability to actually
measure its magnitude. For a discussion and review, see
\cite{Gunion:1996mq} 
(and also \cite{Frampton:2000jt,Cuypers:1997qg}). 
If the $H^{--}$ is heavy {\it and} has significant $W^-W^-$
coupling (requiring $v_L\neq 0$), then it can become broad and the $s$-channel
resonant production cross section is 
suppressed (see, \eg, \cite{Gluza:1997kg}) and might
not be observable. Another production mechanism
sensitive to the $e^-e^-\to H^{--}$ coupling 
that might be useful in such an instance is $e^-e^-\to H^{--} Z$, and
$e^-e^-\to H^{-}W^-$ will be sensitive to the $e^-\nu_e\to H^-$
coupling that would be present for the $H^-$ member of the triplet
representation \cite{Alanakian:1998ii}.
Using just the Majorana
coupling, doubly-charged Higgs bosons can also be produced
via $e^-\gam \to e^+ H^{--}$ and $e^+e^-\to e^+e^+ H^{--}$
\cite{Barenboim:1997pt}  and  the singly-charged
members of the same representation can be produced in 
$e^-e^-\to H^- W^-$ \cite{Alanakian:1998ii}.

Despite loss of $\rho$ predictivity, it could be that
non-zero $v_L$ is Nature's choice.
In this case, the $e^-e^-$ collider option again has some unique advantages.
The neutral, singly-charged and doubly-charged Higgs bosons
of the triplet representation
can {\it all} be produced (via $ZZ$ fusion, $W^- Z$ fusion and $W^-W^-$
fusion, respectively). For example, \cite{Barger:1994wa}
studies $W^-W^-\to H^{--}$ fusion.

\subsection{Pseudo Nambu Goldstone bosons}

In the context of technicolor and related theories,
the lowest-mass states are typically a collection of
pseudo-Nambu-Goldstone bosons, of which the lightest
is very possibly a state $P^0$ which can have mass 
below $200\gev$ and couplings and other properties
not unlike those of a light SM-like Higgs boson.  Typically, its
$WW,ZZ$ coupling is very small (arising via loops or anomalies),
while its $b\anti b$ coupling can be larger. 
The phenomenology of such a $P^0$ was studied in \cite{Casalbuoni:1999fs}.
The best modes
for detection of the $P^0$ at an LC are $e^+e^-\to\gam P^0\to \gam b\anti b$
and $\gam\gam\to P^0\to b\anti b$.  Since the $P^0$ is likely to
be discovered at the LHC in the $\gam\gam$ final state, we will know
ahead of time of its existence, and precision measurements of
its properties would be a primary goal of the LC.

\emptyheads
\blankpage \thispagestyle{empty}
\fancyheads

\setcounter{chapter}{3}

\chapter{Supersymmetry Studies at the Linear Collider}
\fancyhead[RO]{Supersymmetry Studies at the Linear Collider}

\section{Introduction}

The Standard Model (SM) has been tested by a spectacularly large and
diverse set of experiments.  The resulting body of data is consistent
with the matter content and gauge interactions of the SM with a
Higgs boson of mass $m_h \lsim 250~{\rm
GeV}$~\cite{susy-:2001xv}.  If a fundamental Higgs boson exists, it
fits much more naturally into supersymmetric extensions of the SM than
into the SM
itself~\cite{susy-Maiani,susy-Veltman:1981mj,susy-Witten:1981nf,susy-Kaul:1982wp}.
Thus, the study of supersymmetry (SUSY) is among the highest
priorities for future accelerators.

If SUSY exists, many of its most important motivations suggest that at
least some superpartners have masses below about $1\,{\rm TeV}$.
These motivations, ranging from gauge coupling
unification~\cite{susy-Dimopoulos:1981yj,susy-Amaldi:1987fu,susy-Langacker:1991an,susy-Bagger:1995bw,susy-Langacker:1995fk}
to the existence of an excellent dark matter
candidate~\cite{susy-Goldberg:1983nd}, are discussed
in previous chapters and also below.  While none of these is a guarantee
   of  SUSY,  they all provide motivation 
 for the presence of SUSY at the weak-interaction scale.

In the supersymmetric extension of the SM with minimal field content,
hundreds of additional parameters enter the Lagrangian.  If SUSY is
discovered, this discovery will open new questions---to understand
the pattern
of the SUSY  parameters, to determine from them 
the mechanism of SUSY breaking, and
to infer from them the nature of physics at the very highest energy
scales. Such grand goals may be contemplated only if precise and
model-independent measurements of superpartner properties are
possible.

In this chapter, we describe 
the prospects for such measurements at a 0.5--1.0 TeV
$e^+e^-$ linear collider (LC) with longitudinally
polarized electron beams.  The potential of linear colliders
for detailed studies of supersymmetry has been discussed previously in
numerous
reports~\cite{susy-Ahn:1988vj,susy-Zerwas:1900hw,susy-unknown:1992mg,susy-Kuhlman:1996rc,susy-Daniel,susy-Abdullin:1999zp,susy-Heuer:2001qp}.
In this chapter, many well-established results are reviewed, including
the potential for model-independent measurements of superpartner
masses.  In addition, several less well-appreciated topics are
discussed.  These include loop-level effects in supersymmetry, CP
violation, and supersymmetric flavor violation. This discussion
serves
both to illustrate the rich program of supersymmetric studies
available at linear colliders, and to highlight areas that merit
further study.  This chapter concludes with a review of the important
complementarity of the LC and the Large Hadron Collider (LHC) with
respect to supersymmetry studies.

The signatures of supersymmetry are many, ranging from the well-known
missing energy in supergravity with R-parity
conservation~\cite{susy-Nilles:1984ge,susy-Haber:1985rc}  to
exotic signatures appearing in models with
gauge-mediated~\cite{susy-Giudice:1999bp} and
anomaly-mediated~\cite{susy-Randall:1999uk,susy-Giudice:1998xp}
supersymmetry breaking.  Space constraints prevent a complete
review of the considerable work done in each of these, and other,
frameworks.  Instead, this review focuses on supergravity frameworks
leading to the conventional signature of missing energy.  R-parity
violation and alternative  supersymmetry-breaking mechanisms are treated as
variations, and are discussed where they are especially pertinent.

\section{The scale of supersymmetry}

The cleanliness of the linear collider environment implies that
precise, model-independent measurements in supersymmetry are possible, 
but only if
supersymmetric final states are
kinematically accessible.  The mass scale of supersymmetric particles
is therefore of paramount importance.  In this section we review
bounds on superpartner masses from naturalness criteria, dark matter
constraints, Higgs boson searches, and precision electroweak data.  We
also consider the potential of experimental evidence for new physics
to constrain the supersymmetric mass scale; we discuss the muon
anomalous magnetic moment as an example.

\subsection{Naturalness}

In supersymmetric extensions of the SM,
quadratically divergent quantum corrections to the masses of
fundamental scalars are of the order of the superpartner mass scale.  Given a
mechanism for producing sufficiently light superpartners,  the
observed weak scale is obtained without unnaturally large
cancellations in the electroweak potential.  While no analysis of
naturalness can claim quantitative rigor, the importance of
naturalness as a fundamental motivation for supersymmetry has prompted
many studies~\cite{susy-Ellis:1986yg,susy-Barbieri:1988fn,susy-Ross:1992tz,susy-deCarlos:1993yy,susy-Anderson:1995dz,susy-Anderson:1995tr,susy-Anderson:1996cp,susy-Dimopoulos:1995mi,susy-Pomarol:1996xc,susy-Agashe:1997kn,susy-Ciafaloni:1997zh,susy-Bhattacharyya:1997dw,susy-Barbieri:1998uv,susy-Giusti:1999gz,susy-Romanino:2000ut,susy-Chan:1998bi,susy-Chankowski:1998zh,susy-Chankowski:1999xv,susy-Kane:1999im,susy-Bastero-Gil:2000gu,susy-Feng:2000mn,susy-Feng:2000zg,susy-Feng:2001bp}, with important qualitative implications for the
superparticle spectrum.

To study naturalness one must first assume a certain supersymmetric
framework.  Models in this framework are specified by a set of
input parameters, typically defined at some high energy scale.  Together
with experimental constraints and renormalization group equations,
these parameters determine the entire weak-scale Langrangian,
including the $Z$ boson mass, which at tree level is
\begin{equation}
\frac{1}{2} m_Z^2 = \frac{m_{H_d}^2 - m_{H_u}^2 \tan^2\beta }
{\tan^2\beta -1} - \mu^2\ , 
\label{mZ}
\end{equation}
where $m_{H_d}^2$, $m_{H_u}^2$ are the mass parameters of the two
Higgs doublets of the model and $\tan\beta = \VEV{H_u}/\VEV{H_d}$.
Naturalness is then often imposed by demanding that the weak scale be
insensitive to variations in some set of parameters $a_i$, which are
assumed to be continuously variable, independent, and fundamental.
The $a_i$ may be scalar masses, gaugino masses, and other parameters,
but are not necessarily input parameters. The sensitivity is typically
quantified by defining
coefficients~\cite{susy-Ellis:1986yg,susy-Barbieri:1988fn} $c_i \equiv
\left| (a_i/m_Z) (\partial m_Z/ \partial a_i) \right|$
for each parameter $a_i$ and taking some simple combination of the
$c_i$, often $c = \max \{ c_i \}$, as an overall measure of
naturalness.  A naturalness criterion $c < c_{\rm max}$ then implies
upper bounds on supersymmetry parameters and superpartner masses.

Following the early
studies~\cite{susy-Ellis:1986yg,susy-Barbieri:1988fn}, the authors of
\cite{susy-deCarlos:1993yy} stressed the importance of including
one-loop corrections to Eq.~(\ref{mZ}).  They also noted that it is
possible in principle for a given $c_i$ to be large for all possible
choices of $a_i$. In the latter case, the authors of
\cite{susy-Anderson:1995dz,susy-Anderson:1995tr,susy-Anderson:1996cp}
argued that, to avoid misleading results, only {\em unusually} large
sensitivity should be considered unnatural and proposed replacing $c$
by $\tilde{\gamma} \equiv \max \{ c_i / \bar{c}_i \}$, with
$\bar{c}_i$ an average sensitivity. More recently, another alternative
prescription has been
proposed~\cite{susy-Ciafaloni:1997zh,susy-Bhattacharyya:1997dw,susy-Barbieri:1998uv,susy-Giusti:1999gz,susy-Romanino:2000ut}
in which the sensitivity coefficients are replaced by $\left|(\Delta
a_i/m_Z) (\partial m_Z/\partial a_i) \right|$, where $\Delta a_i$ is
the experimentally allowed range of $a_i$. This definition implies
that arbitrarily large but well-measured supersymmetry parameters are
natural, and has been argued to differ sharply from conventional
notions of naturalness~\cite{susy-Feng:2001bp}.

The results of naturalness studies are strongly dependent on the
choice of framework, the choice of fundamental parameters $a_i$, and,
of course, the choice of $c_{\rm max}$ (or the equivalent
$\tilde{\gamma}$ parameter).  The dependence on framework assumptions
is inescapable.  In other studies of supersymmetry there exists, at
least in principle, the possibility of a model-independent study,
where no correlations among parameters are assumed.  In studies of
naturalness, however, the correlations determine the results, and
there is no possibility, even in principle, of an all-inclusive
framework.  We describe here only some of the qualitatively distinct
possibilities.  For alternative analyses, readers are referred to the
original literature~\cite{susy-Ellis:1986yg,susy-Barbieri:1988fn,susy-Ross:1992tz,susy-deCarlos:1993yy,susy-Anderson:1995dz,susy-Anderson:1995tr,susy-Anderson:1996cp,susy-Dimopoulos:1995mi,susy-Pomarol:1996xc,susy-Agashe:1997kn,susy-Ciafaloni:1997zh,susy-Bhattacharyya:1997dw,susy-Barbieri:1998uv,susy-Giusti:1999gz,susy-Romanino:2000ut,susy-Chan:1998bi,susy-Chankowski:1998zh,susy-Chankowski:1999xv,susy-Kane:1999im,susy-Bastero-Gil:2000gu,susy-Feng:2000mn,susy-Feng:2000zg,susy-Feng:2001bp}.

\begin{figure}[t]
\begin{center}
\epsfig{file=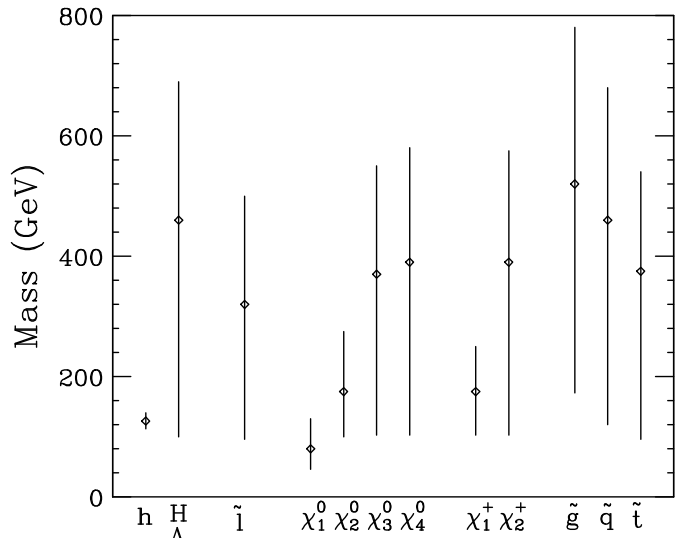,width=0.58\textwidth}
\end{center}
\vspace*{-.4in}
\caption{Natural ranges of superpartner masses in minimal
supergravity.  The upper limits are set by the requirement
$\tilde{\gamma} < 10$ and the diamonds indicate upper bounds
corresponding to $\tilde{\gamma} < 5$.  The lower limits are roughly
those from current collider constraints. Updated from
\protect\cite{susy-Anderson:1995tr}. \label{fig:susy-bounds} }
\end{figure}

In minimal supergravity, one assumes both scalar and gaugino
universality at a high scale.  If one requires insensitivity of the
weak scale with respect to both supersymmetry breaking and Standard
Model parameters, none of the superpartner masses can naturally be far
above the weak scale.  Examples of the resulting naturalness bounds
are given in Fig.~\ref{fig:susy-bounds}. The bounds for
non-strongly interacting superpartners are typically more stringent
than those for colored superpartners.  Similar results are found in
other frameworks where all scalar and gaugino masses are comparable at
some high scale.

Naturalness bounds may be very different in
other frameworks, however, especially for scalars.
  For squark and slepton masses, if no
correlations are assumed, the bounds are highly generation-dependent.
At one-loop, the weak scale is sensitive to sfermion masses only
through renormalization group terms proportional to Yukawa couplings.
Thus, while the scalar masses of the third generation are still
usefully constrained by naturalness criteria, first- and 
second-generation scalars may have masses above $10~\tev$ without requiring
large fine-tuning~\cite{susy-Dimopoulos:1995mi,susy-Pomarol:1996xc},
putting them  far beyond the kinematic reaches of both the LHC and
future linear colliders.  `Superheavy' first and second generation
scalars ameliorate the supersymmetric flavor and CP problems and are
found in many
models~\cite{susy-Drees:1986jx,susy-Dine:1990jd,susy-Dvali:1996rj,susy-Dvali:1998sr,susy-Cohen:1996vb,susy-Zhang:1997vu,susy-Nilles:1997me,susy-Feng:1999iq,susy-Bagger:1999ty,susy-Bagger:2000sy,susy-Shafi:2000vm,susy-Baer:2000gf,susy-Baer:2001xa,susy-Baer:2001vw,susy-Hisano:1999tm,susy-Hisano:2000wy,susy-Kaplan}.

Alternatively, given the possibility that SM couplings are
fixed in sectors separate from supersymmetry breaking, one may
reasonably require only that the weak scale be insensitive to
variations in parameters related to supersymmetry
breaking~\cite{susy-Feng:2000mn,susy-Feng:2000zg,susy-Feng:2001bp}.
With this less stringent criterion, in many simple models, including
minimal supergravity, all scalar partners  may be naturally in the 2--3 TeV
range, as a result of focusing behavior in renormalization group
trajectories~\cite{susy-Feng:2000mn,susy-Feng:2000zg,susy-Feng:2001bp,susy-Feng:2000hg,susy-Agashe:2000ct,susy-Chattopadhyay:2000qa,susy-Allanach:2000ii,susy-Chattopadhyay:2000fj}.  Such ``focus point supersymmetry'' models
also have significant virtues with respect to low-energy constraints,
and predict that even third-generation scalars may have masses well above 1 TeV and be beyond
the reach of linear colliders.

Bounds on the masses of fermionic superpartners are less
framework-dependent.  If the gaugino masses are uncorrelated, the
gluino mass is typically stringently bounded by its indirect influence
on the weak scale through the top squarks.  In this 
general context,  the electroweak gaugino
masses may be significantly
larger~\cite{susy-Kane:1999im,susy-Bastero-Gil:2000gu}.  However, in
most well-motivated models, the gluino is much heavier than the
electroweak gauginos, and so naturalness implies stringent limits on
Bino and Wino masses.  While the scale of the $\mu$ parameter may be
determined~\cite{susy-Giudice:1988yz}, a quantitative theory for the
$\mu$ term is lacking.  The $\mu$ parameter is therefore usually
determined through Eq.~(\ref{mZ}) and is otherwise assumed to be
uncorrelated with other parameters.  Large $\mu$ then necessarily
leads to large fine-tuning, and so heavy Higgsinos are disfavored.  As
a result, given our present understanding, naturalness criteria
typically imply relatively stringent bounds on the masses of all six
chargino and neutralino states, and they encourage the expectation
that all of these particles  will be available for study at linear colliders.

\subsection{Neutralino relic abundance}

An important virtue of many supersymmetric theories is the existence
of a non-baryonic dark matter candidate.  The most straightforward
possibility is the lightest neutralino
$\chi$~\cite{susy-Goldberg:1983nd,susy-Ellis:1983wd}, which is often
the lightest supersymmetric particle (LSP) and so is stable in models
with conserved R-parity.  Current cosmological and astrophysical
measurements prefer $0.1 \lsim \Omega_m h^2 \lsim
0.3$~\cite{susy-omega}, where $\Omega_m$ is the ratio of dark
matter density to critical density, and $h \approx 0.65$ is the Hubble
parameter in units of $100\,{\rm km\,s^{-1}\,Mpc^{-1}}$.  The superpartner
spectrum is then constrained by the requirement that 
 the thermal relic density of the lightest
neutralino satisfy  $\Omega_{\chi} h^2 \lsim
0.3$.

The neutralino relic density is determined by the neutralino pair
annihilation cross section and has been the subject of many
analyses~\cite{susy-Griest:1991kh,susy-Gondolo:1991dk,susy-Nath:1993ty,susy-Mizuta:1993qp,susy-Ellis:1998kh,susy-Gomez:2000dk,susy-Drees:1993am,susy-Olive:1989jg,susy-Griest:1990zh,susy-Olive:1991qm,susy-Lopez:1991bk,susy-Kelley:1993ks,susy-Roberts:1993tx,susy-Kane:1994td,susy-Baer:1996nc,susy-Arnowitt:1998uz,susy-Abdullin:1999nv,susy-Ellis:2000yp,susy-Drees:1997pk,susy-Barger:1998kb,susy-Boehm:2000bj,susy-Bottino:2000gc,susy-Feng:2000gh,susy-Feng:2001zu,susy-Baer:2001jj,susy-Ellis:2001ms,susy-Moroi:2000zb,susy-Murakami:2000me,susy-Birkedal-Hansen:2001is}.
These include refined treatments of
poles~\cite{susy-Griest:1991kh,susy-Gondolo:1991dk,%
susy-Nath:1993ty}, annihilation
thresholds~\cite{susy-Griest:1991kh,susy-Gondolo:1991dk}, and
co-annihilation among Higgsinos~\cite{susy-Mizuta:1993qp} and with
staus~\cite{susy-Ellis:1998kh,susy-Gomez:2000dk}.  The S- and
P-wave contributions to all tree-level processes with two-body final
states are given in \cite{susy-Drees:1993am}.

In general, neutralinos may annihilate through $t$-channel sfermions
to $f\bar{f}$ , through $s$-channel $Z$ and Higgs bosons to
$f\bar{f}$, and through $t$-channel charginos and neutralinos to $WW$
and $ZZ$.  For Bino dark matter, only the sfermion-mediated amplitudes
are non-vanishing.  An upper bound on $\Omega_{\chi} h^2$ then leads
to an upper bound on at least one sfermion mass.  This, together with
the requirement that $\chi$ be the LSP, implies an upper bound on
$m_{\chi}$.  Such reasoning has led to claims of cosmological upper
bounds on superpartner masses with optimistic implications for
supersymmetry at linear colliders~\cite{susy-Olive:1989jg,susy-Griest:1990zh,susy-Olive:1991qm,susy-Lopez:1991bk,susy-Kelley:1993ks,susy-Roberts:1993tx,susy-Kane:1994td,susy-Baer:1996nc,susy-Arnowitt:1998uz,susy-Abdullin:1999nv,susy-Ellis:2000yp}.

These claims must be viewed cautiously, however, as they are true only
in the $\chi \approx \tilde{B}$ limit and are violated even in the
simplest scenarios.  In minimal supergravity, for example, multi-TeV
LSPs are possible for large $m_0$~\cite{susy-Feng:2000gh}, where the
LSP has a significant Higgsino admixture, leading to large
annihilation cross sections to gauge bosons.  Useful upper bounds are
also absent in minimal supergravity at large
$\tan\beta$~\cite{susy-Feng:2000gh,susy-Feng:2001zu,susy-Baer:2001jj,susy-Ellis:2001ms}, where the importance of a small Higgsino admixture
in $\chi$ is amplified and leads to large Higgs boson-mediated
annihilation.  More generally, no guarantee of light superpartners is
possible for Wino-~\cite{susy-Moroi:2000zb,susy-Murakami:2000me,susy-Birkedal-Hansen:2001is} and
Higgsino-like~\cite{susy-Mizuta:1993qp,susy-Drees:1997pk} LSPs, which
annihilate very efficiently to negligible relic densities. Finally, it
is worth recalling that these upper bounds are also inapplicable in
theories with low-energy supersymmetry breaking or R-parity
violation, where the lightest neutralino is no longer stable.

\subsection{Higgs mass and precision electroweak constraints}

As is well known, supersymmetry places severe constraints on the mass
of the lightest Higgs boson.  In the Minimal Supersymmetric Standard
Model (MSSM), one-loop
calculations~\cite{susy-Haber:1991aw,susy-Berger:1990hg,susy-Okada:1991vk,susy-Ellis:1991nz,susy-Barbieri:1991tk,susy-Haber:1993an,susy-Chankowski:1994er,susy-Dabelstein:1995hb,susy-Dabelstein:1995js,susy-Pierce:1997zz} have now been supplemented
with leading two-loop corrections in the Feynman
diagrammatic~\cite{susy-Heinemeyer:1998jw,susy-Heinemeyer:1998kz,susy-Heinemeyer:1999np,susy-Heinemeyer:1999be,susy-Carena:2000dp}, renormalization
group~\cite{susy-Casas:1995us,susy-Carena:1995bx,susy-Carena:1996wu,susy-Haber:1997fp}, and effective potential~\cite{susy-Hempfling:1994qq,susy-Zhang:1999bm,susy-Espinosa:2000df} approaches, leading to an upper bound of $m_h
\lsim 135~\gev$~\cite{susy-Heinemeyer:1999np}.  The consistency of
this bound with precision electroweak fits is a considerable success
of supersymmetry.  At the same time, though, one might expect that the
current lower bound $m_h > 113.5~\gev$ from direct Higgs
searches~\cite{susy-Acciarri:2000ke,%
susy-Barate:2000ts,susy-Abbiendi:2001ac,susy-Abreu:2001fw} and the
success of precision electroweak fits to the SM disfavors
the possibility of light superpartners.

However, closer analysis shows that  light
superpartners are consistent with the current Higgs mass bound.
For example, in general scenarios, the current Higgs mass
limit may be satisfied with large masses only for the top and 
bottom squarks.  Even for these, the constraints are not severe.
Charginos,
neutralinos, and sleptons may be light and within the reach
of linear colliders.  In simpler frameworks, the Higgs limit is
more constraining.  Even in minimal supergravity, however, the
current Higgs mass bound, along with the requirement of a suitable
dark matter candidate, may be satisfied either for chargino masses
above 200 GeV~\cite{susy-Ellis:2000sv} or for large
$m_0$~\cite{susy-Feng:2001bp,susy-Su:2000cb}.  In the latter case,
charginos may be as light as their current LEP bound.  The Higgs mass
bound can also be made consistent with light superpartners if there are large
CP-violating phases, which must necessarily cancel to high accuracy in
electric dipole moments, or new singlets~\cite{susy-Kane:2000kc}.
Thus, the current Higgs mass constraint, although already rather
stringent, does not exclude the possibility of light superpartners.

The supersymmetric spectrum is also constrained by precision
electroweak measurements.  The effects of supersymmetry have been
studied in numerous recent works (see, {\em e.g.},
\cite{susy-Djouadi:1997pa,susy-Djouadi:1998sq,susy-Erler:1998ur,susy-Cho:2000km,susy-Cho:2000sf,susy-Erler:2000jg}).
While there are at present no strong indications for supersymmetry
from these considerations, light superparticles cannot be excluded
either.  This issue is discussed further in Chapter 8, Section 3.2.

\subsection{Evidence for new physics}

Finally, weak-scale supersymmetry has implications for a broad range
of experiments in particle physics and astrophysics.  If deviations
from SM predictions are found, these deviations may also
constrain the scale of superpartner masses. 

\begin{figure}[t]
\begin{minipage}[t]{0.48\textwidth}
\begin{center}
\epsfig{file=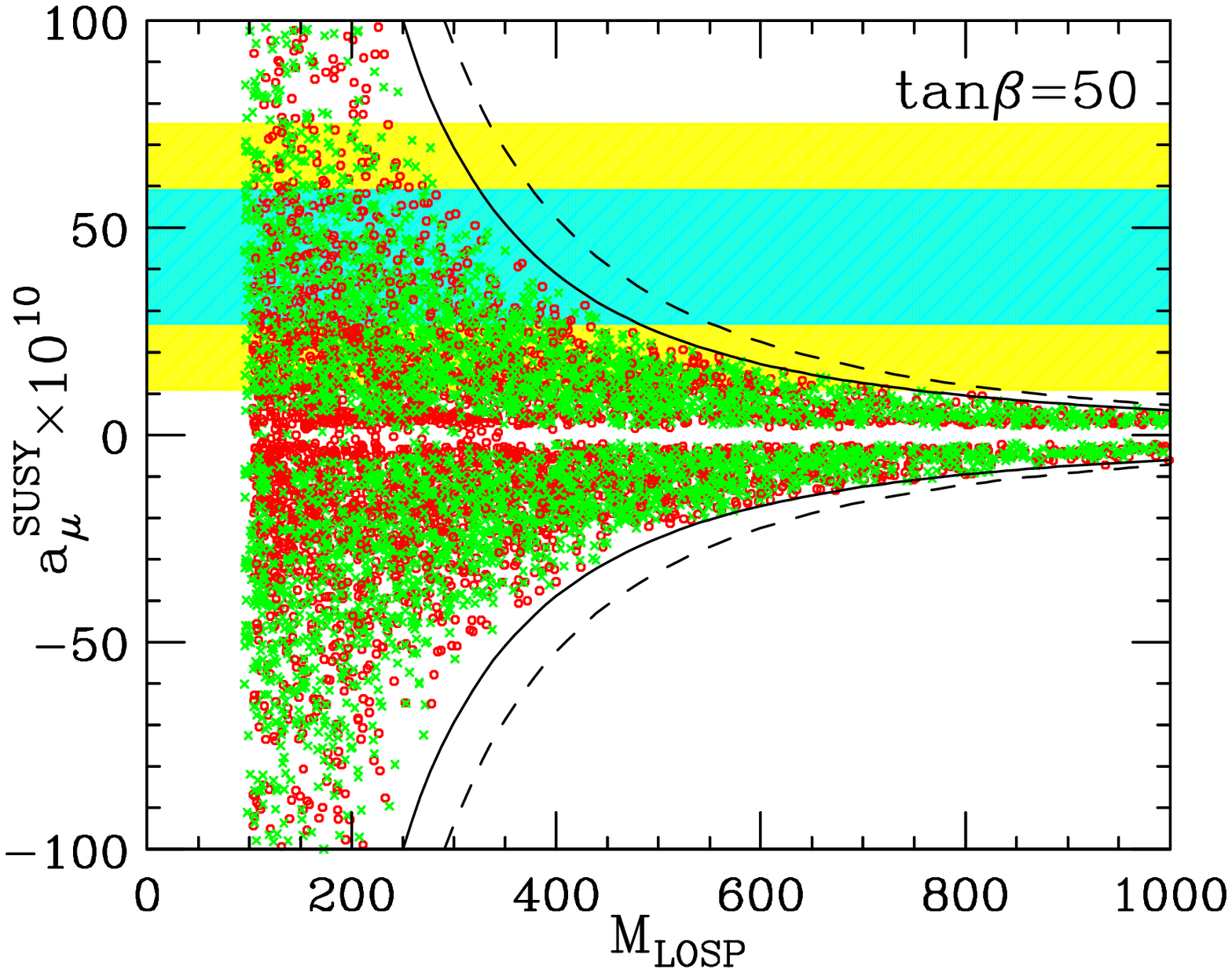,width=\textwidth}
\end{center}
\end{minipage}
\hfill
\begin{minipage}[t]{0.48\textwidth}
\begin{center}
\epsfig{file=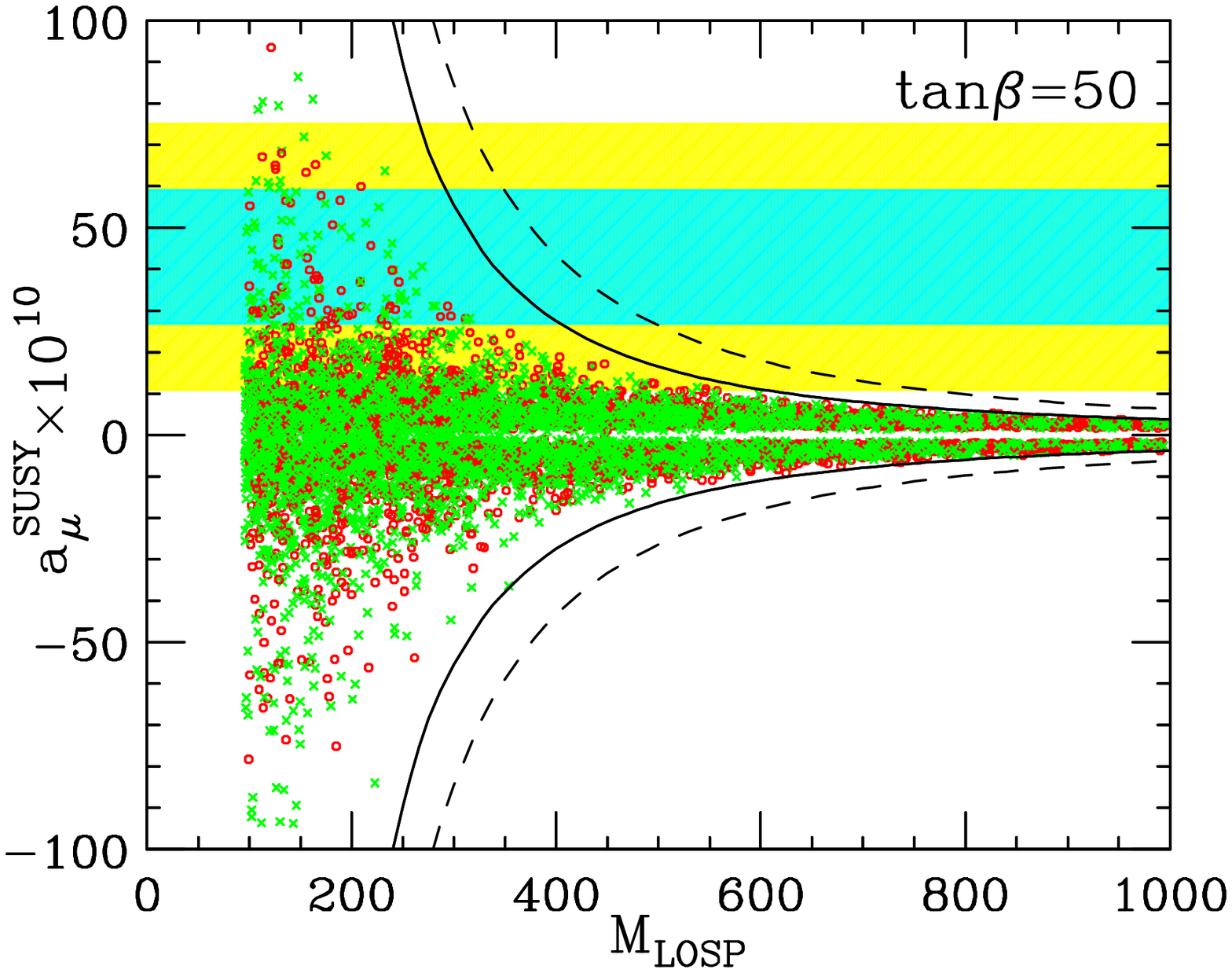,width=\textwidth}
\end{center}
\end{minipage}
\caption{Possible values of the mass of the lightest observable
supersymmetric particle, $M_{\rm LOSP}$, and the supersymmetric
contribution to the muon's anomalous magnetic moment, $a_{\mu}^{\rm
SUSY}$, assuming a stable LSP (left) and a visibly decaying LSP
(right). Crosses (circles) have smuon (chargino/neutralino) LOSPs and
satisfy the parameter constraints $M_2=2M_1$, $A_{\mu} = 0$, and
$\tan\beta = 50$.  Relaxing the gaugino unification assumption leads
to the solid envelope curve, and further allowing arbitrary $A_{\mu}$
leads to the dashed curve. The envelope contours scale linearly with
$\tan\beta$.  The shaded regions are the 1$\sigma$ and 2$\sigma$
experimentally preferred regions. {}From
\protect\cite{susy-Feng:2001tr}. \label{fig:susy-amu} }
\end{figure}

As an example, we consider the recently reported 2.6$\sigma$ deviation
in the anomalous magnetic moment of the muon~\cite{susy-Brown:2001mg}:
$a_{\mu}^{\rm exp} - a_{\mu}^{\rm SM} = (43 \pm 16) \times 10^{-10}$.
Supersymmetric contributions to $a_{\mu}$ are
well known~\cite{susy-Grifols:1982vx,susy-Ellis:1982by,susy-Barbieri:1982aj,susy-Kosower:1983yw,susy-Yuan:1984ww}, and the
measured deviation is naturally explained by
supersymmetry~\cite{susy-Czarnecki:2001pv,susy-Everett:2001tq,susy-Feng:2001tr,susy-Baltz:2001ts,susy-Chattopadhyay:2001vx,susy-Komine:2001fz,susy-Baer:2001kn,susy-Hisano:2001qz,susy-Ibrahim:2001ym,susy-Ellis:2001yu,susy-Choi:2001pz,susy-Martin:2001st}. 
If a supersymmetric interpretation is adopted, the result restricts the
masses of some superpartners.  Highly model-independent upper bounds
on the mass of the lightest observable supersymmetric particle are
given in Fig.~\ref{fig:susy-amu}.  If theory and experiment are required to
agree within 1$\sigma$, at least one observable
superpartner must be lighter than 490 GeV if the LSP is
stable, and lighter than 410 GeV if the LSP decays visibly
in the detector. If agreement only within 2$\sigma$ is required,
these limits weaken to 800 GeV and 640 GeV, respectively.
The bounds are for the case $\tan\beta \le 50$ and scale
linearly with $\tan\beta$.

These results illustrate the power of evidence for new physics to
constrain the scale of supersymmetry.  Of course, many other
experiments may also see supersymmetric effects. Among the areas in
which great experimental progress is expected in the next few years
are searches for new physics at the Tevatron, $B$ physics (CP
violation, rare decays), lepton flavor violation ($\mu$-$e$
conversion, $\mu \to e \gamma$, etc.), electric dipole moments,
searches for dark matter (both direct and indirect), and cosmic ray
physics.  Pre-LHC evidence for supersymmetry is not
guaranteed, but, in simple frameworks like minimal supergravity where
systematic and comprehensive analyses are possible, it is very
likely~\cite{susy-Feng:2001zu}.  Strong evidence for new physics, even
if indirect, will provide important additional constraints on the mass
scale of supersymmetric particles.

\section{Determination of masses and couplings}

The usefulness of a linear collider in the study of SUSY particles lies both
in the simplicity of the production process and in the fact that the electron
can have a large longitudinal polarization. These features
allow one to carry out accurate measurements of the masses and
 the quantum numbers
of the particles being produced, and also to determine their
 gauge coupling constants in a 
model-independent manner~\cite{susy-Tsukamoto,susy-Feng:1995zd}. Such measurements are crucial 
in understanding the nature of the processes being uncovered.

\subsection{Measurement of superpartner masses}

We begin our review of mass  measurements by considering  one particular 
process that illustrates the essential simplicity of the analyses.
The process we will consider is selectron production,
\beq
             \ee \to \s  e_{L,R}^+ \s e_{L,R}^- \ , 
\eeq{susy-selectronrxn}
where $\s e_R^-$, $\s e_L^-$ are the supersymmetry partners of the right-
and left-handed electron.  We assume that both selectrons decay by 
 $\widetilde{e}_{L,R} \rightarrow e \s\chi_1^0$.  The process
has a number of interesting  features.  The masses of the 
 $\widetilde{e}_R$ and $\widetilde{e}_L$ can differ substantially.  
The combinations $\s e_{R}^+ \s e_{R}^-$ and $\s e_{L}^+ \s e_{L}^-$ are 
produced by $s$-channel photon and $Z^0$ exchange,  but all four possible
selectron combinations are produced by $t$-channel neutralino exchange.
Thus, the study of this process can give information on SUSY masses,
quantum numbers, and coupling constants.

In the reaction \leqn{susy-selectronrxn}, the selectrons are produced at
a fixed energy.  Since they are scalars, they decay isotropically in their
own frames.  These distributions of the decay electrons and positrons
 boost to  distributions in the lab
that are flat in energy between the kinematic endpoints.
The electrons and positrons then show box-like distributions.
 The maximum and minimum
energies which form the edges of the box determine the masses of the 
$\s e$ and the  $\s\chi_1^0$ through the relations
\begin{eqnarray*}
M_{\widetilde{e}}^2 &=& 
{E_{\rm cm}^2}\left\{{{E_{e,\rm max}E_{e,\rm min}}\over
{(E_{e,\rm max}+E_{e,\rm min})^2}}\right\}\\
M_{\s\chi_1^0}^2 &=& 
M_{\widetilde{e}}^2\left\{1-2{{E_{e,\rm max}+E_{e,\rm min}}
\over{E_{\rm cm}}}\right\} \ .
\end{eqnarray*} 
If several different combinations of selectrons are produced, the 
electron and positron energy spectra will show a superposition of 
several box-like distributions. Each set of endpoints gives the 
associated selectron masses and an independent
 determination of the  $\s\chi_1^0$
mass.

\begin{figure}[t!]
\begin{center}
\epsfig{file=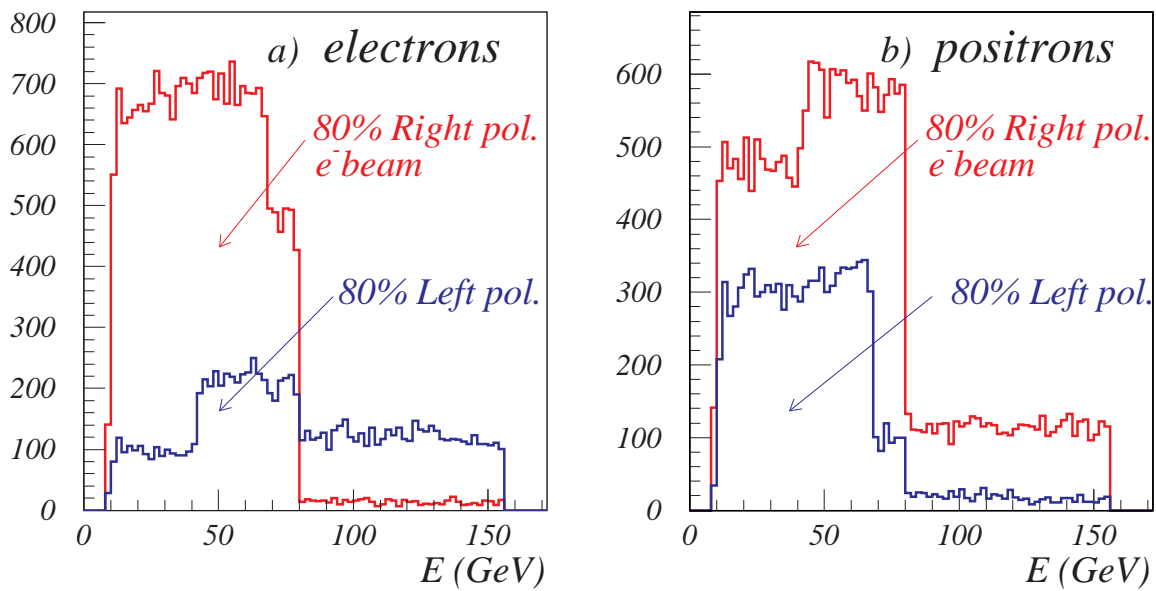, width=6.0in}
\caption{Electron and positron energy distributions for selectron pair production,
with the indicated beam polarizations and integrated luminosity  50 fb$^{-1}$
\cite{susy-Goodman_1}. 
\label{fig:susy-selectron_nlcproc}}
\end{center}
 \end{figure}

 Figure~\ref{fig:susy-selectron_nlcproc} shows the electron and 
positron spectra for a particular set of  MSUGRA parameters constructed
for the  Snowmass `96 summer 
study~\cite{susy-Snowmass}, assuming 50 fb$^{-1}$ of data at 
$\sqrt{s} = 500$ GeV~\cite{susy-Goodman_1}.
 The simulations use the  event generator 
ISAJET~\cite{susy-Baer:1999sp}.  The expected  box-like spectra
appear clearly, with sharp endpoints.  Both the electron and positron
spectra have a strong dependence on polarization, and this allows us to
recognize which components are associated with $\s e_L$ and which with
$\s e_R$.  The electron and positron spectra also differ from each other,
reflecting the different production of $\s e_R^-\s e_L^+$ versus 
 $\s e_L^-\s e_R^+$ from polarized beams.

Figure~\ref{fig:susy-selectron_reco} \ compares the generated electron 
and positron distributions to those reconstructed using energy measurements
from the electromagnetic calorimeter of the L detector described in 
Chapter 15.  The study uses full GEANT simulation of the 
calorimeter~\cite{susy-Dunn}.
 The effect of resolution is clearly observed in 
the upper edge of the energy distribution.
This analysis does not include  beamstrahlung and initial state
radiation, but these effects are not expected to affect significantly the 
determination of the edges in  the energy spectra~\cite{susy-Snowmass}. 

{
\begin{figure}[hbtp]
\epsfig{file=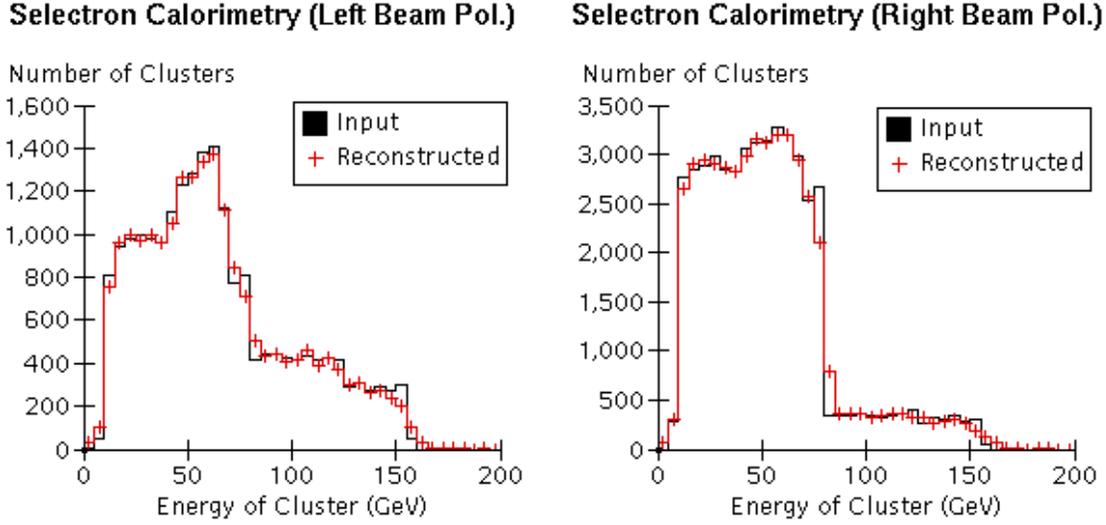, width=6.0in}
\caption{Input and calorimeter-reconstructed $e^\pm$ energy distributions
from selectron pair production for 80\% left-polarized (left) and 80\% 
right-polarized (right) electron beams~\cite{susy-Dunn}. 
The effect of calorimeter resolution is evident at high cluster energies. 
\label{fig:susy-selectron_reco}}
\end{figure}
  
\begin{table} 
\begin{center}
\begin{tabular}{c|c|rr|c|rr}\hline\hline
  Reference          &   Particle  &  Input   & Measured  &    Particle & Input  & Measured \\
\hline
\cite{susy-Goodman_1}& $\elpm$ &   238.2  &  239.4~~~ & $\chionez$ &  128.7 &   129.0~~~ \\
\cite{susy-Goodman_1}& $\erpm$ &   157.0  &  158.0~~~ & $\chionez$ &  128.7 &   129.0~~~ \\
\cite{susy-Staszak}  &$\murpm$ &   157.1  &  143.2~~~ & $\chionez$ &  128.7 &   117.3~~~ \\
\cite{susy-Danielson}&$\sneue$ &   206.6  &  199.4~~~ & $\chionepm$&   96.4 &    96.5~~~ \\
\cite{susy-Tsukamoto}&$\chionepm$& 219.0  &  212.0~~~ & $\chionez$ &  118.0 &   116.5~~~ \\
\cite{susy-Barron_1} &$\chionepm$& 238.0  &  239.8~~~ & $\asneul$  &  220.0 &   221.2~~~ \\
\cite{susy-Baer:1996vd}&$\chitwopm$& 175.2 & 176.5~~~ & $\chionepm$&   85.9 &    86.1~~~ \\ 
\cite{susy-Veeneman} &$\chitwopm$& 290.4  & 282.7~~~  & $\chionepm$&   96.0 &    97.9~~~ \\
\hline\hline
\end{tabular}
\end{center}
\caption{Comparison of the input and measured masses (in GeV) for a few 
supersymmetric particles as determined from the end-point spectrum of the 
observed particles smeared via fast MC techniques. Most of the results are 
based on a 50 fb$^{-1}$ data sample. The pair of masses in each row are 
determined from the end-point measurement in pair-production of the 
first particle listed. \label{tab:masses}}
\end{table}}

Many similar analyses of   the determination
of slepton  masses have been carried out using
 fast Monte Carlo 
techniques~\cite{susy-Goodman_2,susy-Williams,susy-Danielson,susy-Baer:1996vd}.
Some of the results are summarized in
 Table~\ref{tab:masses}.  One can see from the table that we expect
to be able to measure these masses with an 
accuracy of a few percent or less in most cases.  The determination of the mass
of the lighter chargino 
$\chionepm$ has been studied by many groups. Measurements based on an 
analysis using  background 
cuts~\cite{susy-Tsukamoto,susy-Baer:1996vd,susy-Martyn}
indicate that this mass  can be measured 
with accuracies of 1\% or less by this method.
 An interesting signal thast may
be background-free is the case where one $\chionepm$ decays into a lepton 
and a
$\asneul$,  with the $\asneul$ decaying to a $\nu \chionez$, while the
other $\chionepm$ decays into $	q\bar q \chionez$. In this case,
it should be 
possible to remove the $WW$ background completely without affecting the 
signal~\cite{susy-Barron_1}. 
The mass measurement for the heavier chargino $\chitwopm$ has also been 
studied, assuming a CM
energy of 750 GeV. By using the decay of the  $\chitwopm$ 
 into $\chionepm Z^0$, where the $Z$ decays into leptons and the 
$\chionepm$ decays into hadrons, one is able to get quite accurate
 results~\cite{susy-Veeneman}. The conclusions of all these analyses are
also shown in 
Table~\ref{tab:masses}. 
  
\begin{figure}[t]
\begin{center}
\epsfig{file=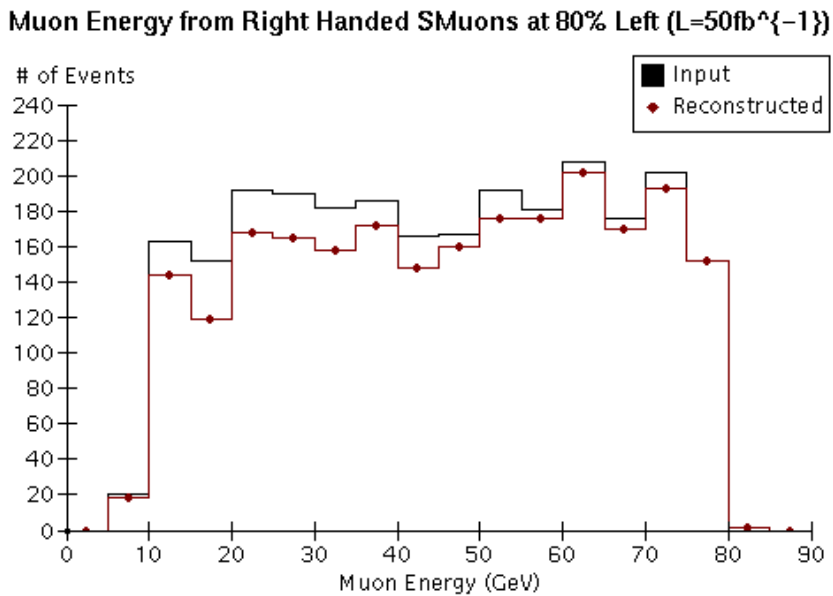, width=4.5in}
\caption{Input and tracker-reconstructed muon energy spectra from
smuon pair production with an 80\% left-polarized electron beam~\cite{susy-Staszak}.
\label{fig:susy-smuon_geant}}
\end{center}
\end{figure}

It is worth reviewing some of the experimental issues that arise in
these measurements.  We have already given an example in which
the calorimeter resolution affects the mass measurements for selectrons 
decaying to  
$e^-$ and $e^+$.  For the case of smuons decaying to $\mu^\pm$, the 
corresponding issue is tracking resolution.  In 
 Fig.~~\ref{fig:susy-smuon_geant}, we show a comparison of generator-level
and reconstructed muon energy in $\s \mu$ pair production.  It is
clear that the tracking reconstruction
 does not significantly  affect the energy edge resolution, 
and hence it does not affect our ability to 
determine supersymmetric masses accurately. For chargino decays, 
both calorimeter and tracking resolution enter the determination of
kinematic endpoints~\cite{susy-Tsukamoto}. 

To examine the supersymmetry signals, it is necessary to 
remove backgrounds events efficiently. The major sources of SM backgrounds are 
the two-photon ($\gamma^\star \gamma^\star$) process, which
gives rise to lepton and quark pairs in the detector, 
$\ee$ annihilation to 
the $W^+W^-$, $Z^0Z^0$, and $Z^0h^0$, and single-$W$ production
($e \gamma^*\to \nu W$).  Methods for removing the annihilation and
single $W$ backgrounds from the supersymmetry sample are explained in 
\cite{susy-Tsukamoto,susy-Dima,susy-Danielson_Goodman}.  The two-photon
background is a problem in reactions whose signatures involve
 missing energy, but it can be
controlled by also requiring  missing transverse momentum.
Methods for measuring the 
two-photon background have been studied in 
\cite{susy-Markiewicz,susy-Danielson_Goodman,susy-Danielson_Newman,susy-Kelly_Takeuchi,susy-Barron_2}.
There may also be backgrounds from the decays of other 
supersymmetric particles but, in most cases, these are either small or have 
distinctive signals that allow one to identify them.

\begin{figure}[htb]
\begin{center}
\epsfig{file=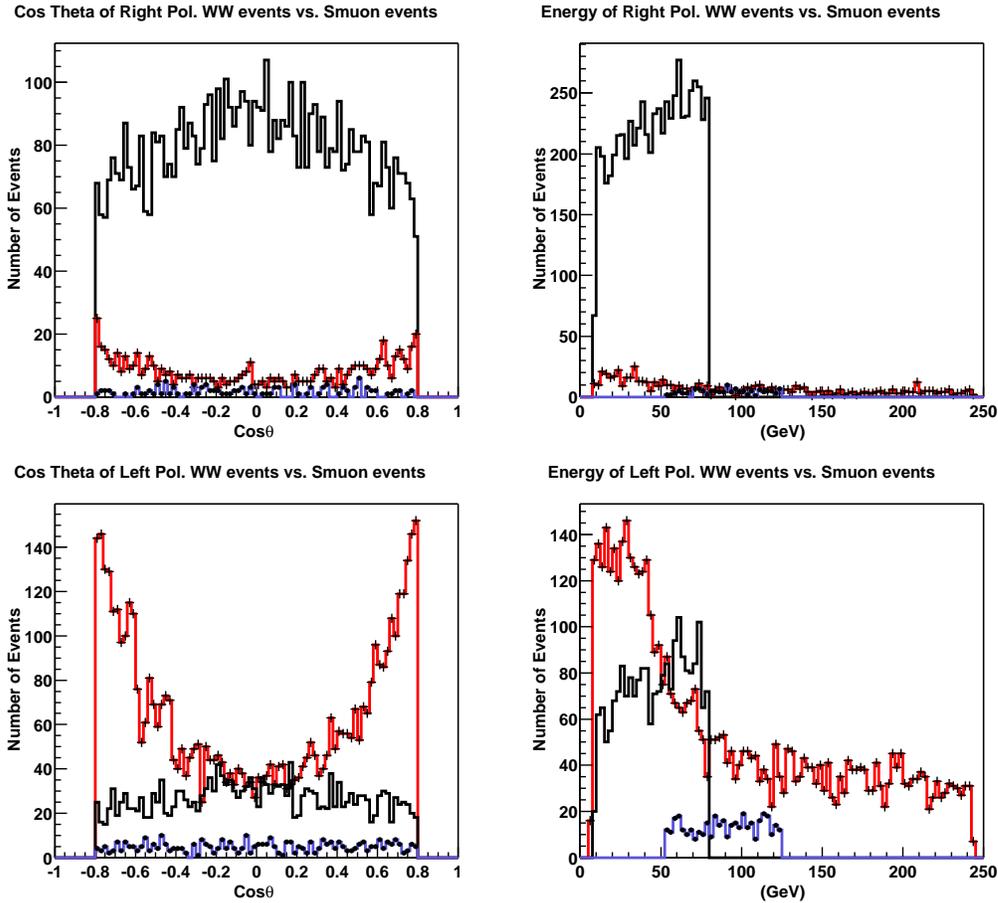, width=5.5in}
\caption{Kinematic distributions of muons from $\s\mu_R$ pair production
(solid), $\s\mu_L$ pair production (dotted), and $W^+W^-$ background
(crossed)~\cite{susy-Staszak}.  An electron beam polarization of  
80\% is assumed.
\label{fig:susy-smuon_background}}
\end{center}
\end{figure}
One case in which $W$ pair production is a serious background is the
study of the muon energy spectrum  $\murlpm$. 
 The cross section for 
$\s\mu$ pair production is small, and the $W$ pair production process
leads to muon pairs with missing transverse momentum from neutrinos.
Figure~\ref{fig:susy-smuon_background} shows the effect of the $W$ pair
background after appropriate cuts~~\cite{susy-Staszak}.  The figure also
shows that  electron polarization can be used to remove this background.
 The $\mur$ signal is most clearly seen with  
a right-handed polarized electron beam, since the $W^+W^-$
production is strongly reduced in this case.
Observing the signal for $\mul$ is 
difficult with either polarization.   If the model parameters are such 
that  the $\widetilde{\ell}_{R,L}$ is
heavier than the $\chitwoz$, this problem can be avoided by 
studying the decay  $\widetilde{\ell}_{R,L}^\pm$$\rightarrow \ell^\pm + 
\chitwoz$, with the $\chitwoz$ decaying to a lepton pair and $\chionez$. Then,
because of the large lepton multiplicity, there are no important SM
backgrounds~\cite{susy-Martyn:1999tc}.

Another kinematic method for determining the
masses of supersymmetric particles is to  exploit the
correlations between the products of the two decaying sparticles in a given
event~\cite{susy-Feng:1994sd}. This technique is especially useful
in cases where low-$p_T$ backgrounds tend to overwhelm the signal. 
Some experimental analyses have been carried out using this 
method~\cite{susy-Wagner_1,susy-Wagner_2}, and it should receive more
attention.

One can also carry out mass measurements using threshold 
scans~\cite{susy-Martyn:1999tc,susy-Martyn}, though in some cases this
requires 100 fb$^{-1}$ of luminosity per threshold.
The method has the potential to measure masses with 
accuracies of  0.1\%.
The effect of backgrounds from SM processes and  other SUSY signals and the 
effects of beamstrahlung and bremsstrahlung need to be understood to 
determine the systematics limits of this method~\cite{susy-Baer_Tata}. 

A special case of spectrum parameters for which SUSY detection and
mass measurement are especially difficult is that of an
almost-degenerate chargino and neutralino.  This situation can occur in the
Higgsino limit of gaugino-Higgsino mixing, and in 
anomaly-mediated supersymmetry breaking (AMSB).
 A recent analysis~\cite{susy-Gunion:2001fu}
shows how to extract the chargino signal in this limit
using the 
reaction e$^+$e$^-\rightarrow\gamma\chionep\chionem$. In some cases,
in particular, those from AMSB,
the $\chionepm$ has a long enough lifetime that, at the linear collider,
one can see the chargino's 
track in the 
vertex detector before it decays. 
One then observes a stiff track turning into a very
soft track,  which would be a dramatic signal.

Table~\ref{tab:masses} makes clear that it is possible to measure the 
first-generation slepton masses with a precision of about
1\%.  This would allow experiments at linear colliders to probe the underlying
 GUT-scale universality of intra-generation slepton masses, with enough
sensitivity to discriminate the MSUGRA framework from other models (\eg,
gaugino-mediation) where small GUT-scale splittings of sleptons are
expected \cite{susy-Baer:2000cb}.   Another important observation from  
Table~\ref{tab:masses}
 is that
the linear collider measurements of SUSY particles will provide multiple
high-accuracy measurements of the mass of the lightest neutralino $\chionez$.
As we will discuss  in Section~7, this information will directly 
complement supersymmetry measurements at the LHC, since this key parameter
will not be well  determined there.

\subsection{Measurement of supersymmetry parameters}

\begin{figure}[hbt]
\begin{center}
\epsfig{file=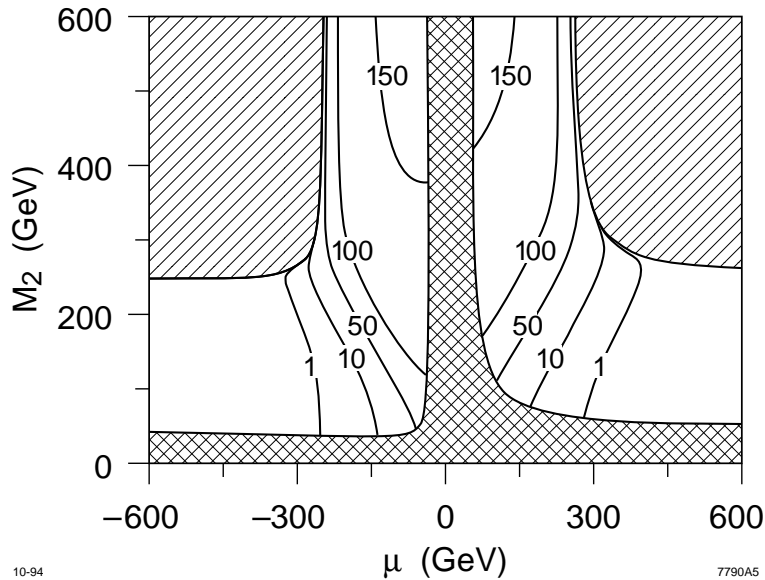, width=4.0in}
\caption{The dependence of the chargino production cross section 
$\sigma(e_R^- e^+ \to \ch1 \chm1)$, in fb, 
on M$_2$ and $\mu$~\cite{susy-Feng:1995zd}.  The value 
 $\tan\beta =4$ is used for this plot,
 but the result is only weakly 
dependent on this parameter.
\label{fig:susy-fengch}}
\end{center}
\end{figure}

Once superpartners are identified and their masses are measured, it is
important to convert the mass and cross section information into 
determinations of the parameters of the SUSY theory.  For the example of 
the MSSM with R-parity conservation, studies have been done to
determine how well one can measure the fundamental parameters. By studying the
production and subsequent decays of $\chionepm$ and $\chitwopm$, the masses and
the gaugino-higgsino mixing angles of these states can be measured and hence 
the values of the MSSM parameters $M_2$, $\mu$, and tan$\beta$ can be 
determined to about  
1\% 
accuracy~\cite{susy-Feng:1995zd,susy-Choi:2000ta,susy-Moortgat-Pick:2000uz}. 
This is illustrated in Fig.~\ref{fig:susy-fengch},  where it is shown that
the value of 
the chargino production cross section from a right-handed polarized
beam allows one to map out whether the lighter chargino is mainly gaugino
or Higgsino. A measurement of both the cross section 
 and the angular distribution allows one to
measure all of  the terms in the chargino 
mass matrix.  It should be noted that the figure shows the tree-level
cross section.  A true determination of parameters to 1\% accuracy should
take account of electroweak and SUSY radiative corrections.

Another method for determining whether the lightest neutralinos and 
chargino are mostly gaugino or Higgsino is to study
slepton pair production with  left-handed and right-handed beam
 polarization.  This is done by measuring the magnitude of the cross section 
and the shape of the production angular distribution~\cite{susy-Tsukamoto}. 
Similarly, measuring the 
cross sections of $\topone$, $\toptwo$, $\tauonepm$ and $\tautwopm$ 
 and $\overline{\nu}_\tau$ with polarized beams allows 
one to determine their mixing 
angles~\cite{susy-Nojiri:1996fp,susy-Bartl:1997yi,susy-Baer:2001hx}. 
Additional measurements associated with polarization in $\s \tau$
reactions are discussed in \cite{susy-Tsukamoto,susy-Nojiri:1996cx}.

By looking at the angular distributions of supersymmetric particles that have 
a $t$-channel exchange involving another supersymmetric particle, 
the mass of the exchanged particle can be 
determined. Similarly, if the decays of the charginos have three-body 
decays because the two-body decay to $W \neu1$ is not allowed kinematically,
decays via $W^*$ can interfere with decays involving a virtual slepton
or squark. This could give useful indirect signals for these particles
in the 
cases where they cannot be produced because they are too 
heavy~\cite{susy-Hikasa:1998qa}.

We should recall that the parameter $\tan \beta$ can be 
determined not only from supersymmetry reactions but also by direct
experimental studies of the extended Higgs sector.  For $\tan\beta < 30$,
one can obtain an accurate value of this parameter by measuring the 
branching ratios for the various possible decays 
of the SUSY Higgs particles: $H^-$ into $\tau\nu$, 
$b \bar t$, and $W^-h$, and  $A^0$  and $H^0$
into $\tau^+ \tau^-$, $b\bar b$, $t\bar t$, 
and $Zh$~\cite{susy-Feng:1997xv,susy-Barger:2000fi}. If the Higgs sector is 
heavy enough, one can include decays into 
lighter supersymmetric particles.  These can
 provide quite sensitive measurements in the high-tan$\beta$ region~\cite{susy-Barger:2000fi}.

Finally, it is important to verify the spin of each supersymmetric partner
experimentally.  This can be done at a linear collider, because the 
simplicity of the production reactions often makes the spin obvious 
from the angular distributions.  For example, the 
$\mur$ signal in Fig.~\ref{fig:susy-smuon_background} exhibits a
 $\sin^2\theta$  distribution that 
is a clear indication that the spin of the $\mur$ is 0.
The spin of supersymmetric particles can also  be 
 determined by measuring the pair-production 
cross section near threshold, which rises as $\beta$
and  $\beta^3$, where $\beta$ is the particle velocity, 
for spin-$\half$ and spin-0 particles, respectively. 

\section{Tests of supersymmetry}

If new particles are discovered with quantum numbers expected
in supersymmetry, it is desirable to
determine whether they are in fact superparticles.  Linear colliders can
verify supersymmetry through highly model-indepedent tests 
accurate at the
percent level.  In addition, since these tests are sensitive to
loop-level effects, they may yield a wealth of additional information.

Supersymmetry may be tested in many ways.  For example, confirmation
that some of the newly discovered particles are scalars, as discussed
at the end of Section 3, constitutes an important, if weak, test
of supersymmetry.  More quantitatively, verification of the
consistency of direct discoveries with the expected indirect supersymmetric
effects in SM processes, as discussed in
Chapter 8, Section 3, also provides a test of supersymmetric
interpretations of new physics. Measurements of the mass differences
between scalar partners in the same SU(2) doublet may also provide quantitative
and rather model-independent checks.

In this section we focus on investigations of supersymmetric coupling
relations, which are among the most incisive and model-independent
tests.   In addition to providing precise
quantitative confirmation of supersymmetry, such tests may also shed
light on otherwise inaccessible superpartners, much as current
precision electroweak measurements bound the Higgs boson mass and
constrain new physics.
 
\subsection{Confirming supersymmetry}

If supersymmetry were an exact symmetry of nature, the properties of
supersymmetric particles would be completely determined by the
properties of their SM partners.  Of course, relations
between masses are broken by soft supersymmetry breaking parameters.
However, supersymmetry also predicts the equivalence of {\em
dimensionless\/} couplings.  For example, supersymmetry implies
\begin{equation}
\label{hard}
g_i=h_i \ , 
\end{equation}
where $g_i$ are the SM gauge couplings, $h_i$ are their
supersymmetric analogues, the gaugino-fermion-sfermion couplings, and
the subscript $i=1,2,3$ refers to the U(1), SU(2), and SU(3) gauge
groups, respectively.  These identities are not broken by soft
supersymmetry-breaking parameters at tree level and are therefore
known as ``hard supersymmetry relations''~\cite{susy-Hikasa:1996bw}.
They are valid in all supersymmetric theories, in contrast to other
predictions such as the universality of scalar or gaugino
masses. Hard supersymmetry relations therefore provide, in principle, a
model-independent method of quantitatively confirming that
newly-discovered particles are indeed
superpartners~\cite{susy-Feng:1995zd,susy-Nojiri:1996fp}.

\subsection{Super-oblique corrections}

At the loop-level, however, even hard supersymmetry relations receive
corrections that would vanish in the supersymmetric
limit~\cite{susy-Chankowski:1990du}.  These corrections are analogous
to the oblique corrections~\cite{susy-Peskin:1990zt} of the Standard
Model.  In the SM,  SU(2) multiplets with custodial
SU(2)-breaking masses, such as the $(t,b)$ multiplet, induce
splittings in the couplings of the $(W,Z)$ vector multiplet at the
quantum level. Similarly, in supersymmetric models, supermultiplets
with soft supersymmetry-breaking masses, such as the $(\tilde{f},f)$
supermultiplets, induce splittings in the couplings of the $({\rm
gauge\ boson}, {\rm gaugino})$ vector supermultiplet at the quantum
level.  This analogy can be made very
precise~\cite{susy-Cheng:1997sq,susy-Cheng:1998vy,%
susy-Nojiri:1998ma,susy-Katz:1998br}.  Corrections to hard
supersymmetry relations are therefore called `super-oblique
corrections', and the splittings are typically written in terms of
`super-oblique parameters.'

If some scalar superpartners $\tilde{f}$ have masses at a high scale
$M$, and all others are light with mass $m \sim M_{\rm weak}$, the
super-oblique parameters are given by
\begin{equation}
\s{U}_{i} \equiv \frac{h_{i}(m)}{g_{i}(m)} - 1 
\approx \frac{g_{i}^{2}(m)}{16\pi^{2}}\Delta b_i
\ln \frac{M}{m} \ ,
\label{delta}
\end{equation}
where $\Delta b_i$ is the one-loop $\beta$-function coefficient
contribution from all light particles whose superpartners are
heavy. Equation~(\ref{delta}) is the leading logarithm contribution to
$\s{U}_i$. The super-oblique parameters for some representative models
are given in Table~\ref{table:t1}.  The super-oblique parameters may
also receive contributions from split exotic supermultiplets, such as
the messengers of
gauge-mediation~\cite{susy-Cheng:1997sq,susy-Katz:1998br}.

\begin{table}[t]
\begin{center}
\begin{tabular}{cccc}\hline\hline
   & $\s{U}_1$ & $\s{U}_2$ & $\s{U}_3$ \\
\hline 
2--1 Models \rule[0.5mm]{0mm}{4mm} &  
$0.35\% \times \ln (M/m)$ & 
$0.71\% \times \ln (M/m)$ & 
$2.5\%  \times \ln (M/m)$ \\
Heavy QCD Models & 
$0.29\% \times \ln (M/m)$ & 
$0.80\% \times \ln (M/m)$ & 
--- \\
\hline\hline
\end{tabular}
\end{center}
\caption{The super-oblique parameters $\s{U}_{i}$ in two 
representative models: `2--1 Models,' with all first and second
generation sfermions at the heavy scale $M$, and `Heavy QCD Models,'
with all squarks and gluinos at the heavy scale.\label{table:t1}}
\end{table}

  From Eq.~(\ref{delta}) we see that, although super-oblique
parameters are one-loop effects, they may be greatly enhanced if many
states are heavy (large $\Delta b_i$).  They also grow logarithmically
with $M/m$: super-oblique parameters are {\em non-decoupling}, and so
are sensitive to particles with arbitrarily high mass.  As noted in
Section 2, the squarks and sleptons of the first and
second generations are only loosely bounded by fine-tuning arguments.
They may have masses far beyond the reach of the LHC, and in fact,
such massive squarks and sleptons considerably ameliorate many
supersymmetric flavor and CP problems.  In these cases, the super-oblique parameters 
are large and provide a
rare window on these heavy scalars. 

\subsection{Measurements at linear colliders}

With respect to super-oblique parameters, the program at a
linear collider consists of two parts: First, one would like to verify
as many hard supersymmetry relations as possible to
determine that newly-discovered particles are in fact
superparticles. Second, if new particles are determined to be 
supersymmetric, small
violations of hard supersymmetry relations may provide the first
evidence for as-yet-undiscovered superparticles.  Precise measurements
of the super-oblique parameters may constrain the mass scales of these
superparticles.
 
The experimental observables that are dependent on super-oblique
parameters have been exhaustively categorized in
\cite{susy-Cheng:1998vy} for both lepton and hadron colliders.
The most promising observables at colliders are cross sections and
branching ratios involving gauginos, and several of these
possibilities have been examined in detailed studies.  The potential
of linear colliders is, of course, highly dependent on the
supersymmetry scenario realized in nature, but we present a brief
synopsis below.

To date, all studies have used tree-level formulae in which the gaugino
couplings are allowed to vary.  Constraints on these gaugino couplings are
then interpreted as measurements of super-oblique parameters. At the level
of precision required, however, it will ultimately be necessary 
to make  a detailed
comparison of cross sections and other observables with full one-loop
predictions.  In chargino pair production, for example, studies of
triangle~\cite{susy-Kiyoura:1998yt,susy-Diaz:1998kv,susy-Diaz:2000hi} and
box~\cite{susy-Blank:2000uc} contributions have been shown to be
important. In addition, beam polarization may enhance the effect of
quantum corrections~\cite{susy-Diaz:1998kv}.  To extract the
non-decoupling effects of very heavy superpartners, one must therefore
control many other effects, including all other virtual effects, either by
including data from direct detection, or by verifying that such effects
are sufficiently suppressed to be negligible.  The study of super-oblique
parameters should be viewed as the first step in the complete program of
one-loop SUSY studies that will be possible at a linear collider.
 
Potential super-oblique parameter measurements at a linear collider should include:
\begin{itemize}
\item \noindent {\em Measurements of $\s{U}_1$.} Selectron pair
production at electron colliders includes a contribution from
$t$-channel gaugino exchange.  In particular, in the reaction $e^+ e^-\to\tilde{e}_R^+ 
\tilde{e}_R^-$, its dependence upon the $\tilde{B}e\tilde{e}$ coupling
$h_1$ has been studied in~\cite{susy-Nojiri:1996fp}. Under the assumption 
that the selectrons decay through $\tilde{e} \to e \tilde{B}$, the selectron
and gaugino masses may be measured through kinematic endpoints.
Combining this information with measurements of the differential cross
section, $\s{U}_1$ may be determined to $\sim 1\%$ with $20~\ifb$ of
data at $\sqrt{s} = 500~\gev$.

This high-precision measurement may be further improved by considering
the process $e^- e^- \to \tilde{e}_R^- \tilde{e}_R^-$.  This process
is made possible by the Majorana nature of gauginos.  Relative to the
$e^+ e^-$ process, this reaction benefits from large statistics for
typical supersymmetry parameters and extremely low backgrounds,
especially if the electron beams are right-polarized.  Depending on
experimental systematic errors, determinations of $\s{U}_1$ at the
level of $0.3\%$ may be possible with integrated luminosities of
$50~\ifb$~\cite{susy-Cheng:1998vy}.

\item \noindent {\em Measurements of $\s{U}_2$.} Chargino pair
production has a dependence on $\s{U}_2$ at lepton colliders through
the $\tilde{\nu}$ exchange amplitude.  This process was first studied
as a way to verify hard supersymmetry
relations~\cite{susy-Feng:1995zd}.  In  \cite{susy-Cheng:1998vy},
estimates of 2--3\% uncertainties for $\s{U}_2$ were obtained from
pair production of 172 GeV charginos with $\sqrt{s} = 400$--500 GeV.
These results are conservative, and are improved in most other regions
of parameter space~\cite{susy-Kiyoura:1998yt}.  Dramatic improvements
may also be possible if both charginos are within kinematic reach and
large luminosities with polarized beams are available, a scenario
studied in~\cite{Choi:2000hb}.

The process $e^+ e^- \to \tilde{\nu}_e \overline{\tilde{\nu}_e}$ also depends
on $\s{U}_2$ through the $t$-channel chargino exchange amplitude.
With a data sample of $100~\ifb$, $\s{U}_2$ may be determined to
$\sim 0.6\%$~\cite{susy-Nojiri:1998ma}.

\item \noindent {\em Measurements of $\s{U}_3$.} The strong
super-oblique parameter may be measured through processes involving
squarks.  The squark pair-production cross sections at lepton colliders
are independent of super-oblique corrections, but the three-body
production processes, such as $\tilde{t}t \tilde{g}$ and $\tilde{b}b
\tilde{g}$,  have been suggested as a
probe~\cite{susy-Cheng:1998vy,susy-Katz:1998br}.

Squark branching ratios are also sensitive to super-oblique
corrections if there are two or more competing
modes~\cite{susy-Hikasa:1996bw}.  In~\cite{susy-Cheng:1998vy},
parameters were studied in which the two decays $\tilde{b}_L \to b
\tilde{g}$ and $\tilde{b}_L \to b \tilde{W}$ were open.  For
parameters where the gluino decay is suppressed by phase space, these
modes may be competitive, and measurements of the branching ratios
yield constraints on $\s{U}_3$.  For example, for $m_{\tilde{b}_L} =
300~\gev$, $\tilde{b}_L$ pair production at a $\sqrt{s} = 1~\tev$
collider with integrated luminosity $200~\ifb$ yields measurements of
$\s{U}_3$ at or below the 5\% level for $10~\gev \lsim m_{\tilde{b}_L}
- m_{\tilde{g}} \lsim 100~\gev$. These measurements are typically
numerically less stringent than those discussed above, but the SU(3)
super-oblique correction is also larger by a factor $\alpha_s/\alpha_w$.

\item \noindent {\em Measurements of Wino-Higgsino mixing.}  The
presence of the $W$ boson mass in the tree-level chargino mixing
matrix is also a consequence of supersymmetry (relating the $WWh$ and
$\tilde{W}\tilde{h}h$ couplings).  Wino-Higgsino mixing receives
non-decoupling corrections, and may be constrained through chargino
pair production~\cite{susy-Feng:1995zd,susy-Kiyoura:1998yt}.

\item \noindent {\em Measurements of trilinear gaugino/gauge boson
couplings.}  Finally, the supersymmetric equivalence of triple gauge
boson and gaugino couplings may also be broken.  In
\cite{susy-Mahanta:1999hx}, splittings of the $WW\gamma$ and
$W\tilde{W}\tilde{\gamma}$ couplings were calculated and found to be 
present at the few-percent level.
Such splittings could be probed in $\tilde{W} \to W
\tilde{\gamma}$ decays.

\end{itemize}

These studies demonstrate the promise of linear colliders for
loop-level studies of supersymmetry.  If charginos or sleptons are
produced at linear colliders, precision tests will be able to verify
that their couplings are as predicted by supersymmetry to the percent
level.  In addition, small corrections to these relations are
sensitive to arbitrarily heavy superpartners, and, if some
superpartners are kinematically inaccessible, precise determination of
the super-oblique parameters may provide a target mass range for
future searches.

\section{Symmetry violating phenomena}

\subsection{R-parity violation}

Up to this point we have considered only R-parity (R$_{\rm{p}}$)-conserving
supersymmetric theories.  R$_{\rm{p}}$ is a multiplicative
discrete symmetry~\cite{susy-Dreiner:1997uz,Bhattacharyya:1997vv,Bhattacharyya:1997nj,Bisset:1998bt} defined for each particle to be
\begin{equation}
{\rm  R}_{\rm{p}}=(-1)^{3B+L+2S}
\end{equation}
where $B$ is baryon number, $L$ is lepton number, and $S$ is the particle's 
spin. This symmetry is not automatic in the MSSM as it is in the SM. We now
consider the possibility that the symmetry is not 
respected~\cite{susy-Hall:1984id}.  

Without R$_{\rm{p}}$ conservation, the most
general gauge-invariant and Lorentz-invariant superpotential is
\begin{eqnarray}
W  & = & \mu H_u H_d +y^e_{ij} H_d L_j e^c_k 
        + y^d_{jk} H_d Q_j d^c_k + y^u_{jk}H_u Q_j u^c_k \nonumber \\[1ex]
   & &  +\lambda_{ijk}L_iL_je^c_k+\lambda'_{ijk}L_iQ_j d^c_k 
        +\lambda''_{ijk}u^c_id^c_jd^c_k + \mu_i H_u L_i . 
\end{eqnarray}
The $\lambda$- and $\lambda'$- terms do not respect lepton number
and the $\lambda''$-terms do not respect baryon number. Proton decay
is unacceptably rapid if all terms are allowed without extreme suppressions;
this requires  $\lambda' \lambda'' \lsim 10^{-36}$. But, since
proton decay requires both lepton and baryon number
violation, it is possible to escape this constraint by forbidding
{\it one or the other} of lepton number violation or baryon number violation.
That is, the constraint on  $\lambda' \lambda''$ can be accomodated
by setting  $\lambda'=0$ (lepton number conservation) or 
$\lambda''=0$ (baryon number conservation). The $\mu_i$ terms also 
violate lepton 
number conservation, although these terms can be defined away at tree level.

In the next few paragraphs, we will describe the signals expected
at a $500\,\gev$ linear collider for a theory with non-zero $\lambda$ 
as the only R$_{\rm{p}}$-violating couplings.  We will then reanalyze the
same theory but this time with only non-zero $\lambda'$ couplings, and
finally with only non-zero $\lambda''$ couplings.  We further assume that the 
R$_{\rm{p}}$-violating couplings are too weak to participate in observables
in any way except to allow the lightest neutralino to decay promptly
in the detector.  Making the couplings stronger usually implies even
more phenomena by which to discover supersymmetry (additional production
modes via R$_{\rm{p}}$ violation). Making the couplings very weak will
cause the phenomenology to asymptotically approach that of the MSSM with
R$_{\rm{p}}$ conservation.

When applicable, we will illustrate 
phenomena with model E of~\cite{susy-Ghosh:1999ix},
which is the heaviest superpartner model considered in this paper. This
model assumes $M_2=2M_1=200\,\gev$, $\mu =-250\,\gev$, $\tan\beta =20$,
and $m_{\tilde e_L}=m_{\tilde e_R}=200\,\gev$.  The chargino masses
are then $173.4$ and $292.1\,\gev$, and the neutralino
masses are $97.7, 173.6, 260.8,$ and $290.1\,\gev$.

\subsubsection{$\lambda_{LLe^c} \neq 0$}

In these theories the LSP always decays into two charged leptons and
a neutrino (missing energy): 
\begin{equation}
\tilde{\chi}^0_1\to \ell^+ + \ell^- +\slashchar{E}.
\end{equation}
When superpartners are produced in pairs, they will cascade-decay
down to two LSPs (plus SM jets or leptons), and the LSPs will then decay
into two leptons plus missing energy.  Therefore, the signal always
includes at least four leptons plus missing energy, and quite often
contains more leptons and additional jets from the cascades.  This
is a spectacular signature that will not go unnoticed.  For example,
the cross section for the $4l+\slashchar{E_T}$ signature for our considered 
example model is approximately $274\xfb$, much higher than the expected 
$0.4\xfb$ background rate~\cite{susy-Ghosh:1999ix}.

\subsubsection{$\lambda'_{\, LQd^c}\neq 0$}

In these theories the LSP always decays into two jets with an
accompanying charged lepton or neutrino:
\begin{equation}
\tilde \chi^0_1\to l^\pm q\bar q'~~~{\rm or}~~~\nu q\bar q.
\end{equation}
All supersymmetry signals must pass through $\tilde \chi^0_1
\tilde \chi^0_1+X_{\rm SM}$, where $X_{\rm SM}$ represents SM
states (jets, leptons, or neutrinos) arising from the cascade decays
of the produced parent superpartners.  In this case the final-state
signatures of all superpartner production processes will be
\begin{equation}
({\rm 0, 1,\, or\, 2~leptons})+{\rm 4\, jets}+X_{SM}.
\end{equation}
Furthermore, all events that do not have 2 leptons will have
some missing energy in them from escaping neutrinos.

Many of the signal events of this type of $R_p$ violation will
be swamped by backgrounds. The two most promising modes to search
are $3l$ and $4l$ final states, where at least one additional
lepton comes from the cascade products in $X_{\rm SM}$.  Another
intriguing possibility is to search for like-sign dilepton
events.  This signature is made possible by each
independent $\tilde \chi^0_1$ decaying into
a lepton of either positive or negative charge.  Approximately one-eighth of
the $\tilde \chi^0_1\tilde \chi^0_1$ decays end in like-sign
dileptons.  The background in this case 
is very small whether $X_{\rm SM}$ contains leptons or not.
Furthermore, it appears that the LSP mass may be obtainable by
analyzing the invariant mass distribution of the hardest lepton combined
with all hadronic jets in the same hemisphere~\cite{susy-Ghosh:1999ix}.

\subsubsection{$\lambda''_{\, u^cd^cd^c}\neq 0$}

In these theories the LSP always decays into three jets:
\begin{equation}
\tilde \chi^0_1\to q'q\bar q.
\end{equation}
All supersymmetry events will then have at least six jets from LSP decays 
in the final state plus the cascade decay products of the parent sparticles.
Although jet reconstruction algorithms will generally not resolve all
six jets, they will usually register at least three in the 
event~\cite{susy-Ghosh:1997wk}.

Perhaps the most important signature for discovery in these theories 
comes from chargino pair production, where each chargino decays
as $\tilde \chi^\pm_1\to l^\pm\nu \tilde\chi^0_1$.  The final state
will then be 2 leptons plus many jets.  Unfortunately the lepton often
finds itself inside one of the many hadronic jets and fails
the isolation requirements.  Nevertheless, the rate is sufficiently 
large that it is a viable signal for our example model.  According
to~\cite{susy-Ghosh:1999ix}, the signal in this mode---including also 
the smaller contribution from $\tilde \chi^0_i\tilde \chi^0_j$
production---is approximately
$40\xfb$ compared to a background of $243\xfb$.  A moderate luminosity
of $10\xfb^{-1}$ would produce a $S/\sqrt{B}$ significance greater than  8.

To determine the LSP mass, one can use strategies similar to ALEPH's four-jet 
analysis~\cite{susy-Barate:1998tz}
to combine jets within same hemispheres to look for matching invariant
mass peaks.  Careful comparisons with background have not yet been performed
to see how accurately the LSP mass can be extracted with this technique.

\subsubsection{$\mu_i \neq 0$}

The parameter space with just $\mu_i\neq 0$ is often called
Bilinear R-Parity Violation (BR$_{\rm{p}}$V).  It has special
theoretical motivations in 
supersymmetry~\cite{susy-deCampos:1995av,susy-Banks:1995by,susy-Smirnov:1996ey,susy-Nilles:1997ij}.
One interesting
phenomenological feature of the model is its ability to
predict the three neutrino 
masses and the three mixing angles by adding to the MSSM only one 
or two extra parameters. This is done in a SUGRA context with radiative 
electroweak symmetry breaking and universality of soft parameters 
at the GUT scale~\cite{susy-Diaz:1998xc}. 
At tree level, one neutrino acquires a mass from neutrino-neutralino mixing. 
The masslessness and degeneracy 
of the other two neutrinos is lifted at one loop, giving masses
and mixings that account for  the solar 
and atmospheric neutrino 
anomalies~\cite{susy-Hempfling:1996wj,susy-Chun:2000bq,susy-Hirsch:2000ef,susy-Romao:2000up}.  The parameters of the model can be measured from the leptonic
branching fractions of the lightest 
neutralino~\cite{susy-Romao:2000up,susy-Porod:2000hv}. Thus, in this model,
crucial information needed to understand neutrino physics comes from 
experiments at the linear collider.

\subsection{Lepton flavor violation}

A linear collider enables the careful study of flavor physics
in supersymmetry.  With the apparent confirmation of neutrino
masses, non-trivial lepton-slepton flavor angles are assumed
to exist.  There are constraints on the magnitude of these angles
from $B(\mu \to e\gamma)$ bounds, for example.  However, the constraints
are weaker if the sleptons are nearly degenerate in mass.  We will make this
assumption here, thereby
invoking a super-GIM suppression to suppress the radiative
flavor-violating lepton decays.

\begin{figure}[t]
\begin{center}
\epsfig{file=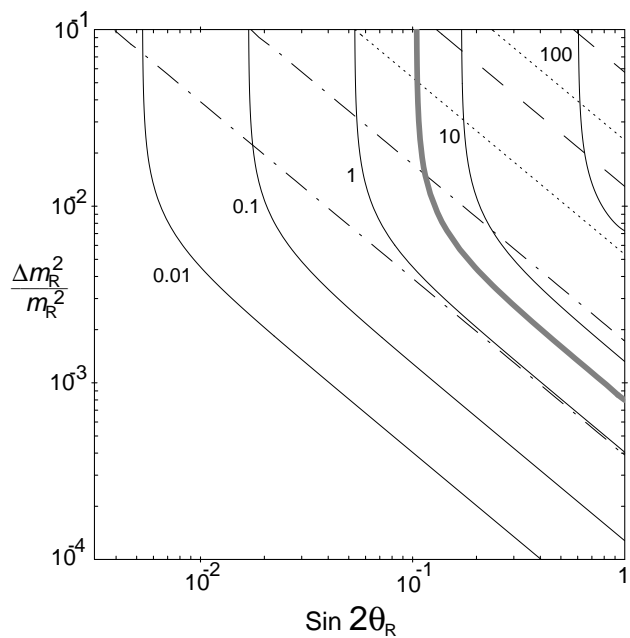, width=3.5in}
\caption{Contours 
of constant $\sigma(\ee \to e^\pm \mu^\mp\tilde\chi^0_1\tilde\chi^0_1)$ in fb
at a $\sqrt{s}=500$ GeV $e^+e^-_R$ collider.  The signal arises
from right-slepton production and subsequent decay to lepton plus lightest
neutralino.  The $\s \ell_R$ masses are approximately 200 GeV and the
lightest neutralino is a Bino with mass 100 GeV.  
The thick gray contour represents
optimal experimental reach with $50\, {\rm fb}^{-1}$ integrated luminosity.
The straight lines (dotted and dashed) represent contours of 
constant $B(\mu\to e\gamma)$. These depend on additional parameters such
as the $\s \ell_L$ mass and the off-diagonal entries of the slepton 
mass matrix. See~\cite{susy-Arkani-Hamed:1996au} for more details.}
\label{fig:susy-acfh}
\end{center}
\end{figure}

Direct production of sleptons and close scrutiny of their decays
allow probing of these flavor angles at
more sensitive 
levels~\cite{susy-Arkani-Hamed:1996au,susy-Arkani-Hamed:1997km,susy-Hisano:1999wn,susy-Cao:1999kh,susy-Nomura:2000zb,susy-Guchait:2001us}. 
The nearly degenerate sleptons will undergo flavor oscillation after being
produced and then decay quickly. Analogous to neutrino oscillations, the
detectability of slepton oscillations is best characterized in the
$(\sin 2\theta ,\, \Delta m^2)$ plane, where $\theta$ is the angle between
the weak eigenstates $|\tilde e\rangle$, $|\tilde \mu \rangle$ and
the mass eigenstates $|1\rangle$, $|2\rangle$:
\begin{eqnarray}
|\tilde e\rangle = +\cos\theta |1\rangle + \sin\theta |2\rangle \\ \nonumber
|\tilde \mu\rangle = -\sin\theta |1\rangle + \cos\theta |2\rangle .
\end{eqnarray}

Figure~\ref{fig:susy-acfh} shows contours 
of constant $\sigma(\ee\to e^\pm \mu^\mp\tilde\chi^0_1\tilde\chi^0_1)$, in fb,
at a $\sqrt{s}=500$ GeV collider with $e^+e^-_R$ collisions.  The signal arises
from $\s\ell_R$ production and subsequent decay to a lepton plus the lightest
neutralino.  The $\s\ell_R$ masses are approximately 200 GeV, and the
lightest neutralino is a Bino with mass 100 GeV. 
From this figure we can see that careful measurement
of the cross section enables probing of flavor-violating couplings to
very small mass splitting and mixing angle.

\subsection{CP violation}

The new mass parameters associated with supersymmetry may not all be real,
and could lead to CP violation effects~\cite{susy-Dugan:1985qf} at high-energy 
colliders. The parameters $\mu$, $M_1$ and $M_2$ can in general be complex.
By rotating the phases of the gauginos we are free to choose 
$M_2$ real, leaving us with
\begin{equation}
\mu = |\mu |e^{i\phi_\mu} ~~~~{\rm and}~~~~ M_1\to |M_1| e^{i\phi_1}.
\end{equation}
In addition to these phases, each of the tri-scalar $A$ terms connecting
the Higgs bosons with left and right scalar superpartners of the fermions
can in principle have its own independent phase.

Generic ${\cal O}(1)$ phases associated with superpartner masses near
the weak scale are ruled out by the electric dipole moments (EDMs) of
the neutron and electron if superpartners are light enough to be 
accessible at a 1 TeV linear collider.
 Therefore,  we assume here that the phases
must be small, ${\cal O}(0.1)$.   We remark that 
tuned cancellations~\cite{susy-Ibrahim:1998je,susy-Brhlik:1999zn} 
may allow ${\cal O}(1)$ phases for light superpartners,
thereby leading to effects much  larger than the estimates given below.

Supersymmetric CP-violating phases have two important effects:  they disrupt
the relations among CP-conserving observables, and they give birth to
non-zero CP-violating observables.  Much work has gone into both types
of analyses.  For example, CP-violating observables in $e^+e^-\to t\bar t$
may be the most promising way to find actual CP violation effects at
the linear collider.  We refer the reader to 
\cite{susy-Atwood:2000tu,susy-Grzadkowski:1993kb,susy-Bartl:1997iq}
for a comprehensive review of this subject, and a description of the
challenges facing experiment to confirm CP-violating effects. Here, we
briefly focus on the effects that small phases have on CP-conserving
observables.

Recently several groups have shown how CP-violating phases affect almost
all interesting MSSM observables at a linear 
collider~\cite{susy-Choi:1999ei,susy-Kneur:2000nx,susy-Barger:2000tn,susy-Choi:2000ta,susy-Barger:2001nu}.  For example,
the chargino mass eigenstates depend non-trivially on the phase of $\mu$:
\begin{equation}
m^2_{\tilde{\chi}^\pm_{1,2}} = \frac{1}{2}\left[ M_2^2+|\mu |^2
   +2m^2_W\mp \Delta_C\right],  
\end{equation}
where
\begin{eqnarray}
\Delta_C &= & \left[(M^2_2-|\mu |^2)^2+4m_W^4\cos^22\beta
       +4m^2_W(M_2^2+|\mu |^2)\right. \nonumber \\
  & &     +\left. 8m_W^2M_2|\mu | \sin 2\beta \cos \Phi_\mu\right]^{1/2}.
\end{eqnarray}

\begin{figure}[hbtp]
\begin{center}
\epsfig{file=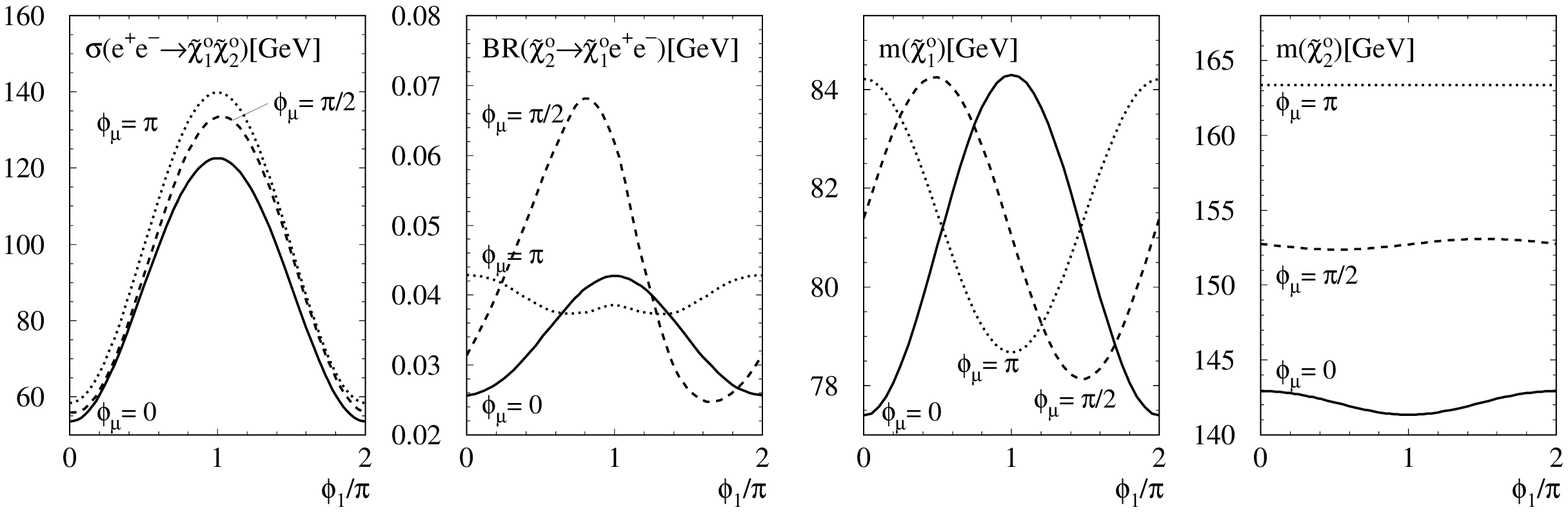, height=1.8in}
\caption{The effects on supersymmetry
observables obtained by varying the phases $\phi_1$ and
$\phi_\mu$ in the example model discussed 
in the text~\cite{susy-Barger:2001nu}.}
\label{fig:susy-cxn_mass}
\end{center}
\vskip.1in
\begin{center}
\epsfig{file=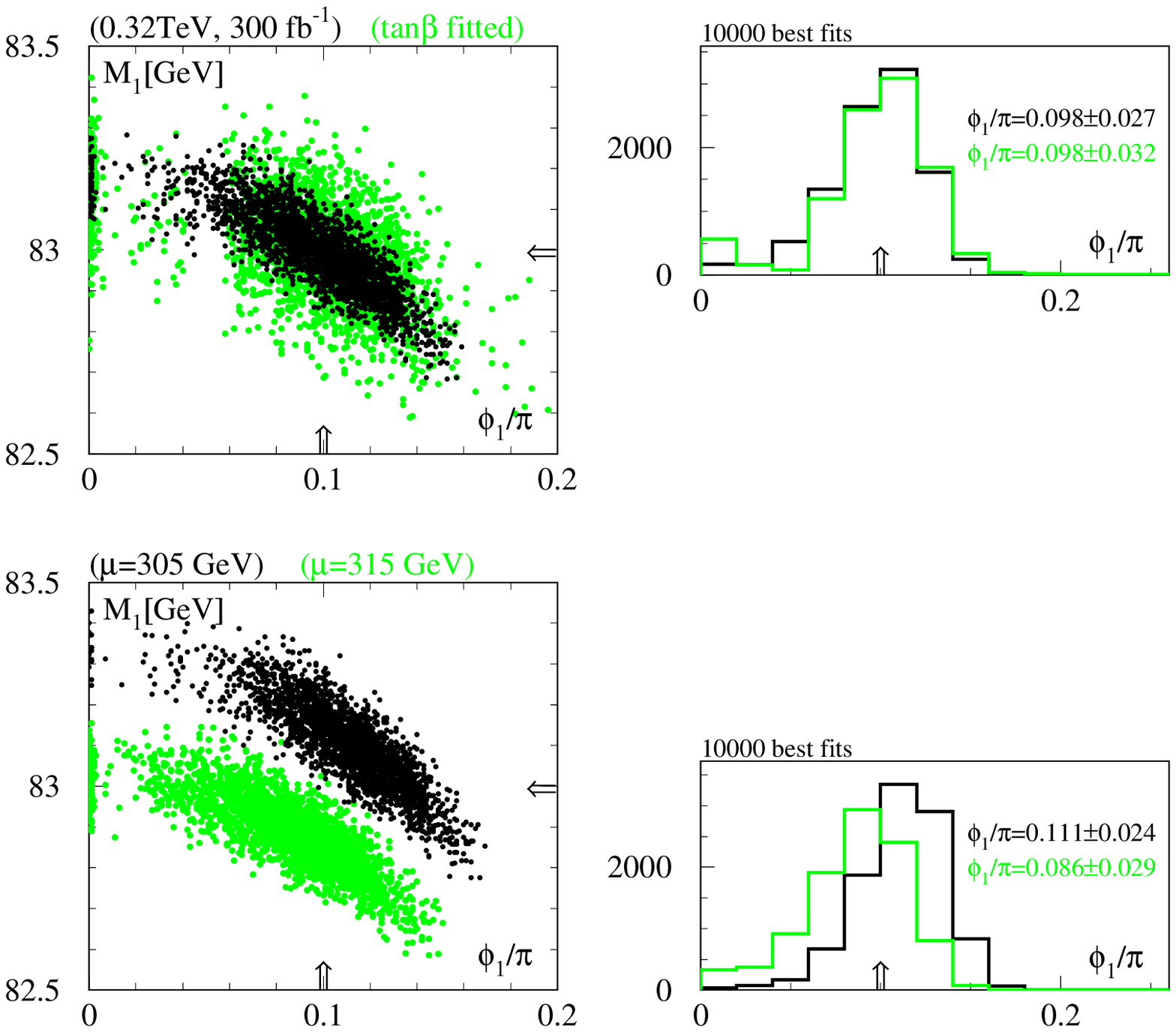, height=4.0in}
\caption{Demonstration of the 
interdependence of parameters in the extraction of CP-violating phases
 from  linear collider SUSY
observables~\cite{susy-Barger:2001nu}.}
\label{fig:susy-cpextract}
\end{center}
\end{figure}

The effects of phases on observables have been illustrated 
in~\cite{susy-Barger:2001nu} with a reference model corresponding to
an mSUGRA point with $m_{1/2}=200~\gev$, $m_0=100~\gev$, $A_0=0$,
$\tan\beta=4$, and $\mu >0$.  This parameter choice corresponds 
to the mass values
$|M_1|=83~\gev$, $M_2=165~\gev$, $\mu =310~\gev$, $m_{\tilde e_L}=180~\gev$,
$m_{\tilde \nu}=166~\gev$, and $m_{\tilde e_R}=132~\gev$.  
In Fig.~\ref{fig:susy-cxn_mass}, the effects of
 varying the phases
$\phi_1$ and $\phi_\mu$ are demonstrated for several observables.

Motivated by the EDM constraints  on the phases
of supersymmetric mass parameters, the authors of~\cite{susy-Barger:2001nu}
set $\phi_\mu=0$ and simulated how evidence for  a small but
non-zero $\phi_1$
phase would be extracted at a linear collider.  They generated 10000
data sets, smeared with respect to the true values by experimental 
resolution.  The input data included three cross sections
($\tilde\chi^0_1\tilde\chi^0_2$,  $\tilde\chi^0_2\tilde\chi^0_2$, and
$\tilde\chi^\pm_1\tilde\chi^\mp_1$) and three masses ($m_{\tilde\chi^0_1}$,
$m_{\tilde\chi^0_2}$, and $m_{\tilde\chi^\pm_1}$).
Figure~\ref{fig:susy-cpextract} demonstrates
the extraction of several different parameters, and their interdependence.
For example, the bottom figures show  the systematic error one
would encounter by having a wrong input for $|\mu |$ given a known
$\tan\beta$.  Perhaps the most interesting conclusion one can draw
from this exercise is that
$\phi_1=0$ is strongly disfavored, incicating that the linear collider
measurements of CP-conserving observables can give a strong signal for
nonzero CP-violating phases if they are present.

\section{Supersymmetry and $e^-e^-$, $e^-\gamma$, and $\gamma \gamma$ 
colliders}

\subsection{Supersymmetry and $e^-e^-$ colliders}
\label{sec:ee}

The features of $e^-e^-$ colliders are reviewed in Chapter 14.
The unique quantum numbers of the $e^-e^-$ initial state forbid
the production of most superpartners.
However, slepton pair production through $t$-channel neutralino
exchange is always possible~\cite{susy-Keung:1983nq}.
The opportunities at $e^-e^-$ colliders for measurements of slepton
masses, mixings, and couplings are unparalleled, and exploit many of
the unique properties of $e^-e^-$ colliders.

\subsubsection{Masses}

As reviewed in Section 2, masses at linear colliders
are most accurately determined through kinematic endpoints and
threshold scans. In $e^+e^-$ mode, the threshold cross section for
pair production of identical scalars rises as $\beta^3$, where $\beta$
is the velocity of the produced particles.  Threshold studies for
identical scalars are therefore far less effective than for fermions,
and consequently require large investments of integrated
luminosity~\cite{susy-Martyn:1999tc}.

\begin{figure}[t]
\begin{center}
\epsfig{file=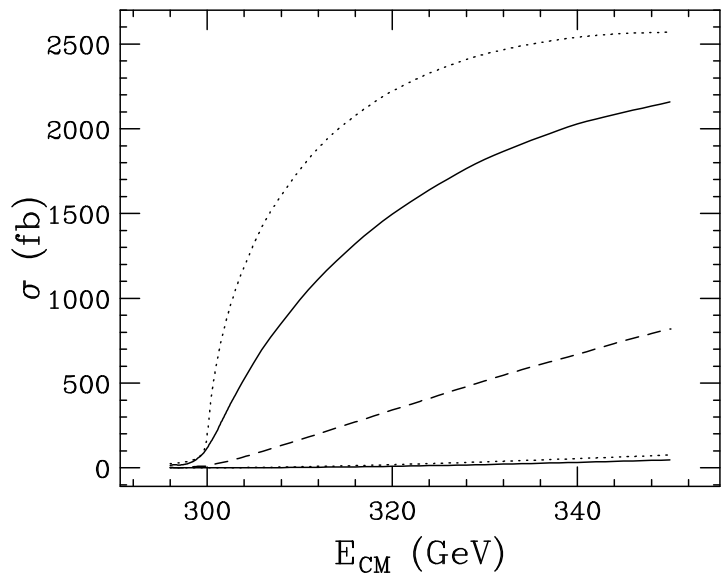,width=0.62\textwidth}
\end{center}
\caption{Threshold behavior for $\sigma(e^-e^- \to \tilde{e}_R^-
\tilde{e}_R^-)$ (upper two contours) and $\sigma(e^+e^- \to \tilde{e}_R^+
\tilde{e}_R^-)$ (lower two contours) for $m_{\tilde{e}_R} = 150 ~{\rm
GeV}$ and $M_1 = 100~{\rm GeV}$~\cite{susy-e-e-inprepFP}.
 In each pair, the dotted curve
neglects all beam effects, and the solid curve includes the initial
state radiation, beamstrahlung, and beam energy spread for flat
beams. Results for $e^-e^-$ round beams (dashed) are also shown.  The
selectron width is included, and beam polarizations $P_{e^-} = 0.8$
and $P_{e^+}=0$ are assumed.
\label{fig:compare}}
\end{figure}

At $e^-e^-$ colliders, however, the same-helicity selectron pair
production cross section has a $\beta$ dependence at
threshold~\cite{susy-Feng:1998ud}. This is easily understood: the
initial state in $e^-_R e^-_R \to
\tilde{e}^-_R \tilde{e}^-_R$ has angular momentum $J=0$, and so the
selectrons may be produced in the S wave state.  Cross sections for
$\tilde{e}_R$ pair production in $e^-e^-$ and $e^+e^-$ modes are
compared in Fig.~\ref{fig:compare}.  For round beams, the increased
beamstrahlung and decreased luminosity of the $e^-e^-$ mode compromise
this advantage.  However, beamstrahlung is reduced for flat
beams~\cite{susy-Thompson:2000ij}, and mass measurements of order 100
MeV can be achieved with two orders of magnitude less luminosity than
required in $e^+e^-$
collisions~\cite{susy-e-e-inprepFP,susy-e-e-inprep}. Incidentally, the
full arsenal of linear collider modes allows one to extend this mass
measurement to the rest of the first-generation sleptons through a
series of $\beta$ threshold scans: $e^-e^- \to \tilde{e}^-_R
\tilde{e}^-_R$ yields $m_{\tilde{e}_R}$; $e^+e^- \to \tilde{e}^{\pm}_R
\tilde{e}^{\mp}_L$ yields $m_{\tilde{e}_L}$; $e^+e^- \to
\tilde{\chi}^+_1
\tilde{\chi}^-_1$ yields $m_{\tilde{\chi}^{\pm}_1}$; and $e^- \gamma
\to \tilde{\nu}_e \tilde{\chi}^-_1$ yields
$m_{\tilde{\nu}_e}$~\cite{susy-Barger:1998qu}.  The process $e^-e^-
\to \tilde{e}^-_R \tilde{e}^-_R$ may also be used to determine the
Bino mass $M_1$ with high accuracy even for very large
$M_1$~\cite{susy-Feng:1998ud,susy-e-e-inprep}.

\subsubsection{Mixings}

Now that neutrinos are known to mix, lepton flavor is no longer a perfect
symmetry.  Sleptons may also have inter-generational mixings.
Such mixing leads to decays $\tilde{e} \to \mu\tilde{\chi}^0_1, \tau\tilde{\chi}^0_1$ 
and may be searched for at either $e^+e^-$ or $e^-e^-$ colliders.

At $e^+e^-$ colliders, the signal is $e^+e^- \to e^{\pm} \mu^{\mp}\tilde{\chi}^0_1 
\tilde{\chi}^0_1,~e^{\pm} \tau^{\mp}\tilde{\chi}^0_1 \tilde{\chi}^0_1$.  
The backgrounds are $e^+e^- \to W^+ W^-$, $e^+e^- \to
\nu \nu W^+ W^-$, $e^+e^- \to e^{\pm} \nu W^{\mp}$, and $\gamma \gamma
\to W^+ W^-$.  The first two backgrounds may be reduced by $e^-_R$
beam polarization; however, the last two are irreducible.

In the $e^-e^-$ case, the signal is $e^-e^- \to e^- \mu^- \tilde{\chi}^0_1
\tilde{\chi}^0_1,~e^- \tau^- \tilde{\chi}^0_1\tilde{\chi}^0_1$.  Among 
potential backgrounds, $e^-e^- \to W^- W^-$ is
forbidden by total lepton number conservation, $e^-e^- \to \nu \nu W^-
W^-$ and $e^-e^- \to e^- \nu W^-$ may be suppressed by
right-polarizing both $e^-$ beams, and $\gamma \gamma \to W^+ W^-$
does not yield two like-sign leptons.  As a result, the sensitivity of
$e^-e^-$ colliders to slepton flavor violation is much greater than at
$e^+e^-$ colliders, and probes regions of parameter space beyond
current and near-future low-energy experiments searching for $\mu$-$e$ and
 $\tau$-$e$ 
transitions~\cite{susy-Arkani-Hamed:1996au,susy-Arkani-Hamed:1997km}.

\subsubsection{Couplings}

The excellent properties of $e^-e^-$ colliders are also ideal for
exploring selectron gauge couplings. As noted in Section 4, 
precise comparisons of the $e \tilde{e} \tilde{B}$ and $e e B$
couplings provide a model-independent test of supersymmetry.  The
$e\tilde e \tilde B$ coupling is a
non-decoupling observable sensitive to arbitrarily heavy
superpartners.  The nearly background-free environment of $e^-e^-$
colliders makes possible extremely precise measurements of selectron
couplings, surpassing those available at $e^+e^-$
colliders~\cite{susy-Cheng:1998vy}, and may help set the scale for 
far-future colliders in scenarios where some superpartners are extremely heavy.

\subsection{Supersymmetry and $e^- \gamma$ colliders}
\label{sec:egamma}

Even if several neutralinos and charginos have light
masses such that they can be produced in pairs at the LC,
the sleptons might be above threshold for pair production
in $e^+e^-$ collisions.
In this case, the sleptons may be accessible in the
$e^-\gamma$ colliding option in the single-slepton plus 
lighter-neutralino
final state $\tilde{\chi}^0_i\tilde e_{L,R}$.

This reaction was studied 
in~\cite{susy-Choudhury:1995vi,susy-Kiers:1996ux,susy-Barger:1998qu}.
For example, the parameters chosen in ~\cite{susy-Barger:1998qu} lead 
to the masses: $m_{\tilde{\chi}^0_1}=65$~GeV,  
$m_{\tilde{\chi}^\pm_1}= 136$~GeV, $m_{\tilde e_L}=320$~GeV,
$m_{\tilde e_R}=307$~GeV, and $m_{\tilde{\nu}_e}=315$~GeV.
With these values,
pair production of charginos is accessible at a 500 GeV linear
collider but slepton pair production is not.

\begin{figure}[t]
\begin{center}
\epsfig{file=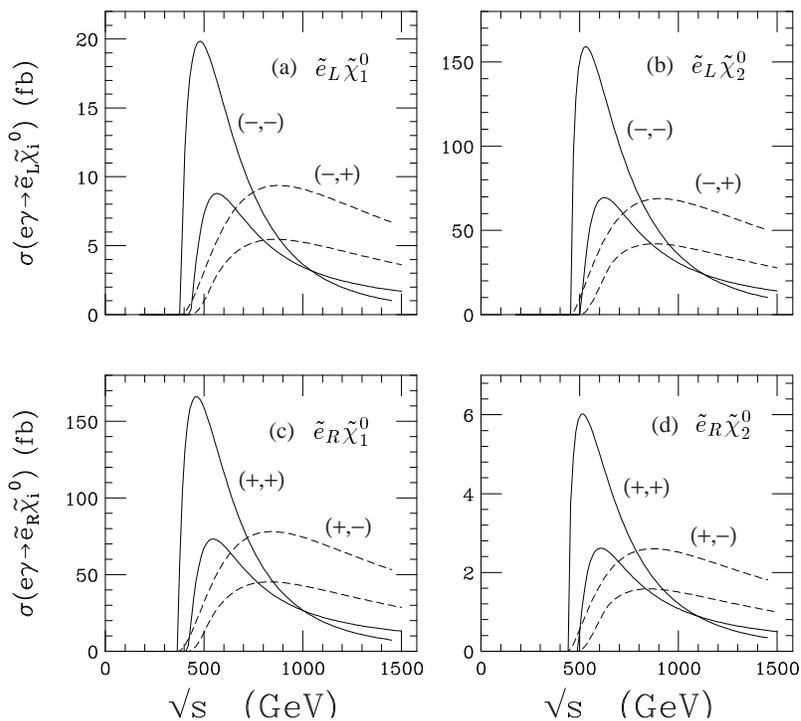, width=0.62\textwidth, angle=90}
\end{center}
\caption{Cross sections for $e^-\gamma\to \s e \s \chi^0$ processes, 
from~\cite{susy-Barger:1998qu}.
The upper two curves show the total cross section (in fb)
for $e^-\gamma\to\tilde e\tilde\chi^0$
versus $\sqrt{s_{e\gamma}}$ (in GeV)
for the SUSY and machine parameters given in the text:
(a)~$\tilde e_L\tilde\chi_1^0$; (b)~$
\tilde e_L\tilde\chi_2^0$; (c)~$\tilde e_R\tilde\chi_1^0$; (d)~$\tilde
e_R\tilde\chi_2^0$.  The solid curves represent $e$, $\gamma$
helicities $(-,-)$ for (a), (b) and $(+,+)$ for (c), (d).  The dashed
curves represent helicities $(-,+)$ for (a), (b) and $(+,-)$ for (c),
(d).  The lower two curves
are corresponding results, convoluted with the backscattered
photon spectrum, versus $\sqrt{s_{ee}}$.
\label{fig:e slepton}}
\end{figure}

Figure~\ref{fig:e slepton} shows the cross sections for slepton-neutralino
production  as a function of the $e^-\gamma$ center-of-mass energy for the 
four different helicity combinations of the incoming electron and photon.  The 
cross section for $\tilde e_R\tilde{\chi}^0_1$ in the $(+,+)$ helicity 
combination is sharply peaked at center-of-mass energies
not far from the threshold.  The signal for this process
is $e^-$ plus missing energy.  
The background~\cite{susy-Choudhury:1995vi,susy-Barger:1998qu}
has a cross section of a  few picobarns and mainly arises
from $W^-\nu\to e^-\nu\nu$. This background
 can be reduced dramatically by using
a polarized $e^-_R$ beam.  With the above parameters, using
polarization and a few judicious kinematic cuts on the final state
particles, the slepton can be discovered and studied.
It has been estimated that both the slepton 
and sneutrino masses can be measured to about 1\% accuracy.

\subsection{Supersymmetry at $\gamma \gamma$ colliders}
\label{sec:gammagamma}

One of the main motivations for the  $\gamma\gamma$ collider option is
to study direct single Higgs production through the $\gamma\gamma h$
coupling.  This motivation is especially powerful in supersymmetry since
most versions of the theory predict a Higgs boson below about 135 GeV.
The motivation is further strengthened by the realization that additional
Higgs states exist in supersymmetry that may not be accessible at the 
LHC or $e^+e^-$ annihilation 
but may be visible in single production from $\gamma\gamma$.  These issues
are discussed in more detail in Chapters 3 and 13.

For direct superpartner pair production, $\gamma\gamma$
collisions also have an important advantage: 
the unambiguous production mode for 
superpartners through photons coupled to charge.  Knowing exactly how a
particle is produced reaps great benefits when analyzing the actual data
recorded by the detectors.  Production cross sections of superpartners
have been calculated most recently by~\cite{Berge:2001cb,Mayer:2000ui}.  
It has been argued~\cite{Mayer:2000ui} that some observables derived from
$\gamma\gamma\to \chi^\pm_1\chi^\mp_1$ production are very useful in
extracting fundamental parameters of the supersymmetric Lagrangian. The
special advantages $\gamma\gamma$ collisions offer supersymmetry deserve
additional careful study.

\section{Comparison with LHC}

\begin{figure}[t]
{\centerline{\epsfxsize=2.9in\epsfbox{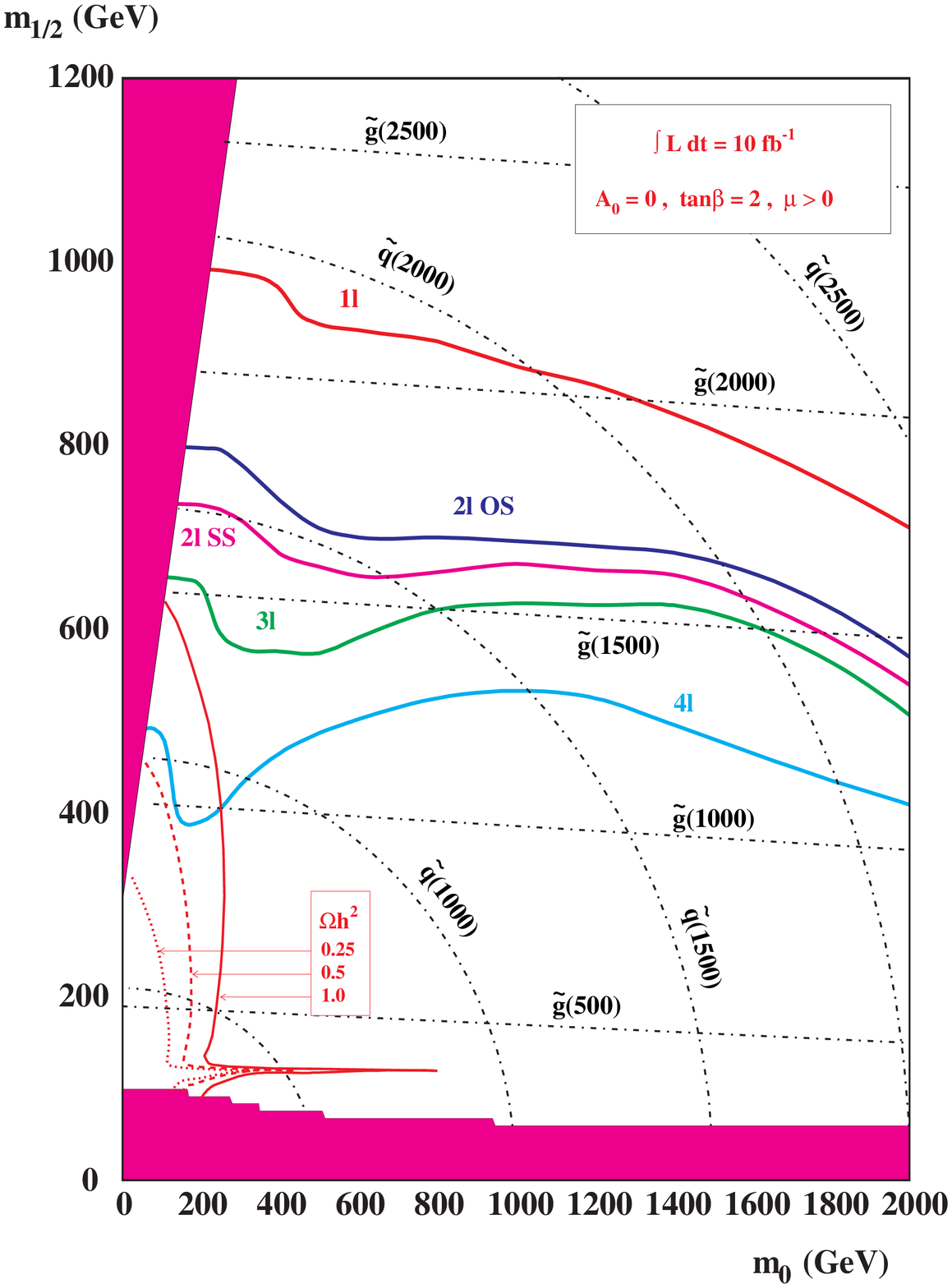}%
             \hfil\epsfxsize=2.9in\epsfbox{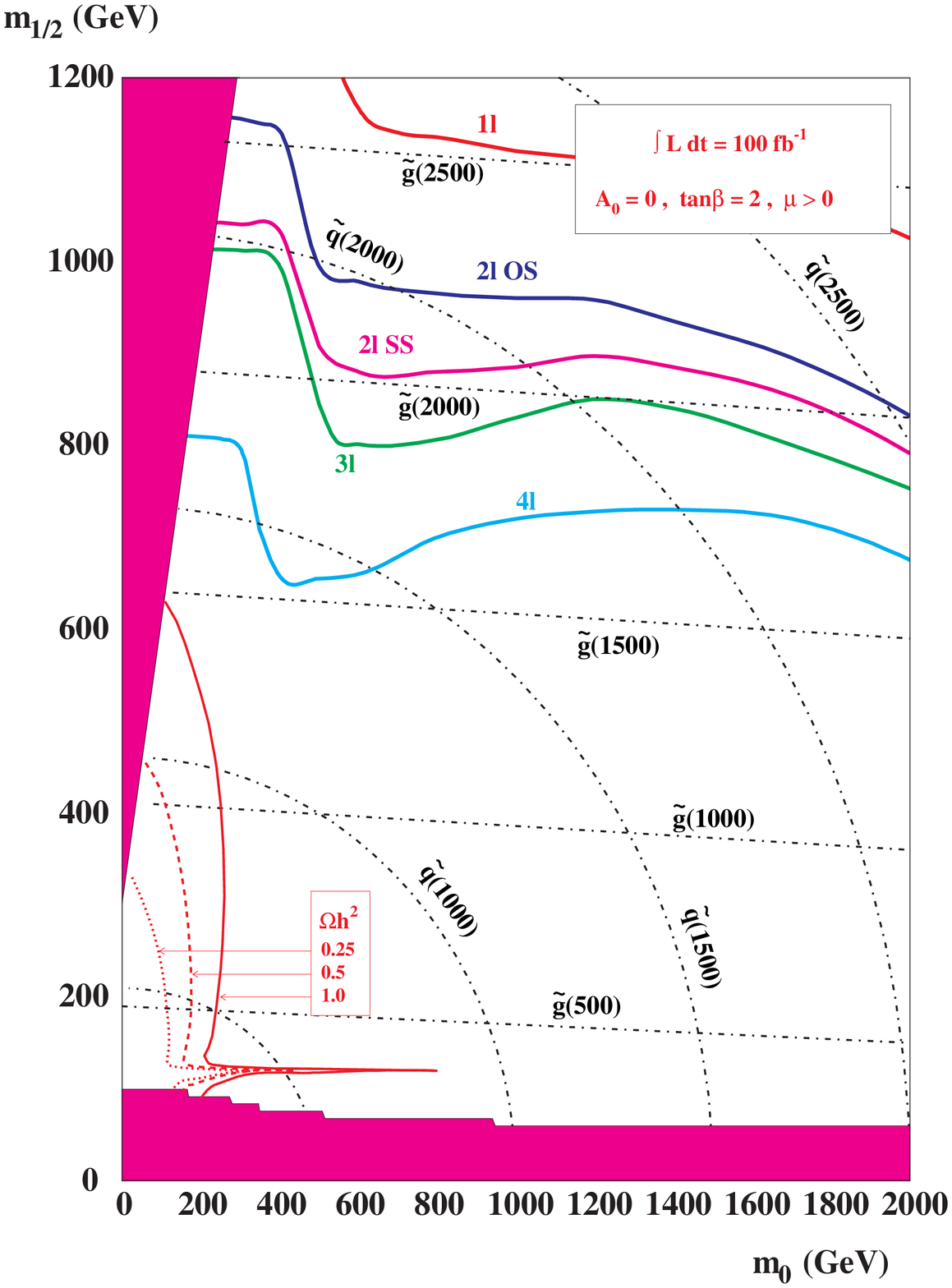}}}
\caption{\label{susy-reach}
Plot of $5\sigma$ reach with multiple jets plus $\etmiss$ plus
leptons in minimal SUGRA model at LHC for $10\,\fbi$ (left) and
$100\,\fbi$ (right)~\cite{susy-Abdullin:1998pm}. 
Also shown are 
contours of the squark and gluino masses and of the cold dark matter density $\Omega h^2$.
}
\end{figure}

If SUSY is relevant to electroweak symmetry breaking, then the arguments
summarized in Section 2 suggest that in many models the gluino and 
some squark masses are less 
than ${\cal O}(1\,\TeV)$. This is also true in most models with SUSY particles 
visible at a $500\,\GeV$ LC. Gluinos and squarks then dominate the LHC SUSY 
cross section, which is of order 10 pb. Since they are strongly 
produced, it is easy to separate SUSY from SM backgrounds provided only that 
the SUSY decays are distinctive. In the MSUGRA model, these decays produce 
multiple jets and $\etmiss$ plus varying numbers of leptons~\cite{susy-Baer:1987au}.  
Figure~\ref{susy-reach} shows the $5\sigma$ reach in this model at the LHC for 
an integrated luminosity of $10\,\fbi$ and $100\,\fbi$~\cite{susy-Abdullin:1998pm}. 
The reach is comfortably more than the expected mass range.

While the reach in Fig.~\ref{susy-reach} has been calculated for a specific 
SUSY model, the multiple jet plus $\etmiss$ signature is generic in most 
R-parity-conserving models. GMSB models can give additional photons or leptons 
or long-lived sleptons with high $p_T$ but $\beta<1$, making the search 
easier~\cite{susy-Baer:1998ve,susy-Baer:2000pe}. R-parity-violating models 
with leptonic $\neu1$ decays also give 
extra leptons and very likely violate $e$-$\mu$ universality. 
R-parity-violating models with $\neu1 \to qqq$ give signals at the LHC with very large 
jet multiplicity, for which the SM background is not well known. 
For such models, it may be necessary to rely on leptons produced in the cascade
decay of the gluinos and squarks. In AMSB models, cascade decays of gluinos and
squarks again lead to a substantial reach for SUSY by the LHC~\cite{susy-Baer:2000bs}.
In all cases, it seems likely that 
SUSY can be discovered at the LHC if the masses are in the expected
range~\cite{susy-AtlasTDR,susy-Baer:1995nq,susy-Baer:1996va}.

\begin{figure}[hbt]
\dofigs{2.9in}{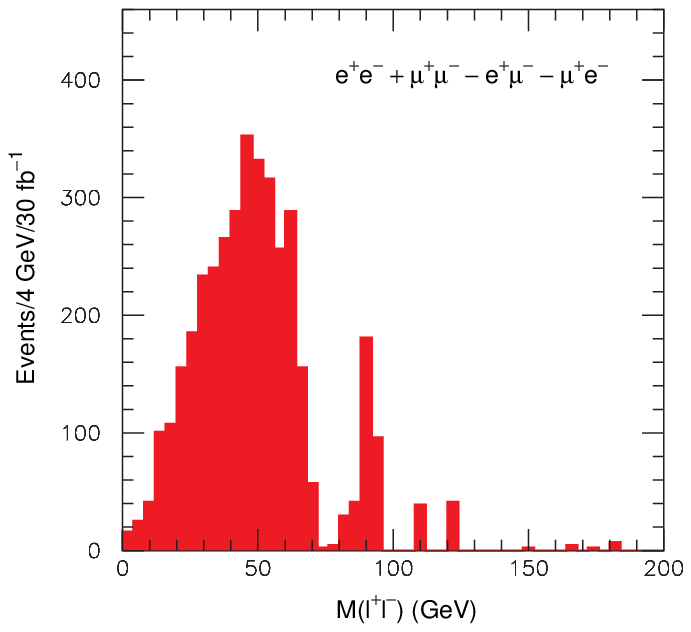}{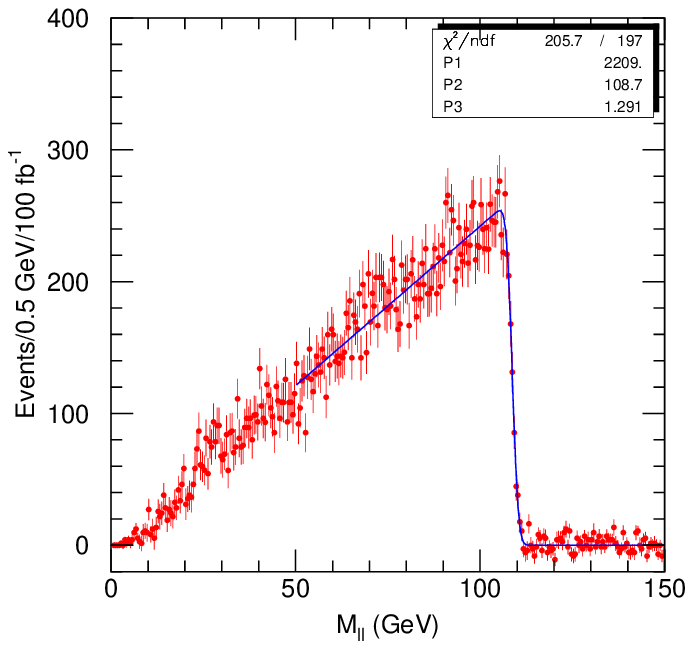}

\caption{Plot of the $e^+e^- + \mu^+\mu^- - e^\pm\mu^\mp$ mass distribution
for LHC SUGRA Point~4 with direct $\neu2 \to \neu1\ell\ell$ decay (left)
and for LHC SUGRA Point~5 with $\neu2 \to \s\ell^\pm\ell^\mp \to
\neu1\ell^+\ell^-$ (right)~\cite{susy-AtlasTDR}.  The event generator
ISAJET is used.  The shape of the peak on the left plot below 70~GeV
should be compared to the shape of the peak in the right plot.  The left plot
also contains a $Z\to\ell^+\ell^-$
signal that comes from heavier gauginos.
\label{susy-mll.eps}}
\end{figure}

The main problem at the LHC is not to observe a signal that deviates from the
SM but to separate the many different channels produced by all the SUSY cascade
decays from the produced squarks~\cite{susy-Baer:1997wa} and gluinos. One 
promising approach is to try to identify particular decay chains and to measure
kinematic endpoints for combinations of the visible particles in
these~\cite{susy-Hinchliffe:1997iu}. For example, the $\ell^+\ell^-$ mass
distribution from $\neu2 \to \neu1\ell^+\ell^-$ has an endpoint that measures
$M_{\neu2}-M_{\neu1}$~\cite{susy-Baer:1994nr}, while the 
distribution from $\neu2 \to \s\ell^\pm
\ell^\mp \to \neu1 \ell^+\ell^-$ has a different shape and measures
$$
M_{\ell\ell}^{\rm max} = \sqrt{ (M_{\neu2}^2 - M_{\s\ell}^2)
(M_{\s\ell}^2 - M_{\neu1}^2) \over M_{\s\ell}^2}  \ .
$$
The flavor-subtraction combination $e^+e^- + \mu^+\mu^- - e^\pm\mu^\mp$
removes backgrounds from two independent decays. Dilepton mass
distributions~\cite{susy-AtlasTDR} after cuts for an example of each
decay are shown in Fig.~\ref{susy-mll.eps}. 

\begin{figure}[hbt]
\dofigs{2.9in}{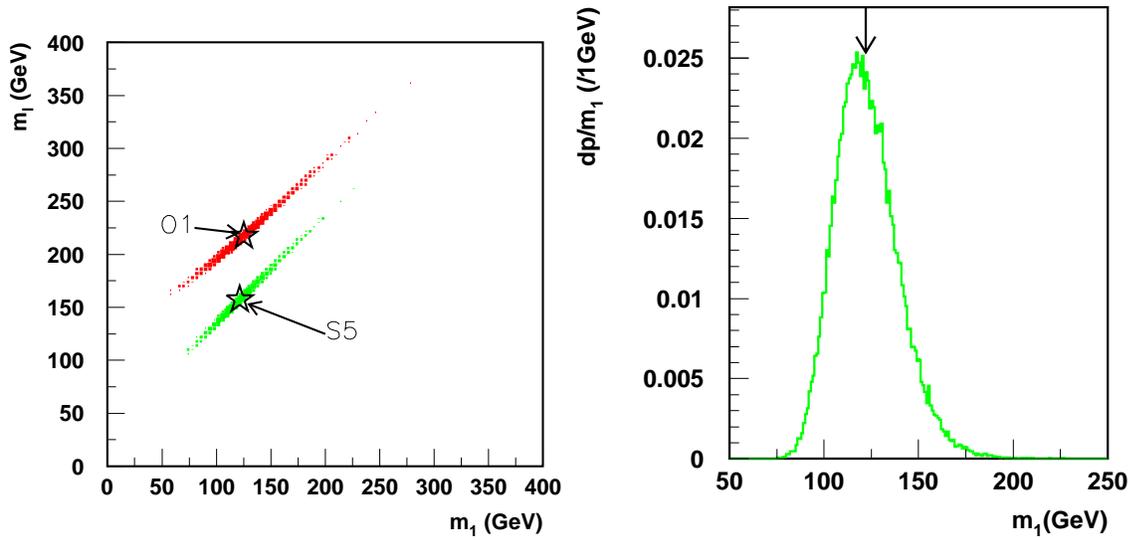}{susy/susy-camb11.epsi}
\caption{Left: Scatter plot of reconstructed values of the $\s\ell_R$ and
$\neu1$ masses for LHC Point~5 (S5) and for a different model (O1) using the
decay chain $\s{q}_L \to \neu2 q \to\s\ell_R \ell q \to \neu1 \ell \ell q$.
Right: Projection of $M_{\neu1}$ for LHC 
Point~5~\cite{susy-Allanach:2000kt}.
\label{susy-Allanach}}
\end{figure}

If a longer decay chain can be identified, then more combinations of masses
can be measured. Consider, for example, the decay chain
$$
\s{q}_L \to \neu2 q \to \s\ell_R^\pm \ell^\mp q \to \neu1\ell^+\ell^-q \ .
$$
For this decay chain, kinematics gives $\ell^+\ell^-$, $\ell^+\ell q$, and two
$\ell q$ endpoints in terms of the masses. If a lower limit is imposed on the
$\ell^+\ell^-$ mass, there is also a $\ell^+\ell^- q$ lower edge. With
suitable cuts all of these can be
measured~\cite{susy-AtlasTDR,susy-Bachacou:2000zb} for the cases considered.
The statistical errors on the measured endpoints are typically comparable to
the systematic limits, ${\cal O}(0.1\%)$ for leptons and ${\cal O}(1\%)$ for jets.
Figure~\ref{susy-Allanach} shows a scatter plot of the resulting $\s\ell_R$
and $\neu1$ masses for LHC SUGRA Point 5 and for a similar point in another
SUSY model with this decay chain~\cite{susy-Allanach:2000kt}.  The relations
between masses are determined with good precision, so these two models are
easily distinguished.  However, the LSP mass is only measured to ${\cal O}(10\%)$.

Analyses such as  these have proved useful for a number of SUSY points in a
variety of SUSY models~\cite{susy-AtlasTDR}. The method seems fairly general:
there is usually at least one distinctive mode --- typically $\neu2 \to \neu1
\ell^+\ell^-$, $\neu2 \to \s\ell_R^\pm\ell^\mp$, or $\neu2 \to \neu1 h \to 
\neu1 b \bar b$ --- from which to start. But some points are much more 
difficult than others. For example, in MSUGRA with $\tan\beta\gg1$ it is 
possible to choose parameters such that the only allowed 2-body decays of 
$\neu2$ and $\s\chi_1^\pm$ are $\s\tau_1^\pm\tau^\mp$ and 
$\s\tau_1^\pm\nu_\tau$~\cite{susy-Baer:1999sz} 
respectively.\footnote{The simple class of such models
considered in \protect\cite{susy-AtlasTDR}, however, gives an 
excessively large
contribution to $g_\mu-2$~\protect\cite{susy-Brown:2001mg}.} 
These modes then have branching ratios 
in excess of 99\%.  While it is possible to identify and to measure hadronic 
$\tau$ decays~\cite{susy-AtlasTDR}, the measurements are much less precise than
those involving leptons. Even if $\tau$ decays are not dominant, they may be 
important, since they can provide information on $\s\tau_L-\s\tau_R$ and 
gaugino-Higgsino mixing.

If SUSY is found at the LHC, the SUSY events will contain much more
information than just endpoints like those described above. For example,
while it is not possible to reconstruct $\s\chi_{1}^\pm$ decays in the
same way because of the missing neutrino, one can get information about
the chargino mass by studying $M_{\ell q}$ and other distributions for
1-lepton events. Cross sections and branching ratios can also be
measured; interpretation of these will be limited by the theoretical
errors on the calculation of cross sections and acceptances. Without
real experimental data, it is difficult to assess such theoretical
systematic errors.

SUSY signatures at the LHC typically come from a combination of many
SUSY particles, so the analysis is considerably more complicated than
that at a LC. However, the initial steps at the LHC are fairly clear.
First, one will look for a deviation from the SM in
inclusive distributions such as multiple jets plus $\etmiss$, perhaps
accompanied by leptons and/or photons. If a signal consistent with SUSY
is found, it should determine both the mass
scale~\cite{susy-AtlasTDR,susy-Tovey:2000wk} and the qualitative
nature of the signal. (As a simple example, in a GMSB model with a
long-lived slepton NLSP, SUSY events would contain two high-$p_T$
particles with $\beta<1$.) Next, one will look for various kinematic
endpoints like those described above and use them further to constrain
the SUSY masses. After this, one will look at more model-dependent
quantities such as kinematic distributions, cross sections, and
branching ratios. These seem difficult to assess without real data.

This program is likely to provide considerable information about
gluinos, squarks, and their primary decay products, including $\neu1$,
$\neu2$, $\s\chi^\pm$, and any sleptons that occur in their decays. It
is more dangerous to predict what cannot be done, but there are
measurements that appear difficult at the LHC and that could be done at
a $500\,\GeV$ LC. For example:
\begin{itemize}

\item While it is possible to measure the $\neu1$ mass at the LHC in
favorable cases, it seems difficult to reduce the error below
${\cal O}(10\%)$. If any visible SUSY particle is produced at a LC, the error
on $M_{\neu1}$ should be ${\cal O}(1\%)$.
\item Sleptons that are not produced in $\neu2$ or $\s\chi_1^\pm$ decays
are difficult to study at the LHC: both the Drell-Yan process and decays
of heavier gauginos typically give very small rates~\cite{susy-Baer:1994ew}. 
They can be precisely measured at a LC.
\item Distinguishing $\s\ell_L$ from $\s\ell_R$ appears very difficult at
the LHC except perhaps for $\s\tau$'s, but this is straightforward at a LC
using the polarized beam.
\item Hadronic $\tau$ decays are easier to identify and to measure at a
LC because there is no underlying hadronic event.
\item Branching ratios currently seem difficult to measure with high
precision at the LHC: both the production cross sections and the
acceptance have theoretical uncertainties of ${\cal O}(10\%)$. In particular,
it seems difficult to make precise tests of SUSY relations among
couplings.
\end{itemize}
More generally, while the LHC seems sure to discover SUSY at the TeV
scale if it exists, the measurements of SUSY that can be made there 
depend on the SUSY model. A LC can provide precise, detailed
measurements of any kinematically accessible SUSY particles. Ultimately,
one will want such measurements for the entire SUSY spectrum.

\end{document}